\documentclass[nofootinbib,
twocolumn,
superscriptaddress,
 amsmath,amssymb,
 aps,prd
]{revtex4-2}
\usepackage[]{hyperref}
\usepackage[nameinlink]{cleveref} 
\usepackage[utf8]{inputenc}
\usepackage{amsmath,amssymb,amsthm,amstext}
\usepackage{mathtools}
\usepackage{siunitx}
\usepackage{braket}
\usepackage{verbatim}
\usepackage{multirow}
\usepackage{dsfont}
\usepackage{siunitx}
\usepackage{float}
\usepackage{graphicx,subfigure}
\usepackage{changebar}
\usepackage{graphicx}
\usepackage{dcolumn}
\usepackage{bm}
\usepackage{orcidlink}
\usepackage{xcolor}
\usepackage{slashed}
\usepackage{feynmp-auto}
\usepackage{crossreftools}
\usepackage{hyperref}
\usepackage{enumitem}
\definecolor{bubbles}{rgb}{0.91, 1.0, 1.0}
\definecolor{brightcerulean}{rgb}{0.11, 0.67, 0.84}
\definecolor{mbcol}{rgb}{1, 0, 1}
\usepackage[colorinlistoftodos]{todonotes}
\usepackage{soul}

\graphicspath{{figs/}}

\newcommand{\vect}[1]{\boldsymbol{#1}}
\newcommand{\Eq}[1]{Eq.~(\ref{#1})}
\newcommand\scalemath[2]{\scalebox{#1}{\mbox{\ensuremath{\displaystyle #2}}}}
\DeclareMathOperator*{\SumInt}{%
	\mathchoice%
	{\ooalign{\raisebox{.15\height}{\scalebox{0.9}{$\textstyle\sum$}}\cr\hidewidth$\displaystyle\int$\hidewidth\cr}}
	{\ooalign{\raisebox{.14\height}{\scalebox{.7}{$\textstyle\sum$}}\cr\hidewidth$\textstyle\int$\hidewidth\cr}}
	{\ooalign{\raisebox{.2\height}{\scalebox{.6}{$\scriptstyle\sum$}}\cr$\scriptstyle\int$\cr}}
	{\ooalign{\raisebox{.2\height}{\scalebox{.6}{$\scriptstyle\sum$}}\cr$\scriptstyle\int$\cr}}
}

\newcommand{\optionaldesc}[2]{%
  \phantomsection
  #1\protected@edef\@currentlabel{#1}\label{#2}%
}

\newcounter{prop}[section]
\renewcommand*{\theprop}{\thesection.\arabic{prop}}

\begin{document}

\title{Renormalization-group consistent treatment of 
color superconductivity in the NJL model}

\author{Hosein Gholami \orcidlink{0009-0003-3194-926X}}
\email{mohammadhossein.gholami@tu-darmstadt.de}
\affiliation{Technische Universität Darmstadt, Fachbereich Physik, Institut für Kernphysik,
Theoriezentrum, Schlossgartenstr.~2 D-64289 Darmstadt, Germany
}
\author{Marco Hofmann \orcidlink{0000-0002-4947-1693}}
\email{marco.hofmann@tu-darmstadt.de}
\affiliation{Technische Universität Darmstadt, Fachbereich Physik, Institut für Kernphysik,
Theoriezentrum, Schlossgartenstr.~2 D-64289 Darmstadt, Germany
}
\author{Michael Buballa \orcidlink{0000-0003-3747-6865}}
\email{michael.buballa@tu-darmstadt.de }
\affiliation{Technische Universität Darmstadt, Fachbereich Physik, Institut für Kernphysik,
Theoriezentrum, Schlossgartenstr.~2 D-64289 Darmstadt, Germany
}
\affiliation{Helmholtz Forschungsakademie Hessen f\"{u}r FAIR (HFHF), 
	GSI Helmholtzzentrum f\"{u}r Schwerionenforschung,
	Campus Darmstadt,
	64289 Darmstadt,
	Germany}

\date{\today}

\begin{abstract}
The Nambu-Jona-Lasinio (NJL) model and specifically its extension to color superconductivity (CSC) is a popular effective model for investigating dense quark matter. However, the reliability of its results is challenged by cutoff artifacts, which emerge if temperature or chemical potential are of the order of the cutoff energy scales. In this work, we generalize an idea from \href{https://scipost.org/SciPostPhys.6.5.056}{[Braun et al. SciPost Phys., 6:056, 2019]}, which is based on the requirement of renormalization-group (RG) consistency and has successfully been applied to the two-flavor Quark-Meson-Diquark model, to the NJL model for electrically and color-neutral three-flavor color-superconducting quark matter. To this end, we analyze the medium divergences of the model and eliminate them by appropriate counterterms, introducing three different schemes. We show that the RG-consistent treatment removes the cutoff artifacts of the conventional regularization and enables the investigation of CSC matter at higher densities by the model. Our studies reveal the emergence of a so-called $d$-quark superconducting (dSC) phase within the melting pattern of the Color-Flavor Locked (CFL) phase at high chemical potentials, consistent with earlier Ginzburg-Landau analyses.
\end{abstract}

\maketitle

\section{Introduction}

The phase structure of strong-interaction matter at densities of a few times nuclear saturation density and temperatures from zero up to about 100~MeV is relevant for understanding the inner structure of neutron stars, proto-neutron stars and neutron-star mergers. 
For very high densities, where Quantum Chromodynamics (QCD) becomes asymptotically free, it was shown long ago in a weak-coupling expansion that the ground state of matter is a color superconductor (CSC)
\cite{Son:1998uk,Schafer:1999jg,Pisarski:1999tv}, i.e., a state in which quarks form Cooper pairs
(``diquark condensates''); 
for a review on CSC, see Ref.~\cite{Alford:2007xm}.
The most attractive pairing channel has spin 0 and is antisymmetric in both color and flavor, implying that the Cooper pairs consist of quarks with unequal colors and flavors. 
Restricting the analysis to three quark flavors, the favored pairing pattern is then the so-called color-flavor locked (CFL) phase~\cite{Schafer:1999fe,Shovkovy:1999mr}, where the quarks of all colors (red, green and blue) and flavors (up, down and strange) are paired~\cite{Alford:1998mk}.

Unfortunately, the densities encountered in neutron stars and mergers are not high enough for the weak-coupling expansion to be valid. 
Qualitatively it is clear that, if we start in the CFL phase and continuously lower the density, eventually the strange-quark mass $M_s$ can no longer be neglected against the quark chemical potential $\mu$. At some point the pairing between strange and non-strange quarks might become energetically disfavored, possibly giving rise to the two-flavor superconducting (2SC) phase, where only up and down quarks are paired~\cite{AlfordUnlocking}. 
In neutron stars additional complications arise from the fact that the matter has to be electrically neutral, resulting in unequal Fermi momenta of up and down quarks, hindering their pairing as well~\cite{Alford:2002kj,Rajagopal_stressed_pairing}.
The resulting phase structure in this regime could thus be rather complex.

In order to investigate this more quantitatively, nonperturbative approaches are required. 
While lattice QCD is not applicable to the regime of low temperatures and nonzero chemical potential because of the sign problem, there are interesting studies of CSC phases in QCD using Dyson-Schwinger equations (DSEs)
\cite{Nickel:2006vf, Nickel:2006kc,Nickel:2008ef,Muller:2013pya,Muller:2016fdr}. 
However, these calculations quickly become very involved and therefore have not yet reached the same sophistication as DSE studies of non-superconducting matter, see, e.g., Ref.~\cite{Gunkel:2021oya} (for a review on DSE studies of the QCD phase diagram see Ref.~\cite{Fischer:2018sdj}).
Moreover, it is generally very difficult to calculate the pressure and, hence, the equation of state from DSEs. 
Therefore many authors have resorted to QCD-inspired models, in particular the Nambu--Jona-Lasinio (NJL) model, to describe CSC matter, see Ref.~\cite{BUBALLAHABIL} for a review and
Refs.~\cite{Ruester:2005jc,Blaschke:2005uj,Abuki:2005ms} for early studies of NJL phase diagrams with CSC under compact-star conditions. 
Although these models have limited predictive power, they are well suited for explorative studies and as a basis for astrophysical applications.
A particularly interesting feature of the NJL model is that both pairing gaps and effective (``constituent'') quark masses are dynamically generated quantities and as such temperature and density dependent~\cite{Buballa:2001gj}. 

On the other hand the NJL model has the well-known drawback that it is non-renorma\-lizable, i.e., the cutoff parameters needed to regularize divergent momentum integrals cannot be removed by absorbing them in renormalized quantities but must be considered as part of the model~\cite{Klevansky:1992qe}. 
The cutoff scale then sets a limit to the energy range in which the model can be applied. 
Hence, while the NJL model is a perfectly valid low-energy model in vacuum, one has to be careful when applying it to temperatures or chemical potentials of the order of the cutoff. 
In practice, however, it turns out that the Fermi momenta in the CSC region are never \textit{much} smaller than typical cutoff values. 
For instance in Refs.~\cite{Ruester:2005jc}
and \cite{Blaschke:2005uj}
the momentum cutoff is roughly 600~MeV, while the onset chemical potential of the CSC regime is at about 350~MeV and the investigations are performed up to 500~MeV
or even 550~MeV, respectively.
In particular at the upper end of this region we should thus expect strong cutoff artifacts. 
This is not just an academic problem but in order to study the inner core of compact stars or neutron-star mergers we need to know the equation of state in this problematic region of the phase diagram.

As we will discuss in more details, there are indeed several
qualitative cutoff artifacts one finds in model calculations of this type: 
First, while at fixed temperature the pairing gaps in a given phase should monotonically increase with density, one finds that, as $\mu$ approaches the cutoff, eventually the gaps decrease again and finally even vanish \cite{Farias:2005cr}.
Closely related to this, also the critical temperatures of a given CSC phase do not rise monotonically with the chemical potential but eventually reach a maximum and decrease again.

A third issue concerns the sequential melting of the diquark condensates in the CFL phase with increasing temperature. 
The melting pattern which has been predicted by a Ginzburg-Landau (GL) analysis for small $M_s^2/\mu^2$ \cite{Iida:2003cc,Iida:2004cj} does not agree with the pattern found in NJL-model calculations ~\cite{Ruester:2005jc,Blaschke:2005uj,Abuki:2005ms}. 
Whether or not this can be explained by higher-order corrections in the GL expansion \cite{Abuki:2005ms} (see also Ref.~\cite{Fukushima:2004zq}), could not be fully resolved, since, due to cutoff artifacts, 
the NJL calculations cannot be extended to density regimes where $M_s^2/\mu^2$ is sufficiently small.

In general cutoff artifacts are not restricted to CSC phases. 
For instance the cutoff can also severely affect the high-temperature thermodynamics, violating the Stefan-Boltzmann limit and even causality \cite{Pasqualotto:2023hho}. 
In non-superconducting matter a standard way to cure this problem is to separate the
effective potential into a vacuum part and a medium part, and to regularize only the former. 
This is possible because the separation is relatively straightforward and the medium part is finite and does not need to be regularized. 
Unfortunately, this procedure does not work for CSC matter. In this case vacuum and medium parts do not separate naturally. 
But even if we isolate a vacuum part by taking the $T=\mu = 0$ - limit of the effective potential, after subtracting it, the remaining medium part is still divergent and needs some  regularization. Most authors so far have therefore decided to regularize the entire expression, hoping that the cutoff artifacts are not relevant in the region the model is applied.

In Ref. \cite{Farias:2005cr}, the authors proposed an implicit regularization scheme for general four-fermion interactions with color superconductivity. There, the divergences in the gap equations were isolated in $\mu-$independent integrals which were then fitted to physical observables in vacuum by the use of scaling relations. This approach was later applied for a color and charge neutral two flavor NJL model \cite{Duarte:2018kfd}. 

Quite recently, a new approach to this issue was proposed in Ref.~\cite{Braun:2018svj}, based on the requirement of renormalization-group (RG) consistency.
The concept originally comes from the field of the functional renormalization group (FRG), where it states that the full quantum effective action of a given theory must not depend on the ultraviolet scale at which the theory is initialized. 
Although much more general, it was shown in Ref.~\cite{Braun:2018svj} that it can also be applied to simple mean-field models.  
Most interesting for us, the authors developed an RG-consistent regularization scheme for the two-flavor Quark-Meson-Diquark model and showed, among others, that the critical temperature of the 2SC phase keeps rising at large chemical potential. In Ref. \cite{Braun:2022olp} the RG-consistent regularization scheme was applied to study a charge- and color-neutral two-flavor diquark model at zero temperature.
In this paper we generalize this idea to the three-flavor NJL model including neutrality constraints and in principle arbitrary pairing patterns. As a specific example we consider the model of Ref.~\cite{Ruester:2005jc} and compare the results of the RG-consistent scheme with those where all momentum integrals were regularized with the same cutoff as in the original publication. In particular we will show that the cutoff artifacts mentioned above are removed by the RG-consistent treatment.
 
 The remainder of the paper is organized as follows. 
In Sec.~\ref{sec:model} we review the three-flavor NJL model for neutral color superconducting matter used in Ref.~\cite{Ruester:2005jc} with conventional cutoff regularization.
The concept of RG consistency and its application to regularized mean-field models is reviewed in Sec.~\ref{sec:RG_consistency}. Here our main goal is to convey the ideas of the original work \cite{Braun:2018svj} to mean-field practitioners without major FRG background. In Sec.~\ref{sec:meddiv} we investigate two-flavor color-superconducting matter. We identify a ``medium divergence'' which emerges in the presence of a diquark condensate and show that it can be eliminated by a prescription proposed in Ref.~\cite{Braun:2018svj}.
Our main theoretical results can be found in Sec.~\ref{sec:renopro}, where we present a general framework for the RG-consistent treatment of the divergences in the full three-flavor model with neutrality conditions. Based on a renormalization picture, we remove the divergences by introducing counterterms. For this, we discuss three different schemes.
In Sec.~\ref{sec:results} we show numerical results of this implementation and compare the phase diagram with the result obtained in Ref.~\cite{Ruester:2005jc} with conventional cutoff regularization.
Our conclusions are drawn in Sec.~\ref{sec:conclusion}.

\section{NJL-type model for color superconductivity}\label{sec:model}

As outlined in the Introduction, we want to generalize the ideas of Ref.~\cite{Braun:2018svj} to the three-flavor NJL model, taking the model of Ref.~\cite{Ruester:2005jc} as a specific example. In this section we therefore summarize the main features of that model as far as necessary to follow our discussion. For further details we refer the reader to Ref.~\cite{Ruester:2005jc}. 

The Lagrangian of the model can be written as 
\begin{equation}
\mathcal{L}=\mathcal{L}_0 + \mathcal{L}_{\bar q q}+\mathcal{L}_{qq},
\label{eq:Lagrangian}
\end{equation}
where 
\begin{equation}
\label{eq:L0}
\mathcal{L}_0 = \bar{\psi}(i\slashed{\partial}+\gamma^0\hat{\mu}-\hat{m})\psi 
\end{equation}
describes the free propagation of quarks. Here $\psi$ denotes a spinor field with three flavor ($u,d,s$) and three color ($r,g,b$) degrees of freedom. 
$\hat{m}=\text{diag}_f(m_u,m_d,m_s)$ is the matrix of bare quark masses, which is diagonal in flavor space. In addition $\mathcal{L}_0$ includes a matrix of chemical potentials $\hat{\mu}$, which is diagonal in color and flavor space.

The quarks interact with each other via local vertices which are defined by the two other terms on the right-hand side of Eq.~(\ref{eq:Lagrangian}). The term
\begin{eqnarray}
\mathcal{L}_{\bar q q} &=&
G_S\sum_{a=0}^8\left[(\bar{\psi}\tau_a\psi)^2 + (\bar{\psi} i \gamma_5 \tau_a \psi)^2\right]
\nonumber\\&&-K [\text{det}_{\text{f}}(\bar{\psi}(\mathds{1}+\gamma_5)\psi) + \text{det}_{\text{f}}(\bar{\psi}(\mathds{1}-\gamma_5)\psi)]
\label{eq:Lagrangian_qbarq}
\end{eqnarray}
corresponds to the standard three-flavor NJL-model interaction, see, e.g., Ref.~\cite{Rehberg:1995kh}.
It consists of a $U(3)_L \times U(3)_R$ invariant combination of scalar and pseudoscalar four-point vertices with coupling constant $G_S$ and a six-point interaction, which is $SU(3)_L \times SU(3)_R$ symmetric but explicitly breaks $U_A(1)$, scaled with the coupling constant $K$ \cite{Kobayashi:1970ji,tHooft:1976rip}. 
Here and throughout the paper $\tau_a$ denote the Gell-Mann matrices in flavor space for $a=1,..,8$ complemented by $\tau_0=\sqrt{2/3}\mathds{1}_f$.

The last interaction term,
\begin{equation}
\label{eq:Lqq}
\mathcal{L}_{qq}= G_D\sum_{\gamma,c}(\bar{\psi}^a_\alpha i\gamma_5 \epsilon^{\alpha\beta\gamma} \epsilon_{abc}(\psi_C)^b_\beta)
    ((\bar{\psi}_C)^r_\rho i\gamma_5 \epsilon^{\rho\sigma\gamma} \epsilon_{rsc} \psi^s_\sigma),
\end{equation}
is introduced to allow for quark pairing in the Hartree approximation. 
Here $\psi^a_\alpha$ denotes the flavor ($\alpha=u,d,s$) and 
color ($a=r,g,b$) components of the quark spinor.
The charge-conjugate spinors are defined as $\psi_C=C\bar{\psi}^T$ and $\bar{\psi}_C=\psi^TC$
with the charge conjugation operator $C=i\gamma^2\gamma^0$.
The Levi-Civita tensors ensure antisymmetry in color and flavor space.
The quantum numbers of the bilinears then correspond to scalar diquarks in the color and flavor antitriplet channels, which are relevant for both 2SC and CFL pairing. 
In principle we should also have the corresponding pseudoscalar terms to preserve chiral symmetry. Here we omit these terms for simplicity, since they do not contribute in the Hartree approximation, which we apply in this work.

\subsection{Mean-field effective potential}

The thermodynamic properties of the model are encoded in the grand potential per volume $\Omega(\vect{\mu},T)$, where $T$ is the temperature and $\vect{\mu}$ denotes a vector whose components are a set of independent chemical potentials to be specified later.
In the following we want to calculate $\Omega(\vect{\mu},T)$ in the mean-field (Hartree) approximation. To that end we consider the three chiral condensates
\begin{equation}
\phi_f=\langle \bar{\psi}_f \psi_f \rangle
\end{equation}
for flavor $f = u, d, s$
and the three diquark condensates 
\begin{equation}
\label{eq:DeltaA}
\Delta_A=-2G_D\langle \bar{\psi}^a_\alpha i\gamma_5 \epsilon^{\alpha\beta A} \epsilon_{abA}(\psi_C)^b_\beta \rangle
\end{equation}
with $A = 1,2,3$.
The latter are the order parameters for the different color-superconducting phases listed in Table \ref{tab:gaps}. 
The former are related to the dynamical quark masses, which are given by
\begin{equation}
M_\alpha = m_\alpha - 4G_s \phi_\alpha + 2K \phi_\beta \phi_\gamma,
\end{equation}
where $(\alpha,\beta,\gamma)$ is any permutation of $(u,d,s)$.
For convenience, we combine the six condensates to a vector $\vect\chi=(\phi_u,\phi_d,\phi_s,\Delta_1,\Delta_2,\Delta_3)$. We also define two vectors for the physical gaps $\vect M=(M_u,M_d,M_s)$ and $\vect\Delta=(\Delta_1,\Delta_2,\Delta_3)$.

\begin{table*}[t]
\caption{Nonzero diquark condensates and flavor-color structure of the corresponding pairs in different color-superconducting phases.}
                \label{tab:gaps}
				\begin{tabular}{|c | c | c |}
					\hline
					\textbf{Phase} & \textbf{Nonzero diquark condensates}& \textbf{Pairing}  \\
					\hline
      2SC$_{ds}$ & $\Delta_1$ & $(d_g,s_b),(d_b,s_g)$ \\
     2SC$_{us}$ & $\Delta_2$ & $(u_r,s_b),(u_b,s_r)$  \\
     2SC & $\Delta_3$& $(u_r,d_g),(u_g,d_r)$  \\
     uSC & $\Delta_2, \Delta_3$ & $(u_r,s_b),(u_b,s_r),(u_r,d_g),(u_g,d_r)$ \\
     dSC & $\Delta_1, \Delta_3$ & $(d_g,s_b),(d_b,s_g),(u_r,d_g),(u_g,d_r)$ \\
     sSC & $\Delta_1, \Delta_2$ & $(d_g,s_b),(d_b,s_g),(u_r,s_b),(u_b,s_r)$ \\
					CFL & $\Delta_1, \Delta_2, \Delta_3$ & $(d_g,s_b),(d_b,s_g),(u_r,s_b),(u_b,s_r),(u_r,d_g),(u_g,d_r)$ \\
					\hline
				\end{tabular}
			\end{table*}

For a clear notation, we distinguish between the physical grand potential defined above and the \textit{effective} grand potential (per volume) 
$\Omega_\text{eff}(\vect{\mu},T,\vect\chi)$ that depends, besides on $\vect{\mu}$ and $T$, also on the mean fields summarized in $\vect\chi$. 
The relation between the two potentials is given by
\begin{equation}
    \Omega(\vect{\mu},T) = \Omega_\text{eff}(\vect{\mu},T,\vect\chi = \bar{\vect\chi})
\end{equation}
where $\bar{\vect\chi}=(\bar{\phi}_u,\bar{\phi}_d,\bar{\phi}_s,\bar\Delta_1,\bar\Delta_2,\bar\Delta_3)$ are the condensates which minimize $\Omega_\text{eff}$ at the given chemical potentials and temperature, i.e., they are the physical solutions of the gap equations
\begin{align}\label{eq:gapeq}
    \frac{\partial \Omega_\text{eff}}{\partial \phi_f}\bigg|_{\phi_f=\bar{\phi}_f}=\frac{\partial \Omega_\text{eff}}{\partial \Delta_A}\bigg|_{\Delta_A=\bar{\Delta}_A}=0 .
\end{align}

In order to take the diquark condensates into account in a convenient way, we use the Nambu-Gorkov (NG) formalism, i.e., we artificially double the degrees of freedom by combining $\psi$ and $\psi_C$ to a NG spinor $\Psi$.
The mean-field effective potential per volume can then be brought into the general form 
\begin{equation}\label{eq:Omega_eff}
\Omega_{\text{eff}}(\vect{\mu},T,\vect\chi)
=-\int\frac{d^3p}{(2\pi)^3}\mathcal{A}(\vect p;\vect{\mu},T,\vect\chi)\,+\, \mathcal{V}(\vect\chi)
\end{equation}
with the potential term for the condensates
\begin{equation}\label{eq:V_def}
    \mathcal{V}(\vect\chi)=2G_S(\phi_u^2+\phi_d^2+\phi_s^2)-4K\phi_u\phi_d\phi_s+\frac{1}{4G_D}\sum_{A=1}^3\vert\Delta_A\vert^2
\end{equation}
and the momentum integrand
\begin{equation}
\mathcal{A}(\vect p;\vect{\mu},T,\vect\chi)=
\frac{1}{2}T\sum\limits_n \text{tr}\ln\frac{S^{-1}(P;\vect{\mu},\vect\chi)}{T}.
\label{eq:Adef}
\end{equation}
Here $S^{-1}$ denotes the inverse dressed quark propagator, which is a function of $P=(i\omega_n,\vect p)$ with the three-momentum $\vect{p}$ and the fermionic Matsubara frequency
$\omega_n = (2n+1)\pi T$. It is a 
$72 \times 72$-matrix in flavor, color, Dirac and Nambu-Gorkov space, and the trace is to be taken in this space.
The factor $1/2$ in front corrects for the  artificial doubling of the degrees of freedom in NG formalism. To evaluate the trace logarithm, $S^{-1}$ can be decomposed into a block-diagonal structure of seven block matrices and further diagonalized numerically (see Appendix \ref{sec:appendix_pairing_pattern} and Ref.~\cite{Ruester:2005jc} for details). 
This yields
\begin{eqnarray}
&&\mathcal{A}(\vect p;\vect{\mu},T,\vect\chi)
=\nonumber\\&& \sum_{j=1}^{18}\left(
    \epsilon_j(\vect{p};\vect{\mu},\vect\chi) + 2T\ln\left(
   1+e^{-\frac{\epsilon_j(\vect{p},\vect{\mu},\vect\chi)}{T}}\right)\right),
   \label{eq:Aform}
\end{eqnarray}
where the quasiparticle dispersion relations $\epsilon_j(\vect{p};\vect{\mu},\vect\chi)$ are eigenvalues of the block matrices. These come in pairs of positive and negative signs, which was taken into account in Eq.~\eqref{eq:Aform} by restricting the sum to the 36 positive eigenvalues. In addition, the eigenvalues have a twofold spin degeneracy, which was used to reduce the number of terms by another factor of 2. Note that, depending on the thermodynamic phase, there can be further degeneracies.

\subsection{Choice of the chemical potentials}
\label{subsec:chempot}

In the NJL-type model defined above both flavor and color are globally conserved quantum numbers, so that in principle we can define nine independent chemical potentials $\mu_{fc}$ for the three flavors $f$ and the three colors $c$. In fact, the formalism of the RG-consistent regularization we will discuss in this paper will be valid for this most general case. 
On the other hand, for the numerical applications we will follow Ref.~\cite{Ruester:2005jc} and consider the special set of chemical potentials appropriate for the description of cold compact stars. 
In this case, because of weak decays, quark flavor is no longer conserved, but the total quark number $n$ (or baryon number $n_B = n/3$), the quark color charges and the total electric charge $n_Q$ are conserved quantities. 
The chemical potential matrix in Eq.~(\ref{eq:L0}) can then be written as
\begin{equation}\label{eq:mu_matrix}
    \hat{\mu}^{\alpha\beta}_{ab}=(\mu\delta^{\alpha\beta}+\mu_Q Q^{\alpha\beta})\delta_{ab}+[\mu_3 (\lambda_3)_{ab}+ \mu_8 (\lambda_8)_{ab}]\delta^{\alpha\beta}
\end{equation}
with the electric charge operator $Q=\text{diag}_f(2/3,-1/3,-1/3)$ and the third and eighth Gell-Mann matrices in color space, $\lambda_3$ and $\lambda_8$. Here, $\mu=\mu_B/3$ is the quark number chemical potential and we refer to $\mu_Q$ as the electric chemical potential and $\mu_3$ and $\mu_8$ as color chemical potentials.

For consistency, we add a leptonic part $\Omega_L$ to the grand potential,
\begin{equation}
\Omega_\text{eff}^\text{total}(\vect{\mu},T,\vect{\chi})
=
\Omega_\text{eff}(\vect{\mu},T,\vect{\chi})
+ 
\Omega_L(\vect{\mu},T).
\end{equation}
Specifically we consider a free Fermi gas of electrons and muons,
\begin{eqnarray} &&\Omega_L(\vect{\mu},T)=\nonumber\\
&&-2T\sum_{l=e,\mu}\int \frac{d^3 p}{(2\pi^3)}\left(
    \ln(1+e^{-\frac{E-\mu_l}{T}})+ \ln(1+e^{-\frac{E+\mu_l}{T}})
    \right),\nonumber\\
\end{eqnarray}
assuming that the neutrinos remain untrapped and do not contribute to the thermodynamics.\footnote{Note that we have dropped an infinite vacuum energy in $\Omega_L$. This is possible because, unlike the quark effective potential, this term is a pure constant and therefore does not have any observable consequences.}
This means that lepton number is not conserved either, and the electron and muon chemical potentials are solely determined by their electric charge, $\mu_e=\mu_{\mu}=-\mu_Q$.

Finally we require electric and color neutrality of the system, i.e., we impose the neutrality conditions 
\begin{equation}\label{eq:neutrality}
    \frac{\partial \Omega_\text{eff}^\text{total}}{\partial \mu_Q}\bigg|_{\mu_Q=\bar{\mu}_Q}=\frac{\partial \Omega_\text{eff}^\text{total}}{\partial \mu_3}\bigg|_{\mu_3=\bar{\mu}_3}=\frac{\partial \Omega_\text{eff}^\text{total}}{\partial \mu_8}\bigg|_{\mu_8=\bar{\mu}_8}=0,
\end{equation}
corresponding to vanishing total electric charge and equal number densities of red, green and blue quarks. (Note that there are caveats related to the fact that color $SU(3)$ is only a global symmetry in NJL, while it is a gauge symmetry in QCD \cite{Buballa:2005bv}.)
We are then left with a single independent chemical potential, namely the quark chemical potential $\mu$. 
In general, there may be multiple solutions to Eqs.~(\ref{eq:gapeq}) and \eqref{eq:neutrality} for a given quark chemical potential and temperature. Then, the physical solution is the solution with the largest pressure $P=-\Omega^{\text{total}}(\mu,T)$.

\subsection{Divergences and conventional regularization}
\label{subsec:conreg}

The momentum integral in Eq.~(\ref{eq:Omega_eff}) diverges in the ultraviolet. This can be seen most easily for normal-conducting quark matter ($\Delta_A = 0$, $A=1,2,3$), for which the quasiparticle dispersion relations entering \Eq{eq:Aform} reduce to expressions
of the form
$\epsilon_{j\mp} = |\sqrt{M_j^2 + p^2} \mp \mu_{j}|$
for quarks ($-$) and antiquarks ($+$) of mass $M_j$ and chemical potential $\mu_j$. 
Combining these two modes and making use of the identity
\begin{equation}
|x| + 2T\ln\left(1+e^{-\frac{|x|}{T}}\right)
=
x + 2T\ln\left(1+e^{-\frac{x}{T}}\right)
\label{eq:trick}
\end{equation}
one arrives at the well-known decomposition
\begin{eqnarray}\label{eq:vac_med_decomposition}
&&\int\frac{d^3p}{(2\pi)^3}\sum_{\sigma=\pm}
\left(\epsilon_{j\sigma} + 2T\ln\Big(1+e^{-\frac{\epsilon_{j\sigma}}{T}}\Big)
\right)
=\nonumber\\&& 
\int\frac{d^3p}{(2\pi)^3} \, \left( f_\text{vac}(p) + f_\text{med}(p) \right)
\end{eqnarray}
with the vacuum part
\begin{equation}
\label{eq:fvac}
f_\text{vac}(p) 
=
\sqrt{M_j^2 + p^2} - \mu_j + \sqrt{M_j^2 + p^2} + \mu_j 
=
2\,\sqrt{M_j^2 + p^2} 
\end{equation}
and the medium part
\begin{eqnarray}
f_\text{med}(p) 
=
2T
\,\bigg(&&
\ln\Big(1+e^{-\frac{\sqrt{M_j^2 + p^2} -\mu_j}{T}}\Big)
\nonumber\\+&& \ln\Big(1+e^{-\frac{\sqrt{M_j^2 + p^2} +\mu_j}{T}}\Big)
\bigg) .
\end{eqnarray}
From this one sees immediately that the momentum integral over $f_\text{vac}$ is quartically divergent, i.e., if we integrate up to momenta $|p| < k$, it behaves as $k^4$ to leading order.
On the other hand, the integral over $f_\text{med}$ is convergent. 

Because of the UV divergence of the vacuum contribution, the model needs to be regularized. 
As the NJL-model is non-renormalizable, the regularization is a part of the model and the results depend on the used regularization scheme. Common regularization methods for the NJL model employed in the literature are three- and four-momentum cutoff, Pauli-Villars, and proper time regularization, see Ref.~\cite{Klevansky:1992qe} for an overview and Ref. \cite{Pannullo:2024sov} for a recent discussion in the context of inhomogeneous phases. The regularization methods have different advantages and drawbacks and the regularization method is chosen depending on the quantities that are calculated. 
We follow again Ref.~\cite{Ruester:2005jc} and employ a sharp three-momentum cutoff, which is the most simple regularization method and therefore commonly used in the literature. Here, the integral is restricted to momenta below a cutoff scale $\Lambda'$:
\begin{equation}\label{eq:effective_pot_cutoff}
\Omega_{\text{eff}}^{\Lambda'}(\vect{\mu},T,\vect\chi)= -\int_{\vert p \vert < \Lambda^{'}} \frac{d^3p}{(2\pi)^3}\mathcal{A}(\vect p;\vect{\mu},T,\vect\chi)+\mathcal{V}(\vect\chi).
\end{equation}
From here on we drop the explicit dependence on momenta and Matsubara frequencies. As in  Ref.~\cite{Ruester:2005jc} we adopt the model parameters of Ref.~\cite{Rehberg:1995kh}
with a cutoff $\Lambda'=602.3\,$MeV. 
This value, together with other model parameters, which we list at the beginning of the results section \ref{sec:results}, have been fitted to vacuum observables (pion decay constant and pseudoscalar meson spectrum). As discussed in the Introduction, this rather low cutoff scale may lead to severe artifacts if one applies the model at finite temperature or chemical potential of the same order. In normal-conducting matter this problem can be avoided rather easily by regularizing only the divergent integral over $f_\text{vac}$ while leaving the convergent medium part unregularized. Unfortunately, this does not work for color-superconducting quark matter. In fact, as one can see from \Eq{eq:fvac}, even in normal-conducting quark matter, the separation into vacuum and medium parts is almost an accident due to the cancellation of the chemical potentials in $f_\text{vac}$. In CSC matter this cancellation does no longer occur. For instance in a 2SC phase with up and down quarks with equal masses $M$ and pairing gap $\Delta$ at a common chemical potential, the dispersion relations of the paired quarks are given by $\epsilon_{2SC,\mp} = \sqrt{(\sqrt{M^2 + p^2} \mp \mu)^2 + |\Delta|^2}$, and there is no way that the chemical potentials cancel by combining these terms.
In this case, it is a common procedure to regularize all momentum integrals of the model, convergent and divergent ones, with the same cutoff $\Lambda'$ \cite{BUBALLAHABIL}. This conventional regularization of the action with the three-momentum cutoff $\Lambda'$ for all momentum integrals is practical, but, as already pointed out in the Introduction, leads to unphysical results of the model in the temperature and chemical potential region of phenomenological interest. 

In the remaining parts of this paper, we show how these cutoff artifacts can be removed by utilizing the principle of RG consistency \cite{Braun:2018svj}.

\section{RG-consistency}\label{sec:RG_consistency}

In a functional renormalization group (FRG) calculation, one introduces a scale-dependent effective action $\Gamma_k$, which includes all physics above the momentum scale $k$. The goal is to calculate the full quantum effective action $\Gamma$ that includes all
momentum scales $\Gamma=\lim_{k\to 0}\Gamma_k$.
The evolution equation (flow equation) connecting the scale-dependent effective action at different scales is given by the Wetterich equation \cite{Wetterich:1992yh}
\footnote{We use the notation $\SumInt_P f(P)=\int \frac{d^3 p}{(2\pi)^3}\; T\!\!\sum\limits_{n=-\infty}^\infty  f(p_0=i\omega_n,\bm{p})$ for working in a medium with nonzero temperature.}
\begin{align}\label{eq:wetterich}
    \partial_k \Gamma_k = -\frac{1}{2}\SumInt_P\text{tr}\left[
    \partial_k R_k (\Gamma_k^{(2)}+ R_k)^{-1}\right].
\end{align}
The flow equation is a functional differential equation for the scale-dependence of $\Gamma_k$ where $\Gamma_k^{(2)}$ denotes the second functional derivative with respect to the fermionic fields and $R_k(P)$ is a regulator function. 
In a typical FRG calculation, the form of the UV-action $S_{\text{cl}}$ (also called \textit{classical action}) is fixed at a UV scale $\Lambda$, i.e., $\Gamma_{\Lambda}=S_{\text{cl}}$. Thus, in order to obtain the full quantum effective action, the flow equation \eqref{eq:wetterich} is integrated from $k=\Lambda$ to $k=0$. To do this in practice, one has to introduce a truncation for $\Gamma_k$ and must specify the regulator function.
For a pedagogical introduction to FRG flow equations, see Ref.~\cite{Gies:2006wv}.

The authors of Ref.~\cite{Braun:2018svj} define a theory to be RG consistent, if the resulting effective action does not depend on the initial UV scale, i.e.,
\begin{equation}\label{eq:RG_consistency}
     \Lambda \frac{d\Gamma}{d\Lambda}=0.
\end{equation}
This condition can only be true for scales $\Lambda$ that are much higher than the thermodynamic scales, e.g., the chemical potential and temperature at which the model is evaluated.
In fundamental theories, this can easily be achieved by choosing the UV scale sufficiently high.
In models, on the other hand, one often encounters the situation that the classical action which defines the model is given at a fixed finite scale. 
Specifically, as we will see below, for the NJL model introduced in the previous section, the UV scale can be identified with the three-momentum cutoff $\Lambda'$ we employed for the regularization of the effective potential in \Eq{eq:effective_pot_cutoff}. 
As $\Lambda'$ was fitted to vacuum properties, leading to a rather low value of about 600~MeV \cite{Rehberg:1995kh}, the model, as it stands, violates RG consistency if the temperatures or chemical potentials are of this  order. This is closely related to the cutoff artifacts discussed in the previous section. 

Our goal is therefore to construct a quantum effective action $\Gamma$ for the mean-field model of section II in a way that it fulfills the RG-consistency criterion at large scales $\Lambda$. 
To be precise, we require
\begin{equation}\label{eq:crit1}
    \Lambda\frac{\partial\Gamma}{\partial \Lambda}\to0 \text{ for } \Lambda\to\infty.
\end{equation}
The basic idea of Ref.~\cite{Braun:2018svj} is to construct a new UV effective action $\Gamma_\Lambda$ at a large scale $\Lambda$ from the original one given at scale $\Lambda'$ by integrating \Eq{eq:wetterich} in vacuum ($\vect{\mu} = T = 0$) from $\Lambda'$ to $\Lambda$. 
Thus, by construction, $\Gamma_\Lambda$ and $\Gamma_{\Lambda'}$ are equivalent in vacuum.
However, the essential point is that $\Lambda$ can be chosen arbitrarily high, so that RG consistency in the form of \Eq{eq:crit1} can now be achieved even when the model is applied in medium at temperatures or chemical potentials of the order of the original UV scale $\Lambda'$ or higher.  

In the following, we work out this idea more explicitly. To this end we first note that, in general, the effective potential and the effective action are related by \footnote{In the convention used here, the effective potential and the effective action have the same sign, i.e. both quantities are bounded from below. Note that in an older preprint-version of this article \href{https://arxiv.org/abs/2408.06704v1}{arXiv:2408.06704v1} the effective action was defined with a relative minus sign, such that it was bounded from above. The sign convention was changed to be consistent with the standard used in the FRG community.}
\begin{equation}
\label{eq:Omeff_Gamma}
  \Omega_{\text{eff}}=\frac{\Gamma}{V_4} ,  
\end{equation}
where $V_4=V_3/T$ and $V_3$ denotes the Euclidean three-volume. 
Hence, in order to guarantee the vacuum equivalence of a model defined via the effective action at the UV scale $\Lambda$ to the effective potential \Eq{eq:effective_pot_cutoff}, regularized  with the three-momentum $\Lambda'$, we require that $\Gamma$ reduces to this standard cutoff regularization in the vacuum, i.e.,
\begin{equation}\label{eq:crit2}
\frac{\Gamma}{V_4}(\bm{\mu}=0,T=0,\vect\chi)=-\int_{\vert p \vert < \Lambda'} \frac{d^3p}{(2\pi)^3}\mathcal{A}_\text{vac}(\vect\chi)+\mathcal{V}(\vect\chi),
\end{equation}
where $\mathcal{A}_\text{vac}(\vect\chi)=\mathcal{A}(\vect\mu=0,T=0,\vect\chi)$.
Thus, equations \eqref{eq:crit1} and \eqref{eq:crit2} are the two criteria which an RG-consistent effective action for our model must fulfill.

For the purpose of this work, we are only interested in the evolution of the scale-dependent effective action in the mean-field approximation. There, quantum fluctuations of the bosonic condensates are neglected and it is sufficient to study $\Gamma_k$ up to 1-loop order. In this case, $\Gamma_k^{(2)}$ gets replaced by the corresponding functional derivative of the UV-action $S_{\text{cl}}^{(2)}$ in the flow equation which then simplifies to \cite{Gies:2006wv}
\begin{align}
    \partial_k\Gamma_k=&-\frac{1}{2}\SumInt_P\text{tr}((S_\text{cl}^{(2)}+R_k)^{-1}\partial_k R_k))
    \nonumber\\
    =&-\frac{1}{2} \SumInt_P\text{tr}\,\partial_k\ln(S_\text{cl}^{(2)} + R_k).
\end{align}
In the mean-field approximation, the euclidean classical action has the form 
$S_\text{cl}=\SumInt_P \bar{\psi}S^{-1}\psi+V_4\mathcal{V}$, thus $S_\text{cl}^{(2)}$ is simply given by the euclidean inverse propagator $S^{-1}$. 
Now consider two momentum scales $k_1$ and $k_2$. Integrating the flow equation from $k_2$ to $k_1$ gives
\begin{eqnarray}
    \Gamma_{k_1}-\Gamma_{k_2}&=&\int_{k_2}^{k_1} dk\,\partial_k\Gamma_k \nonumber\\&=&- \frac{1}{2}\SumInt_P \text{tr}\left(
    \ln(S^{-1} + R_{k_1})-\ln(S^{-1} + R_{k_2})
    \right).\nonumber\\
\end{eqnarray}
To arrive at a result that we can directly compare with 
the cutoff-regularized action in mean field, we choose the 3d sharp cutoff regulator function $R_k=a^2\frac{k^2}{p^2}\theta(k^2-p^2)$ with 
$p = |\vect{p}|$ and
$a\to\infty$ \cite{Pawlowski:2015mlf} and obtain
\begin{align}
    \Gamma_{k_1} &= \Gamma_{k_2} - \frac{1}{2}\SumInt_P \text{tr}\Big(
    \ln(S^{-1}(P))\theta(p^2 - k_1^2) \nonumber\\
    &\qquad\qquad\qquad - \ln(S^{-1}(P))\theta(p^2 - k_2^2)
    \Big) \nonumber\\
    &= \Gamma_{k_2} - \frac{V_4}{2}T\sum_n \int \frac{d^3p}{(2\pi)^3} \text{tr} \ln\left(\frac{S^{-1}(i\omega_n,\bm{p})}{T}\right) \nonumber\\
    &\qquad\qquad\qquad \times \left(\theta(p - k_1) - \theta(p - k_2)\right) \nonumber\\
    &= \Gamma_{k_2} - V_4 \int \frac{d^3p}{(2\pi)^3}\, \mathcal{A}(\bm{\mu},T,\vect\chi)\, \nonumber\\
    &\qquad\qquad\qquad \times\left(\theta(p - k_1) - \theta(p - k_2)\right),
    \label{eq:solution_flow_equation}
\end{align}

where in the last step we used \Eq{eq:Adef}. By choosing $k_1=0$ and $k_2=\Lambda'$, one obtains
\begin{equation}\label{eq:mean-field-result}
\Gamma\equiv\Gamma_0=\Gamma_{\Lambda'}-V_4\int\limits_{|p|< \Lambda'} \frac{d^3p}{(2\pi)^3}\mathcal{A} \,(\bm{\mu},T,\vect\chi).
\end{equation}
Comparing this with \Eq{eq:effective_pot_cutoff} together with \Eq{eq:Omeff_Gamma}, we see that indeed the UV RG scale $\Lambda'$ can be identified with the three-momentum cutoff regularizing the effective potential (at least for our particular choice of the regulator $R_k$), while the effective action at this scale corresponds to the potential term,
$\Gamma_{\Lambda'} = V_4 \mathcal{V}(\vect\chi)$.

From now on we restrict our discussion to homogeneous matter. 
This is not really essential, but it allows us to write the integrals above in a way that the relation to the RG scale becomes more transparent.
For homogeneous matter the function $\mathcal{A}$ depends only on the modulus $p$ of the three-momentum $\vect{p}$ but not on its direction, so that the angular parts of the momentum integrals can be performed trivially. 
\Eq{eq:solution_flow_equation} then takes the form 
\begin{equation}
\label{eq:flow_equation_k2-k_1}
    \Gamma_{k_1}= \Gamma_{k_2} 
     + \frac{V_4}{2\pi^2} \int_{k_2}^{k_1} dp\,p^2\, \mathcal{A}(\bm{\mu},T,\vect\chi), 
\end{equation}
making the momentum flow from the scale $k_2$ to $k_1$ explicit.
Again choosing $k_1=0$ and $k_2=\Lambda'$ we get for the full (mean-field) quantum effective action
\begin{align}\label{eq:MF_effective_pot}
\Gamma(\vect\mu,T,\vect\chi)=
\Gamma_{\Lambda'} 
+\frac{V_4}{2\pi^2}\int_{\Lambda'}^0dp\,p^2\,\mathcal{A}(\vect\mu,T,\bm{\chi}).
\end{align}
Here we kept the ordering of the bounds of integration according to the FRG interpretation that the integral describes the flow from the UV scale $\Lambda'$ down to the infrared. This should be contrasted with the standard cutoff interpretation of $\Lambda'$ as an upper limit of the momentum integral.

As pointed out before, if temperature or chemical potentials become comparable to the size of the cutoff $\Lambda'$, equation \eqref{eq:MF_effective_pot} is plagued with cutoff artifacts. In the language of FRG, this is because $\Gamma_{\Lambda'}=V_4\mathcal{V}(\chi)$ is only a good ansatz for the classical action as long as the model is evaluated at parameters 
that are small compared to $\Lambda'$. 
In the following we therefore write
\begin{equation}
\label{eq:GLpvac_V}
  \Gamma_{\Lambda'}^\text{vac}\equiv +V_4\mathcal{V}(\vect{\chi}),  
\end{equation}
in order to stress that $\Gamma_{\Lambda'}$ in medium may be different in an RG-consistent framework.\footnote{We note that $\Gamma_{\Lambda'}^\text{vac}$ still depends on the temperature via the four-volume $V_4$. This trivial factor drops out in the effective potential, see \Eq{eq:Omeff_Gamma}.}

However, we are now in the position to pursue the idea outlined earlier in this section: Sticking to $\vect{\mu}= T = 0$ for a moment, we can employ the flow equation 
(\ref{eq:flow_equation_k2-k_1}) to calculate the effective action at a higher scale $\Lambda$ from the one at the original scale $\Lambda'$:
\begin{equation}
\label{eq:Gamma_Lambda}
    \Gamma_{\Lambda}= \Gamma_{\Lambda'}^\text{vac} 
     + \frac{V_4}{2\pi^2} \int_{\Lambda'}^{\Lambda} dp\,p^2\, \mathcal{A}_\text{vac}(\vect\chi).
\end{equation}
By construction, $\Gamma_{\Lambda}$ and $\Gamma_{\Lambda'}^\text{vac} $ are equivalent in vacuum, i.e., both lead to the same vacuum effective action after flowing down to the IR scale $k=0$.
However, unlike $\Lambda'$, which was fixed by fitting vacuum properties, $\Lambda$ can now be chosen much larger than all other physical scales of interest, including temperature and chemical potential. Assuming such a choice, $\Gamma_{\Lambda}$ should then be a good starting point also for calculating the full quantum effective action in medium. 
Analogously to \Eq{eq:MF_effective_pot} we have
\begin{align}
\Gamma(\vect\mu,T,\vect\chi)=&
\Gamma_{\Lambda} 
+\frac{V_4}{2\pi^2}\int_{\Lambda}^0dp\,p^2\,\mathcal{A}(\vect\mu,T,\bm{\chi})
\\
\label{eq:Gamma_RG}
=&
\Gamma_{\Lambda'}^{\text{vac}}  
     +\frac{V_4}{2\pi^2} \int_{\Lambda'}^{\Lambda} dp\,p^2\, \mathcal{A}_{\text{vac}}(\vect\chi)
\nonumber\\
&+\frac{V_4}{2\pi^2}\int_{\Lambda}^0dp\,p^2\,\mathcal{A}(\vect\mu,T,\bm{\chi}),
\end{align}
where in the second step we have inserted \Eq{eq:Gamma_Lambda} for $\Gamma_{\Lambda}$.  

Let us compare this result with the original cutoff model where $\Gamma_{\Lambda'}^\text{vac}$ is taken as the UV-scale effective action also in medium,
\begin{equation}
\label{eq:Gammabar}
\bar\Gamma(\vect\mu,T,\vect\chi)
=
\Gamma_{\Lambda'}^\text{vac}  
+\frac{V_4}{2\pi^2}\int_{\Lambda'}^0dp\,p^2\,\mathcal{A}(\vect\mu,T,\bm{\chi}).
\end{equation}

Here the bar in $\bar\Gamma$ is used in order to distinguish it from $\Gamma$ in \Eq{eq:Gamma_RG}. 
It is easy to see that the two agree in vacuum where $\mathcal{A}$ is equal to $\mathcal{A}_\text{vac}$ and, thus, in \Eq{eq:Gamma_RG}
the two integrals cancel each other on the interval between $\Lambda'$ and $\Lambda$. 
In medium, however, this is in general not the case. Explicitly we get from the above equations

\begin{equation}
\Gamma(\vect\mu,T,\vect\chi)
=
\bar\Gamma(\vect\mu,T,\vect\chi)
+
\Gamma_{\Lambda'}^\text{med}  
\end{equation}
with the medium contribution to the effective action at the scale $\Lambda'$
\begin{eqnarray}
\label{eq:GLpmed}
\Gamma_{\Lambda'}^\text{med}&\equiv& \Gamma_{\Lambda'}-\Gamma_{\Lambda'}^{\text{vac}}  \nonumber\\
&=&-\frac{V_4}{2\pi^2} \int_{\Lambda'}^{\Lambda} dp\,p^2\, \left(\mathcal{A}(\vect{\mu},T,\vect\chi)- \mathcal{A}_\text{vac}(\vect\chi)
   \right), 
\end{eqnarray}
which is non-negligible if there are still considerable medium effects to the function $\mathcal{A}$ above the cutoff scale $p = \Lambda'$.
The fact that this term is missing in \Eq{eq:Gammabar} then leads to the violation of RG consistency.

\begin{figure*}[t]
					\fontsize{12pt}{9pt}\selectfont
					\def\svgwidth{0.9\columnwidth}
					
     \includegraphics[width=0.9\textwidth]{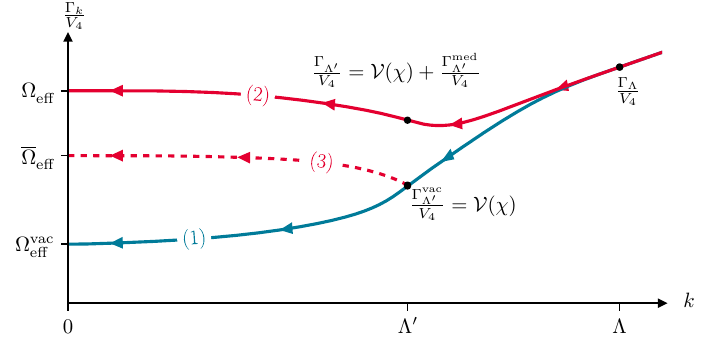}
					\caption{Schematic illustration of the procedure proposed in Ref.~\cite{Braun:2018svj} for the construction of an RG-consistent scheme. The lines represent the RG flows of the scale dependent effective action in vacuum (blue) and in medium (red), initialized at different scales $\Lambda'$ and $\Lambda$. See text for details. 
                    Figure adapted from Fig.~2.4 in Ref.~\cite{Steil:2024phd}.}
     \label{fig:flow}
     
				\end{figure*}

This is illustrated in Fig.~\ref{fig:flow}, which schematically shows the RG flows in vacuum (blue) and in medium (red). In order to eliminate the trivial $T$ dependence contained in $V_4$, we plot $\Gamma_k/V_4$, corresponding to the scale dependent effective potential.
Initializing the flow equation in vacuum at the scale $\Lambda'$ with the ansatz $\Gamma_{\Lambda'}^{\text{vac}}/V_4 = \mathcal{V}(\vect{\chi})$ leads to the full vacuum quantum effective potential $\Omega_\text{eff}^\text{vac}=+{\Gamma}^{\text{vac}}/V_4$ (line 1 in Fig.~\ref{fig:flow}).
Initializing the flow equation in medium with the same ansatz $\bar{\Gamma}_{\Lambda'}=\Gamma_{\Lambda'}^{\text{vac}}$ and integrating the flow equation in medium (line 3) gives the full in-medium quantum effective potential $\bar{\Omega}_{\text{eff}}$. 
If instead the flow is initialized at a higher scale $\Lambda>\Lambda'$, fixing $\Gamma_\Lambda$ according to \Eq{eq:Gamma_Lambda}, the full quantum vacuum potential ${\Omega}_\text{eff}^{\text{vac}}$ stays the same but the full quantum in-medium potential at $k=0$ changes to $\Omega_\text{eff} = \bar{\Omega}_\text{eff} + \Gamma_{\Lambda'}^\text{med}/V_4$, which is in general different from $\bar{\Omega}_\text{eff}$ (line 2). This means, that initializing the theory at scale $\Lambda'$, corresponding to the ordinary regularization with cutoff $\Lambda'$, is not RG consistent: changing the scale at which the flow is initialized changes the result for the effective action in medium at $k=0$, and \Eq{eq:RG_consistency} is not fulfilled. 

On the other hand, if we employ \Eq{eq:Gamma_Lambda} to initialize the theory at a scale $\Lambda$ much larger than all internal scales (such as quark masses and diquark condensates) and external scales (such as $\mu$, $T$), we may expect that further increasing $\Lambda$ does not lead to further changes, or, more precisely, that RG consistency is realized in the form of \Eq{eq:crit1}.

For a more detailed investigation of this issue, we go back to \Eq{eq:Gamma_RG}.
Rearranging the terms and integration bounds, we can write the effective potential as 
\begin{align}
\Gamma(\vect\mu,T,\vect\chi)
&= \Gamma_{\Lambda'}^\text{vac}  
-\frac{V_4}{2\pi^2}\Bigg(\int_0^{\Lambda} dp\, p^2\, \mathcal{A}(\vect\mu,T,\bm{\chi}) \nonumber\\
&\qquad - \int_{\Lambda'}^{\Lambda} dp\, p^2\, \mathcal{A}_\text{vac}(\vect\chi)\Bigg)
\label{eq:RG_consist_eff_action_1}    
\\
&= \Gamma_{\Lambda'}^\text{vac}  
-\frac{V_4}{2\pi^2}\Bigg(\int_0^{\Lambda'} dp\, p^2\, \mathcal{A}_\text{vac}(\vect\chi) \nonumber\\
&\qquad + \int_0^{\Lambda} dp\, p^2\, \Big(\mathcal{A}(\vect\mu,T,\bm{\chi}) - \mathcal{A}_\text{vac}(\vect\chi)\Big)\Bigg).
\label{eq:RG_consist_eff_action} 
\end{align}

Here one can see what the above procedure means in practice: Instead of regularizing the integral over the entire function $\mathcal{A}$ with a single cutoff, the vacuum part $\mathcal{A}_\text{vac}$ is regularized with the cutoff $\Lambda'$, while the medium part $\mathcal{A}_\text{med} \equiv \mathcal{A} - \mathcal{A}_\text{vac}$ is integrated to the (much) larger scale $\Lambda$,\footnote{As stressed by the authors of Ref.~\cite{Braun:2018svj}, it is not required that the model itself is valid at those large scales, which is certainly not the case for the NJL model in the regime of perturbative QCD.} which ideally could be sent to infinity.
Of course, this is only possible if the medium integral is convergent for $\Lambda \rightarrow \infty$. As we will discuss in greater detail in Sec.~\ref{sec:meddiv}, this is not the case in color-superconducting matter. 

For the remainder of this section we therefore restrict ourselves to normal-conducting quark matter, for which the medium part is convergent, see Sec.~\ref{subsec:conreg}. 
For this case we will show that the effective action of \Eq{eq:RG_consist_eff_action} indeed fulfills the RG-consistency criterion \Eq{eq:crit1}.

To this end, 
we set $G_D=0$ and choose the scale $\Lambda$ much larger than the initial cutoff-scale $\Lambda'$. The logarithmic derivative of the effective action with respect to $\Lambda$ reads
\begin{align}
    \Lambda\frac{\partial\Gamma}{\partial \Lambda}
    &= -V_4\, \frac{\Lambda^3}{2\pi^2} \Big(\mathcal{A}(\vect{\mu},T;{\vect\chi}) - \mathcal{A}_\text{vac}({\vect\chi})\Big)\Big\vert_{p=\Lambda} \\
    &= -V_4\, \frac{\Lambda^3}{2\pi^2} \sum_{j=1}^{18} \Bigg(\epsilon_j(\vect{\mu},{\vect\chi}) - \epsilon_j(\vect{\mu}=0,{\vect\chi}) \nonumber\\
    &\qquad + 2T\ln\left(1 + e^{-\frac{\epsilon_j(\vect{\mu},\vect\chi)}{T}}\right)\Bigg)\Bigg\vert_{p=\Lambda},
    \label{eq:RG_derivative}
\end{align}
where in the second line we inserted the explicit expression \Eq{eq:Aform} for $\mathcal{A}$ in terms of the quark dispersion relations.
For ungapped modes, these reduce to pairs
$\epsilon_{j\mp} = |\sqrt{M_j^2 + p^2} \mp \mu_{j}|$,
which can be replaced by $\epsilon_{j\mp} = \sqrt{M_j^2 + p^2} \mp \mu_{j}$ for $p=\Lambda \gg \mu_j$.
Summing over the $\mp$-pairs, the chemical potentials from the first term in parentheses thus drop out, while the remainder is canceled by the second term.
Hence
\begin{align}
    \Lambda\frac{\partial\Gamma}{\partial \Lambda}=&-V_4 T\,\frac{\Lambda^3}{\pi^2}\sum_j\sum_{\sigma=\pm1}\ln\left(
   1+e^{-\frac{\epsilon_{j\sigma}}{T}}\right) \to 0 
\end{align}
for $\Lambda\to\infty$, in agreement with the RG-consistency criterion $\eqref{eq:crit1}$. Moreover, by construction, \Eq{eq:RG_consist_eff_action} reduces to the form \eqref{eq:crit2} in vacuum, so $\Gamma$ fulfills both criteria $\eqref{eq:crit1}$ and $\eqref{eq:crit2}$.\\
In the case of a nonzero diquark interaction $G_D>0$, however, the effective potential includes gapped quasiparticle dispersion relations due to color superconductivity. These gapped modes induce additional divergences in the effective potential, which need to be subtracted from $\Gamma$ while ensuring that both \eqref{eq:crit1} and \eqref{eq:crit2} remain satisfied. This will be discussed in the next section.

\section{Medium Divergences in Color-Superconducting Matter}
\label{sec:meddiv}

The RG-consistency of the effective action \eqref{eq:RG_consist_eff_action} relies on the premise that $\Gamma$ stays finite for $\Lambda\to\infty$. This, however, is not the case for color-superconducting matter.
For simplicity, we 
discuss this first for 2SC-like pairing of quarks with the same chemical potential $\mu$ and mass $M$ and neglect all unpaired quarks, i.e., we only consider a system of red and green up and down quarks (see Table~\ref{tab:gaps}).
Then, having $\Delta\equiv\Delta_3$, the excitation spectrum includes the dispersion relations
\begin{equation}{\label{eq:2scDispersion}} 
\epsilon_{\pm}=\sqrt{(E\pm\mu)^2+\Delta^2}
\end{equation}
with a 4-fold degeneracy. The expression with the negative (positive) sign describes quasiparticles (quasi-antiparticles) with $E=\sqrt{p^2+M^2}$ and the gap $\Delta$, which, without loss of generality, we choose to be real and positive.   
Plugging the explicit expression \eqref{eq:Aform} for the momentum integrands into \Eq{eq:RG_consist_eff_action}, the effective action is
\begin{equation}\label{eq:Gamma_2sc}
    \frac{\Gamma}{V_4}(\mu,T,\bm{\chi})=f_0(\mu,\bm{\chi})+f_1(\mu,T,\bm{\chi})+\mathcal{V}(\bm{\chi})
\end{equation}
with the terms
\begin{align}
    f_0(\mu,\bm{\chi})
    &= -\frac{2}{\pi^2}\Bigg(\int_0^{\Lambda} dp\,p^2 \sum_{j=\pm}
    \left(\epsilon_j(\mu,\vect\chi) - \epsilon_j(\mu=0,\vect\chi)\right) \nonumber\\
    &\qquad + \int_{0}^{\Lambda'} dp\,p^2 \sum_{j=\pm} \epsilon_j(\mu=0,\vect\chi)\Bigg), \label{RGnonNormGamma} \\
    f_1(\mu,T,\bm{\chi}) 
    &= -\frac{4T}{\pi^2}\int_0^{\Lambda} dp\,p^2 \sum_{j=\pm} \ln\left(
   1 + e^{-\frac{\epsilon_j(\mu,\vect\chi)}{T}}\right).
    \label{f1}
\end{align}
Evaluating $f_0$ in the presence of the gapped modes \eqref{eq:2scDispersion},
\begin{align}\label{2scdiv1}
    f_0(\mu,\bm{\chi})
    &= -\frac{2}{\pi^2}\int_0^{\Lambda} dp\,p^2\; \left(\epsilon_{+} + \epsilon_{-} - 2\sqrt{E^2+\Delta^2}\right) \nonumber\\
    &\qquad - \frac{4}{\pi^2}\int_{0}^{\Lambda'} dp\,p^2\; \sqrt{E^2+\Delta^2}.
\end{align}
and Taylor-expanding the integrand of the first term in powers of $\mu$,
\begin{equation}\label{2scdiv2}
(\epsilon_{+}+\epsilon_{-}-2\sqrt{E^2+\Delta^2})=\frac{\Delta ^2}{(\Delta ^2+M^2+p^2)^\frac{3}{2}}\,\mu^2+\mathcal{O}(\mu^4),
\end{equation}
the momentum integration gives 
\begin{eqnarray}{\label{2scDiv}}
f_0(\mu,\bm{\chi})=&&-\frac{2}{\pi^2}\mu^2\Delta ^2 
\ln \left(\frac{\sqrt{ \Delta^2+M^2+\Lambda ^2}+\Lambda }{\sqrt{\Delta^2+M^2}}\right)
\nonumber\\&&
+\text{finite contributions}.
\end{eqnarray}
Here, \textit{finite contributions} denotes terms that stay finite in the limit $\Lambda\to\infty$. This also includes the second integral in \Eq{2scdiv1}, corresponding to the vacuum contribution. As discussed earlier, the cutoff $\Lambda'$ is fitted to vacuum properties and will be kept finite. The term that explicitly depends on temperature, $f_1(\mu,T,\bm{\chi})$, also stays finite for $\Lambda\to\infty$.

Hence, in total, the effective action in \Eq{eq:Gamma_2sc} has a logarithmic divergence of the form
\begin{eqnarray}\label{2SCnochemdiv}
\frac{\Gamma}{V_4}(\mu,T,\vect\chi)&\simeq&-\frac{2}{\pi^2} \mu^2\Delta^2\ln\Lambda.
\end{eqnarray}
This is a divergence which is only present in a medium (nonzero $\mu$) with nonzero diquark condensates $\Delta$.
Evaluating the RG-consistency criterion \eqref{eq:crit1} in the limit $\Lambda\to\infty$ now gives
\begin{eqnarray}\label{eq:medium_divergence}
   \frac{1}{V_4}\lim_{\Lambda\to\infty} \Lambda\frac{d}{d\,\Lambda}\Gamma=-\frac{2}{\pi^2}\mu^2\Delta^2.
\end{eqnarray}
This means this model is not RG consistent. 

The authors of Ref.~\cite{Braun:2018svj} addressed the issue of this medium divergence in a mean-field calculation of a Quark-Meson-Diquark model. Their proposed solution involved the addition of the term
\begin{equation}\label{eq:counterterm_toycase}
\mathcal{Y}(\mu,\vect\chi)=-\frac{1}{2}\mu^2\frac{\partial^2}{\partial\mu^2}\Gamma_{\Lambda'}(\mu,T=0,\vect\chi)\bigg\vert_{\mu=0}
\end{equation}
to the effective action, which leads to the renormalized effective action
\begin{equation}{\label{eq:h:1}}
\Gamma^R=\Gamma+\mathcal{Y}(\mu,\vect\chi).
\end{equation}
As we will see, the same prescription also works in our case. 
For  $\Gamma_{\Lambda'} = \Gamma_{\Lambda'}^\text{vac} + \Gamma_{\Lambda'}^\text{med}$ we obtain from Eqs.~\eqref{eq:GLpvac_V} and \eqref{eq:GLpmed}:
\begin{align}\label{gammalambdaprime}
    \frac{\Gamma_{\Lambda'}}{V_4}=&\,\mathcal{V}(\vect\chi)\\
    &-\frac{1}{2\pi^2}\int_{\Lambda'}^{\Lambda} dp\,p^2 \left(\mathcal{A}(\mu,T,\vect\chi)-\mathcal{A}_\text{vac}(\vect\chi)\right).
\end{align}
Inserting this into \Eq{eq:counterterm_toycase} gives
\begin{align}
\frac{\mathcal{Y}}{V_4}(\mu,\vect\chi) 
&=
\frac{\mu^2 }{4\pi^2} \int_{\Lambda'}^{\Lambda} dp\,p^2 \left(\frac{\partial^2}{\partial\mu^2}\mathcal{A}(\mu,T=0,\vect\chi)\right)\bigg|_{\mu=0}
\label{eq:YoV}
\\
&=
\frac{\mu^2 }{4\pi^2}\int_{\Lambda'}^{\Lambda} dp\,p^2 \sum_j\left(\frac{\partial^2}{\partial\mu^2}\,\epsilon_{j}(\mu,\vect\chi)\right)\bigg|_{\mu=0},
\label{eq:yNJL}
\end{align}
where for the second equality we used again \Eq{eq:Aform}. 
This equation proves particularly useful in the cases where an analytic expression for the dispersion relation is at hand. 
Evaluating it for our NJL model dispersion relations \eqref{eq:2scDispersion} at asymptotically large $\Lambda$ yields
\begin{align}\label{h-1-}
\frac{\mathcal{Y}}{V_4}(\mu,\vect\chi)
=& \frac{2}{\pi^2}\mu^2\Delta^2\ln\Lambda+\text{finite contributions}.
\end{align}

Therefore, on the right-hand side of \Eq{eq:h:1} the divergent medium contributions  from $\Gamma$, \Eq{2SCnochemdiv}, and from the additional term $\mathcal{Y}$, \Eq{h-1-}, exactly cancel. Since all remaining contributions are finite, this also ensures that $\Gamma^R$ is RG-consistent:
\begin{eqnarray}
   \lim_{\Lambda\to\infty}\Lambda \frac{d}{d \,\Lambda}\Gamma^R=0.
\end{eqnarray}

We can generalize the above procedure to the slightly more involved case of electrically and color neutral 2SC quark matter. In this case, the paired quarks have unequal chemical potentials. However, if their masses are equal, their dispersion relations are still known analytically and given by~\cite{Bedaque:1999nu}
\begin{eqnarray}
\epsilon_\pm^+ &=&  \epsilon_\pm + \frac{\delta\mu}{2},\nonumber\\
\quad
\epsilon_\pm^- &=&   \epsilon_\pm - \frac{\delta\mu}{2}\nonumber\\
\quad \text{with} \quad
\epsilon_{\pm}&=&  \sqrt{(E\pm\bar{\mu})^2+\Delta^2},
\end{eqnarray}
each with two-fold degeneracy. 
Here $\bar\mu$ and $\delta\mu$ are the average chemical potential and the chemical potential difference of the paired quarks, respectively, which can be expressed in terms of $\mu$, $\mu_Q$ and $\mu_8$ (see Sec.~\ref{subsec:chempot}) and turn out to be the same for both the $(ur-dg)$ and the $(ug-dr)$ pairs. 
Hence, proceeding in the same way as before, $\delta\mu$ drops out upon summing over the dispersion relations, and the effective action diverges as in \Eq{2SCnochemdiv} with $\mu$ replaced by $\bar\mu$: 
\begin{eqnarray}\label{2SCchemdiv}
\frac{\Gamma}{V_4}(\bar\mu,\delta\mu,T,\vect\chi)&\simeq&-\frac{2}{\pi^2} {\bar\mu}^2\Delta^2\ln\Lambda.
\end{eqnarray}
Consequently, the divergence can be eliminated by substituting $\mu \rightarrow \bar\mu$ in \Eq{eq:counterterm_toycase},
\begin{equation}
\label{eq:counterterm_2SC_neutral}
\mathcal{Y}(\bar\mu,\vect\chi)=-\frac{1}{2}\bar\mu^2\frac{\partial^2}{\partial\bar\mu^2}\Gamma_{\Lambda'}(\bar\mu,\delta\mu,T=0,\vect\chi)\bigg\vert_{\vect{\mu}=0},
\end{equation}
and adding this to the effective action. Again the result is then RG consistent.

For the examples above the cancellation of the divergences by \Eq{eq:counterterm_toycase} or \eqref{eq:counterterm_2SC_neutral} could have been anticipated, even without explicit calculations. 
For this let us go back to our first example and plug Eqs.~\eqref{eq:RG_consist_eff_action} and \eqref{eq:YoV} into \Eq{eq:h:1}. After rearranging terms the renormalized effective action can be written as
\begin{align}{\label{h:3}}
    \frac{\Gamma^R}{V_4}(\mu,T,\vect\chi)
    &= \mathcal{V}(\vect\chi) - \frac{1}{2\pi^2} \bigg(\int_0^{\Lambda} dp\,p^2 \mathcal{A}(\mu,T,\vect\chi) \nonumber\\
    &\quad - \int_{\Lambda'}^{\Lambda} dp\,p^2 \mathcal{A}_\text{vac}(\vect\chi) \nonumber\\
    &\mkern-36mu - \int_{\Lambda'}^{\Lambda} dp\,p^2 \frac{1}{2}\mu^2 \left(\frac{\partial^2}{\partial\mu^2} \mathcal{A}(\mu,T=0,\vect\chi)\right)\bigg|_{\mu=0} \bigg).
\end{align}
This means that in a Taylor expansion of the effective potential in powers of the chemical potential, we effectively regularize the (zero-temperature part of the) $\mu^2$ coefficient, in addition to the vacuum term, which was already regularized in \Eq{eq:RG_consist_eff_action}. Since these are the only divergent parts, see \Eq{2scdiv2}, this procedure yields a finite result.
Obviously the same arguments also hold for the second example, where \Eq{eq:counterterm_2SC_neutral} corresponds to the regularization of the quadratic term of a Taylor expansion in $\bar\mu$. From this perspective it is perhaps not surprising that \Eq{eq:counterterm_2SC_neutral} cancels the divergence, even if we consider 2SC pairing with unequal up and down quark masses and chemical potentials, for which no analytic form of the dispersion relation is known. 

In the next section we motivate these additional terms from a renormalization picture and then apply this to more complicated pairing patterns.

\section{Generalized treatment of divergent medium contributions}\label{sec:renopro}
As shown in the previous section, adding the terms 
\eqref{eq:counterterm_toycase} or
\eqref{eq:counterterm_2SC_neutral}
to the effective action
overcomes the issue of a divergent medium. Here, we generalize
this to more complex cases by an approach inspired by the procedure in renormalizable theories. Of course, the NJL model remains a non-renormalizable model. Therefore the approaches we will describe in the following remain to some extent ad-hoc, similar to other regularization procedures used in the literature. However, as we will show, they have the advantages of being RG consistent and do not suffer from the cutoff artifacts of the standard momentum-cutoff regularization. 

Consider a renormalizable quantum field theory defined by a Lagrangian $\mathcal{L}[\{\phi_i^\text{bare}\}, \{g_\alpha^\text{bare} \}]$, which depends on the bare fields $\phi_i^\text{bare}$ and the bare couplings $g_\alpha^\text{bare}$. 
Alternatively, the Lagrangian can be rewritten identically in terms of ``physical'' (``renormalized'') quantities  $\phi_i^R$ and $g_\alpha^R$ as
$\mathcal{L}[\{\phi_i^\text{bare}\}, \{g_\alpha^\text{bare} \}] \equiv \mathcal{L}^R[\{\phi_i^R\}, \{g_\alpha^R \}]$ with 
 \begin{eqnarray}\label{CT}
    \mathcal{L}^R&=&\mathcal{L}[\{\phi_i^R\}, \{g_\alpha^R \}]+\delta\mathcal{L}[\{\phi_i^R\}, \{g_\alpha^R \}],
   \end{eqnarray}
where here $\mathcal{L}$ has the same form as the original Lagrangian, but now depends on the renormalized quantities instead of the bare ones. This is corrected for by a set of counterterms summarized by $\delta\mathcal{L}$.
The values of these counterterms are fixed by imposing renormalization conditions. 
In a renormalizable theory this eliminates all divergences, so that the total effective action 
$\Gamma^R = \Gamma + \delta\Gamma$ is finite, while the individual pieces $\Gamma$ (corresponding to $\mathcal{L}$ on the RHS of \Eq{CT}) and $\delta\Gamma$ (corresponding to the counterterms) are divergent. 

In non-renormalizable models, like our NJL model, it is not possible to absorb all divergences in the existing bare couplings of the model. Nevertheless, in the spirit of an effective field theory, we can add counterterms to the original Lagrangian, which are designed in such a way that they cancel the divergences arising in the effective action. Here, rather than all divergences, we only consider the medium divergences, while, as before, the vacuum divergences are regularized by a fixed momentum cutoff $\Lambda'$. 
   
In order to illustrate the idea, we first apply it again to the simplified case of 2SC pairing with a single chemical potential $\mu$, as discussed in Sec.~\ref{sec:meddiv}. For this model, the divergence
   \eqref{2SCnochemdiv} can be cancelled by a counterterm of the form
 \begin{eqnarray}
 \label{eq:deltaL}
    \delta\mathcal{L}&=&-\frac{1}{2} Y^{C.T.}(\vect\chi){\mu}^2.
   \end{eqnarray}
 Here $Y^{C.T.}(\vect\chi)$ represents a condensate-dependent renormalization factor. In mean-field calculations, the term above does not contribute to any fermionic loop contributions in the effective potential. Therefore, the renormalized effective action derived from \Eq{CT} is
 \begin{eqnarray}\label{CT2}
    \Gamma^R&=&\Gamma-V_4\,\delta\mathcal{L},
   \end{eqnarray}
where $\Gamma$ is the divergent effective potential given in \Eq{eq:Gamma_2sc}.

From $\Gamma^R$ we can derive a renormalized coupling constant $Y^R$, defined as 
\begin{equation}
Y^R(\vect\chi) \equiv\frac{1}{V_4}\frac{\partial^2 \Gamma^R}{\partial {\mu}^2}\bigg |_{\mu,T=0}\label{eq:YR}.    
\end{equation}
Plugging in Eqs.~\eqref{CT2} and \eqref{eq:deltaL} one finds
\begin{equation}
\label{eq:YRsum}
Y^R(\vect\chi) = \frac{1}{V_4}\frac{\partial^2 \Gamma}{\partial {\mu}^2}\bigg |_{\mu,T=0} + Y^{C.T.}, 
\end{equation}
showing that $Y^{C.T.}$ can be interpreted as the counterterm correction to this coupling constant. 
The first term can in general be written as
\begin{eqnarray}
    \frac{1}{V_4}\,\frac{\partial^2 \Gamma}{\partial{\mu}^2}\bigg |_{\mu,T=0}= Y^0(\vect\chi)+\text{loop contributions},
    \label{eq:Y0loops}
\end{eqnarray}
where $Y^0$ denotes a possible tree-level coupling constant from the Lagrangian $\mathcal{L}$,
\begin{equation}
   Y^0 = -\frac{1}{2} \frac{\partial^2\mathcal{L}}{\partial{\mu}^2}\bigg |_{\mu,T=0}\label{eq:Y0}.
\end{equation}
The second term on the RHS of \Eq{eq:Y0loops} corresponds to the fermionic 1-loop (mean-field) contributions, which lead to the medium divergence \eqref{2SCnochemdiv}.
Thus, in total, the renormalized coupling constant $Y^R$ from 
\Eq{eq:YR} reads 
\begin{eqnarray}
    Y^R=Y^0+Y^{C.T.}+\text{loop contributions}.
\end{eqnarray}
As the Lagrangian \eqref{eq:Lagrangian} does not contain any $\mu^2$ terms, the three-level coupling is $Y^0 = 0$ in our model. This means that divergences from the loop contributions cannot be absorbed into any existing bare coupling constant. Nevertheless they can be cancelled by the counterterms. Since $Y^0 = 0$, a natural renormalization condition seems to be that the renormalized coupling $Y^R$ vanishes as well. However, as known from standard textbook examples, this condition can only be imposed at a given renormalization point. Hence, there is some flexibility in choosing the renormalization scheme, corresponding to different eligible choices of the counterterm, which all cancel the divergences but differ by finite contributions. We will address this in Sec.~\ref{sec:schemes}.

For now, we stay with the ansatz \eqref{eq:counterterm_toycase} proposed in Ref.~\cite{Braun:2018svj}. From \Eq{CT2} we get for the corresponding counterterm
\begin{eqnarray}\label{generalansatzCT}
\delta\mathcal{L}
= -\frac{\mathcal{Y}}{V_4}
=\frac{1}{V_4}\frac{1}{2}\mu^2\,\frac{\partial^2 \Gamma_{\Lambda'}^{}}{\partial{\mu}^2}\bigg |_{\mu,T=0}.
\end{eqnarray}
Comparing this with \Eq{eq:deltaL} and plugging the resulting $Y^{C.T.}$ into \Eq{eq:YRsum}, one finds
\begin{equation}
\label{eq:YRexpl}
    Y^R = \frac{1}{V_4} \left(\frac{\partial^2 \Gamma}{\partial{\mu}^2}  - \frac{\partial^2 \Gamma_{\Lambda'}}{\partial{\mu}^2} \right)\bigg |_{\mu,T=0}. 
\end{equation}
Recalling that $\Gamma \equiv \Gamma_{0}$ is the scale dependent effective action $\Gamma_k$ at the infrared scale $k=0$, this can be interpreted as fixing the scale dependent renormalized coupling $Y^R_k =0$ at the UV scale $k=\Lambda'$ in vacuum. 
Using Eqs.~\eqref{eq:RG_consist_eff_action} and \eqref{gammalambdaprime} one finds
\begin{equation}
 Y^R(\vect\chi)=-\frac{1}{2\pi^2}\int_{0}^{\Lambda'} dp\,p^2\,\left( \frac{\partial^2}{\partial\mu^2}\mathcal{A}(\mu,T=0,\vect\chi)\right)\bigg |_{\mu=0} .
\end{equation}
This expression shows explicitly that the subtraction of the counterterm $Y^{C.T.}$ corresponds to a regularization of the $\mu^2$ contributions with a sharp momentum cut-off $\Lambda'$.

\subsection{Three-flavor case}
The divergence in the neutral three-flavor case is associated with various diquark condensates coupled to distinct chemical potentials. In this case, there is no analytic expression for the quasiparticle dispersion relations. Thus, unlike the simplified cases discussed in Sec.~\ref{sec:meddiv}, the analysis of Eqs.~\eqref{2scdiv1} and \eqref{2scdiv2} cannot directly be applied here.
However, through a detailed procedure outlined in Appendix \ref{app:medium_divergences}, we provide an approach to analytically extract the exact form of the medium divergences. 
Following the calculations there, the form of the divergences in the case of three flavors with different chemical potentials
$\mu_{fc}$ for different quark species with flavor $f$ and color $c$ is
\begin{align}\label{cflgammainfinite}
\frac{1}{V_4}\Gamma \simeq -\frac{1}{4\pi^2}\ln\Lambda \bigg(
& ((\mu_{ur} + \mu_{dg})^2 + (\mu_{ug} + \mu_{dr})^2)\Delta_{3}^2 \nonumber\\
 +& ((\mu_{ur} + \mu_{sb})^2 + (\mu_{ub} + \mu_{sr})^2)\Delta_{2}^2 \nonumber\\
+& ((\mu_{dg} + \mu_{sb})^2 + (\mu_{sg} + \mu_{db})^2)\Delta_{1}^2
\bigg) \nonumber\\
\equiv\; - \frac{1}{\pi^2} \ln\Lambda\bigg(
& (\mu_{ur,dg}^2 + \mu_{ug,dr}^2)\Delta_{3}^2 \nonumber\\
+&  (\mu_{ur,sb}^2 + \mu_{ub,sr}^2)\Delta_{2}^2 \nonumber\\
+&  (\mu_{dg,sb}^2 + \mu_{sg,db}^2)\Delta_{1}^2
\bigg),
\end{align}
where in the last line we introduced the notation $\mu_{\alpha a,\beta b}\equiv\frac{1}{2}(\mu_{\alpha a}+\mu_{\beta b})$.
Comparing the above expression with the definition of $\Delta_A$ in \Eq{eq:DeltaA}, we see that each diquark condensate is multiplied with the average chemical potentials of the quark pairs involved in this condensate (see also Table~\ref{tab:gaps}).

The task is now to construct appropriate counterterms to cancel these divergences. For the consistency of a general formulation, 
the counterterms must meet the following requirements
\begin{enumerate}
    \item[{\crtcrossreflabel{(R1)}[req:1.1]}] For $\vect\Delta=0$ or $\vect\mu=0$ the counterterms should vanish. 
    \item[{\crtcrossreflabel{(R2)}[req:1.2]}] For any $\Delta_A = 0$ the associated counterterm should vanish.
\end{enumerate}
The second requirement implies that for the 2SC phase, i.e., $\Delta_1=\Delta_2=0$, we reproduce the results of Sec.~\ref{sec:meddiv}.

Expressing the different chemical potentials in terms of $\mu$, $\mu_Q$, $\mu_3$ and $\mu_8$ (see Sec.~\ref{subsec:chempot}) it is easy to see that for quark matter in weak equilibrium the two diquark pairs which are related to the same $\Delta_A$ have identical average chemical potentials. Then, using the notation $\mu_{\alpha \beta;ab}\equiv\mu_{\alpha a,\beta b}=\mu_{\beta a,\alpha b}$, \Eq{cflgammainfinite} is simplified to
\begin{equation}{\label{h:5}}
\frac{1}{V_4}\Gamma\simeq -\frac{2}{\pi^2}\ln\Lambda \left( {\mu}_{ud;rg}^2\Delta_{3}^2+
      {\mu}_{us;rb}^2\Delta_{2}^2+
      {\mu}_{ds;gb}^2\Delta_{1}^2
      \right).
\end{equation}
Comparing this with \Eq{2SCnochemdiv}, this resembles three decoupled two-flavor pairings, with each diquark gap and the corresponding average chemical potential contributing to an independent divergence. From this we conclude that we need three independent counterterms of the form of \Eq{generalansatzCT} to cancel these divergences.

However, here we want to keep the problem more general, for situations where $\mu_{\alpha a,\beta b}$ and $\mu_{\beta a,\alpha b}$
are not identical. We therefore treat them as independent and introduce a total of six counterterms, one for each of the six terms in \Eq{cflgammainfinite}:
 \begin{eqnarray}
 \label{3f:CT}
  \delta \mathcal{L}=-\frac{1}{2}\big( &y_{ur,dg}\mu^2_{ur,dg}+&y_{ug,dr}\mu^2_{ug,dr}\nonumber\\
  +&y_{ur,sb}\mu^2_{ur,sb}+&y_{ub,sr}\mu^2_{ub,sr}\nonumber\\\
  +&y_{dg,sb}\mu^2_{dg,sb}+&y_{db,sg}\mu^2_{db,sg}\,\big).
  \end{eqnarray}
For the renormalization factors we take
\begin{equation}\label{eq:RenoFacMass}
    y_{\alpha a,\beta b}(\Delta_{\alpha a,\beta b},\vect{M})=
    -\frac{1}{V_4}
    \frac{\partial^2\Gamma_{\Lambda'}}{\partial\mu_{\alpha a,\beta b}^2}\Bigg|_{T=\vect\mu=0;\Delta_{\alpha a,\beta b}\neq0},
\end{equation}
where $\Delta_{\alpha a,\beta b}$ denotes the gap which is associated with the diquark pairing of flavor $\alpha$ and color $a$ with flavor $\beta$ and color $b$:
\begin{eqnarray}
    \Delta_{ur,dg}&=\Delta_{ug,dr}=&\Delta_3\\
    \Delta_{ur,sb}&=\Delta_{ub,sr}=&\Delta_2\\
    \Delta_{dg,sb}&=\Delta_{db,sg}=&\Delta_1.
\end{eqnarray}
The notation $\Delta_{\alpha a,\beta b}\neq0$ is to be understood that all \textit{other} diquark gaps are set to zero. 
This condition can be motivated by the observation above that we have independent divergences for each gap. 
Therefore we require that each renormalization factor is also independent of the diquark condensates it is not associated with. In fact, this requirement turns out to be necessary to cancel the divergences.

The renormalized effective action is then obtained in complete analogy to \Eq{h:3} and reads:
\begin{align}\label{massgamma}
\frac{\Gamma^R}{V_4}(\vect\mu,T,\vect\chi) 
&= \mathcal{V}(\vect\chi) 
- \frac{1}{2\pi^2} \bigg( \int_0^{\Lambda} dp\, p^2 \mathcal{A}(\vect{\mu},T,\vect\chi) \nonumber\\
&\qquad - \int_{\Lambda'}^{\Lambda} dp\, p^2 \mathcal{A}_{\text{vac}}(\vect\chi) \nonumber\\
&\qquad - \int_{\Lambda'}^{\Lambda} dp\, p^2 \sum \frac{1}{2}\mu_{\alpha a,\beta b}^2\nonumber\\
&\times  
\left(\frac{\partial^2}{\partial \mu_{\alpha a,\beta b}^2} \mathcal{A}(\vect\mu,0,\vect\chi)\right) \bigg|_{\vect\mu=0;\Delta_{\alpha a,\beta b}\neq 0} \bigg).
\end{align}
The summation in the above equation is taken over the six average chemical potentials $\mu_{\alpha a,\beta b}\in\{\mu_{ur,dg},\mu_{ug,dr},\mu_{ur,sb},\mu_{ub,sr},\mu_{dg,sb},\mu_{db,sg}\}$. 
\Eq{massgamma} is our main result.
Using again the technique described in appendix \ref{app:medium_divergences}, we calculated the asymptomatic behavior of the counterterm in \Eq{3f:CT}. The result is
\begin{eqnarray}
    \delta\mathcal{L} \simeq -  \frac{1}{\pi^2} \ln (\Lambda) \bigg( 
     &(\mu_{ur,dg}^2 &+ \mu_{ug,dr}^2)\Delta_{3}^2 \nonumber\\
    + &(\mu_{ur,sb}^2 &+ \mu_{ub,sr}^2)\Delta_{2}^2 \nonumber\\
    + & (\mu_{dg,sb}^2 &+ \mu_{sg,db}^2)\Delta_{1}^2 \bigg).
\end{eqnarray}
Plugging this into \Eq{CT2} cancels the divergence \Eq{cflgammainfinite}. Thus \Eq{massgamma} ensures
\begin{equation}
\lim_{\Lambda\to\infty}\frac{\Gamma^R}{V_4}  = \textit{finite}  
\end{equation}
and makes the model RG consistent
\begin{equation}
\lim_{\Lambda\to\infty}\Lambda\frac{\partial}{\partial\Lambda}\Gamma^R=0.
\end{equation}

Moreover, the requirements \eqref{req:1.1} and \eqref{req:1.2} are also fulfilled, as verified in Appendix~\ref{app:massschemeproof}.
In particular, in agreement with the second requirement \eqref{req:1.2}, 
\Eq{massgamma} reduces to \Eq{h:3} for $\Delta_1 = \Delta_2 =0$ and is thus a generalization of the 2SC case. 

The prescription \eqref{3f:CT} for the counterterms is not unique. There are alternative choices which are also RG consistent and meet the requirements \eqref{req:1.1} and \eqref{req:1.2}. This issue will be addressed next.

\subsection{Alternative schemes}\label{sec:schemes}

As obvious from \Eq{cflgammainfinite}, even in the presence of quark masses (both dynamical and bare), the medium divergence does not scale with the quark masses. Given this, the counterterm \eqref{3f:CT} with the renormalization factors \eqref{eq:RenoFacMass} includes nonzero mass-dependent terms. By subtracting it in \Eq{massgamma}, these mass-dependent terms are eliminated from the effective action. 
One may argue that these subtractions are excessive.
Given the freedom to choose the renormalization scheme, we propose alternative renormalization schemes that avoid unnecessary subtractions. We refer to the renormalization scheme defined in \Eq{eq:RenoFacMass} as the \textit{massive scheme} and introduce two new schemes in this section, which we call the \textit{massless scheme} and the \textit{minimal scheme}.

\subsubsection{Massless scheme}
To avoid the unnecessary subtraction of mass-dependent terms in \Eq{massgamma}, we set the quark masses in the renormalization factors \Eq{eq:RenoFacMass} to zero,
\begin{alignat}{2}\label{CT4}
    y_{\alpha a,\beta b}(\vect{\chi})&=& -\frac{1}{V_4}\frac{\partial^2\Gamma_{\Lambda'}}{\partial\mu_{\alpha a,\beta b}^2}\Bigg|_{T=\vect\mu=\vect M=0;\Delta_{\alpha a,\beta b}\neq0},
\end{alignat}
and hence the counterterm as
\begin{align}{\label{masslessCT}}
   \delta\mathcal{L}=&-\int_{\Lambda'}^{\Lambda} dp\,p^2 \sum\frac{1}{2} \mu_{\alpha a,\beta b}^2\nonumber\\&\times\left(\frac{\partial^2}{\partial\mu_{\alpha a,\beta b}^2}\mathcal{A}(\vect\mu,0,\vect\chi)\right)\bigg|_{\vect\mu=\vect M=0;\Delta_{\alpha a,\beta b}\neq0}.
\end{align}
We call this choice the \textit{massless scheme}\footnote{Note that the condition $\vect M=0$ is different from setting the quark-antiquark condensates $\phi_i$ to zero.}. With this choice it is possible to get an analytic expression for the counterterms which is provided in Appendix~\ref{app:masslessscheme}. This choice satisfies both requirements \eqref{req:1.1} and \eqref{req:1.2} and makes the effective action RG consistent.

\subsubsection{Minimal scheme}

Although the massless scheme avoids unnecessary mass-dependent subtractions, it still includes subtractions of non-divergent contributions from the diquark condensates. A Taylor expansion of the integrand in \Eq{masslessCT} around  $\vect\Delta=0$ yields
\begin{align}
&\frac{1}{2} \mu_{\alpha a,\beta b}^2\left(\frac{\partial^2}{\partial \mu_{\alpha a,\beta b}^2} \mathcal{A}(\vect\mu, 0, \vect\chi)\right)\bigg|_{\vect\mu = \vect M = 0; \Delta_{\alpha a,\beta b} \neq 0} \nonumber\\
&= \frac{1}{2} \mu_{\alpha a,\beta b}^2\left(\frac{\partial^2}{\partial \mu_{\alpha a,\beta b}^2} \mathcal{A}(\vect{\mu}, 0, \vect\chi)\right)\bigg|_{\vect\mu = \vect M = \vect\Delta = 0} \nonumber\\
&\quad + \frac{1}{4} \mu_{\alpha a,\beta b}^2 \Delta_{\alpha a,\beta b}^2\nonumber\\&\qquad\quad\times\left(\frac{\partial^4}{\partial \Delta_{\alpha a,\beta b}^2 \partial \mu_{\alpha a,\beta b}^2} \mathcal{A}(\vect{\mu}, 0, \vect\chi)\right)\bigg|_{\vect\mu = \vect M = \vect\Delta = 0} \nonumber\\
&\quad + \mathcal{O}(\Delta_{\alpha a,\beta b}^4).
\end{align}

According to requirement \eqref{req:1.1},
the first term in the RHS above is zero.
Inserting the remaining terms into \Eq{masslessCT} and comparing this with the divergence \eqref{cflgammainfinite} we find
that it is enough to subtract only the $\Delta_{\alpha a,\beta b}^2$ terms, i.e., neglect the $\mathcal{O}(\Delta_{\alpha a,\beta b}^4)$ contributions. This scheme, which we call the \textit{minimal scheme}, is thus defined by the counterterm 
\begin{eqnarray}\label{eq:minimaldef}
    \delta\mathcal{L}&=&-\int_{\Lambda'}^{\Lambda} dp\,p^2 \sum\frac{1}{4} \mu_{\alpha a,\beta b}^2\Delta_{\alpha a,\beta b}^2\nonumber\\&&\qquad\times\left(\frac{\partial^4}{\partial\Delta_{\alpha a,\beta b}^2\partial\mu_{\alpha a,\beta b}^2}\mathcal{A}(\vect{\mu},0,\vect\chi)\right)\bigg|_{\vect\mu=\vect M=\vect\Delta=0}.
\end{eqnarray}
Due to the simplicity of the dispersion relations for $\vect M=0$, the integral can be evaluated analytically. One finds 
\begin{eqnarray}\label{eq:minimalexpl}
    \delta\mathcal{L}=-\frac{1}{\pi^2}\ln\left(\frac{\Lambda}{\Lambda'}\right) \bigg(&
(\mu_{ur,dg}^2&+\mu_{ug,dr}^2)\Delta_{3}^2\nonumber\\+&
 (\mu_{ur,sb}^2&+\mu_{ub,sr}^2)\Delta_{2}^2\nonumber\\+&
      (\mu_{dg,sb}^2&+\mu_{sg,db}^2)\Delta_{1}^2
      \bigg),
\end{eqnarray}
which exactly eliminates the divergence \eqref{cflgammainfinite} through equation \Eq{CT2}. 
It is also obvious that this choice is consistent with the two requirements \eqref{req:1.1} and \eqref{req:1.2}. 

Before closing this section, we come back to the general renormalization procedure. As pointed out before, the NJL model is not renormalizable and therefore the medium divergences cannot be absorbed into existing bare vertices.
However, the condensates $\Delta_A$ can be identified with the constant expectation values of diquark fields $\Delta_A(x)$, which can be introduced to our model via bosonization techniques (Hubbard-Stratonovich transformations). We may then consider a kinetic term,
\begin{equation}
\label{eq_LDeltakin}
\mathcal{L}_{\Delta,\text{kin}} =
    \sum_{A} Z_{\Delta_A}
    \Delta_A^*\Big((i\partial_0+\mu_{\Delta_A})^2+\nabla^2-m_{\Delta_A}^2\Big)\Delta_A,
\end{equation}
for these fields, with diquark masses $m_{\Delta_A}$ and diquark chemical potentials $\mu_{\Delta_A}$. The latter correspond to the total chemical potentials of the given pairs, i.e., twice their average chemical potentials $\mu_{\alpha a, \beta b}$ we defined previously. Furthermore we introduced the renormalization factors $\sqrt{Z_{\Delta_A}}$ for the fields.

Unlike the original NJL model, $\mathcal{L}_{\Delta,\text{kin}}$ contains a term quadratic in both $\Delta_A$ and $\mu_{\Delta_A}$. Hence, adding this term would allow us to perform the renormalization procedure outlined at the beginning of this section. The corresponding ``coupling'', obtained by taking  
the 4th derivative $\frac{\partial^4}{\partial\Delta_A^*\partial\Delta_A\partial\mu_{\Delta_A}^2}$, can be identified with $Z_{\Delta_A}$.\footnote{Note that in contrast to the coupling $Y^0$ defined in \Eq{eq:Y0} via a second derivative w.r.t.\ the chemical potential, $Z_{\Delta_A}$ is dimensionless, in agreement with general expectations for a renormalizable model.} 
The counterterm \eqref{eq:minimalexpl} then naturally arises from writing $Z_{\Delta_A} = 1 + \delta Z_{\Delta_A}$ and the renormalization condition that the scale dependent renormalization constant $Z_{\Delta_A}^k$ vanishes at $k=\Lambda'$  
(see discussion around \Eq{eq:YRexpl}).

In the NJL model, while diquark masses and field renormalization factors are generated dynamically at the one-loop level, the tree-level Lagrangian $\mathcal{L}_{\Delta,\text{kin}}$ does not exist, and therefore the arguments above can only be taken as a motivation for the introduction of \Eq{eq:minimalexpl} by hand. From this perspective the minimal scheme appears to be the most natural choice. On the other hand, $\mathcal{L}_{\Delta,\text{kin}}$ exists in the Quark-Meson-Diquark model.  Furthermore, in a  
fully renormalized meanfield Quark-Meson-Diquark model, there is no regularization dependence, even in the vacuum (contrary to the case studied here). This possibility is presently under investigation and will be published elsewhere \cite{QMDFUTURE}.\footnote{After submission of our paper to the arXiv, Ref.~\cite{Andersen:2024qus} appeared, where the renormalization of the Quark-Meson-Diquark model is discussed as well.}

\section{Results}
\label{sec:results}

In this section we present numerical results obtained within our RG-consistent framework and compare them with the conventional cutoff scheme. 
We employ the parameters used in Ref.~\cite{Ruester:2005jc}, given by $\Lambda'=602.3\,$MeV, $G_S\Lambda'^2=1.835$, $K\Lambda'^5=12.36$ and the bare quark masses $m_{u,d}=5.5$\,MeV and $m_s=140.7$\,MeV. They have been fitted to the pseudoscalar meson octet in vacuum \cite{Rehberg:1995kh}. Furthermore we set $G_D=G_S$, corresponding to the ``strong diquark coupling'' of Ref. \cite{Ruester:2005jc}.

\begin{figure*}[t]
		\begin{minipage}[t]{0.32\textwidth}		 		
  \includegraphics[width=\textwidth]{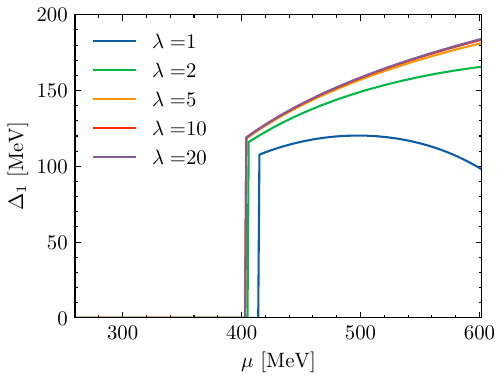}
			 	\end{minipage}
		 		\begin{minipage}[t]{0.32\textwidth}
			 		\includegraphics[width=\textwidth]{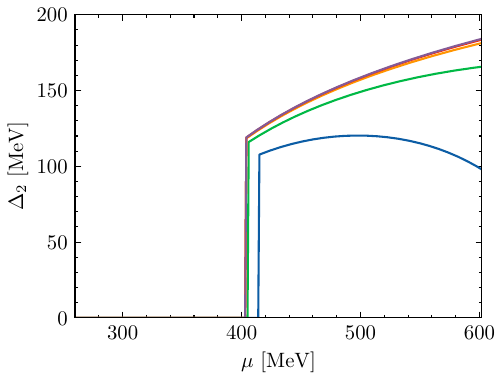}
			 	\end{minipage}
		 	\begin{minipage}[t]{0.32\textwidth}
		\includegraphics[width=\textwidth]{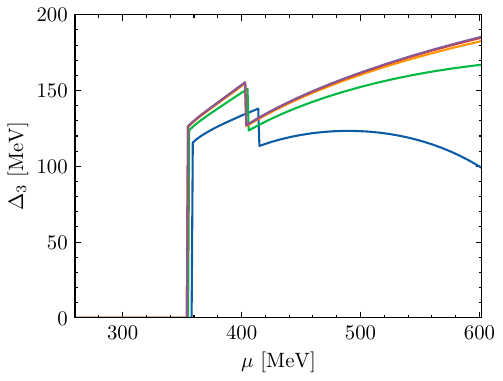}
			 	\end{minipage}
			 			\caption{Diquark condensates $\Delta_1$ (left), $\Delta_2$ (middle) and $\Delta_3$ (right) as functions of the chemical potential at $T=0$ for different ratios $\lambda=\Lambda/\Lambda'=1,2,5,10,20$ in the massless scheme.}
		\label{fig:convergence}	 	
     \end{figure*}	

     \begin{figure*}[t]
		\begin{minipage}[t]{0.32\textwidth}		 		
  \includegraphics[width=\textwidth]{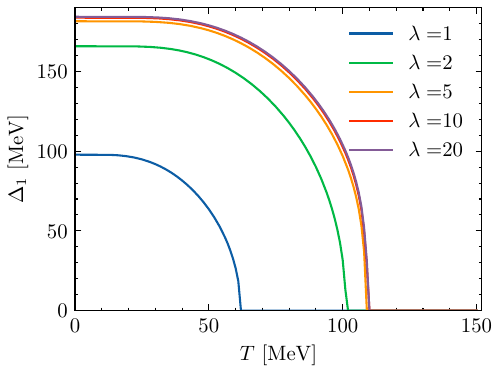}
			 	\end{minipage}
		 		\begin{minipage}[t]{0.32\textwidth}
			 		\includegraphics[width=\textwidth]{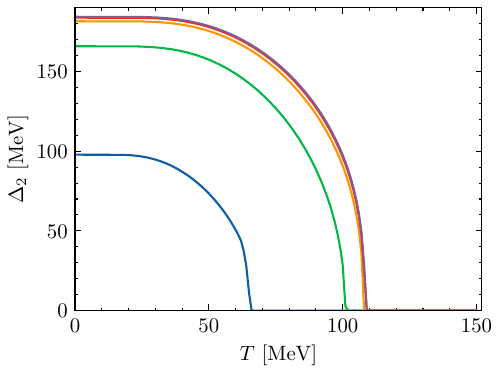}
			 	\end{minipage}
		 	\begin{minipage}[t]{0.32\textwidth}
		\includegraphics[width=\textwidth]{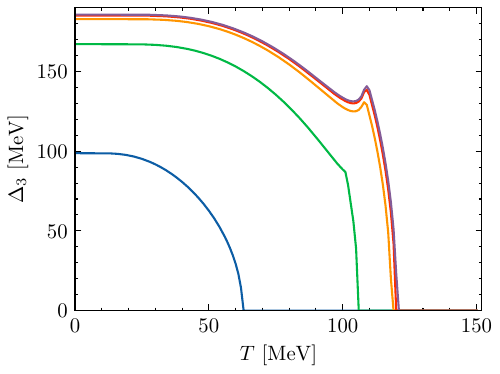}
			 	\end{minipage}
			 			\caption{Diquark condensates $\Delta_1$ (left), $\Delta_2$ (middle) and $\Delta_3$ (right) as functions of the temperature at $\mu=\Lambda'=602.3$\,MeV  for different ratios $\lambda=\Lambda/\Lambda'=1,2,5,10,20$ in the massless scheme.}
		\label{fig:convergence_at_cutoff}	 	
     \end{figure*}	

For these parameters we solve the gap equations \eqref{eq:gapeq} and apply the neutrality constraints \eqref{eq:neutrality}. In the massive scheme, the counterterm (last term in \Eq{massgamma}) contains second derivatives of $\mathcal{A}$ which in general have to be taken numerically. Hence in order to obtain the gap equations and neutrality conditions, we need to calculate third derivates of $\mathcal{A}$. The analytic expressions needed for this are calculated using the method outlined in Appendix \ref{app:gap}. For the massless and the minimal schemes the situation is simpler because we have analytic expressions for the counterterms, see Eqs.~\eqref{TempdeltaL} and \eqref{eq:minimalexpl}.

\subsection{Test of the implementation}

In order to fulfill the RG-consistency criterion \eqref{eq:crit1}, the effective action $\Gamma$ and thus all results in this paper, must not depend on the UV scale $\Lambda$ if $\Lambda$ is chosen sufficiently large. We test this for the different schemes introduced in the previous section by solving the gap equations numerically for different UV scales. In general we find fast convergence of the results with increasing $\Lambda$. For the massless scheme, this is demonstrated in Figs.~\ref{fig:convergence} and \ref{fig:convergence_at_cutoff}, where the diquark condensates are displayed for different ratios $\lambda=\Lambda/\Lambda^{'}$ as functions of the quark chemical potential at vanishing temperature and as functions of the temperature for a quark chemical potential of $\mu=602.3\,$MeV (the value of $\Lambda '$), respectively. Increasing the UV scale strongly changes the results for $1<\lambda<5$, showing the appearance of cutoff-artifacts already at chemical potentials well below the original vacuum cutoff. 
 
For $\lambda\gtrsim5$, however, the results become independent of a further increase of the UV scale, as required for an RG-consistent model. More precisely, we find that a doubling of the UV-scale from $\lambda=10$ to $\lambda=20$ changes the value of the diquark condensates by less than 1\% over the whole range of chemical potentials and temperatures shown in Figs.~\ref{fig:convergence} and \ref{fig:convergence_at_cutoff}.
Thus, with choosing $\Lambda=10\,\Lambda'$, the model is sufficiently close to the RG-consistent point. For the calculation of the phase diagram, which will be discussed in  Sec.~\ref{sec:phasediagram}, we have therefore chosen this value.
The cutoff artifacts at low values of $\lambda$, as well as the fast convergence to the RG-consistent model are both in accordance with the results obtained for a Quark-Meson-Diquark model in Ref.~\cite{Braun:2018svj}.

Comparing the converged results of the RG-consistent calculations with those obtained with conventional cutoff regularization ($\lambda=1$), we see that in Fig.~\ref{fig:convergence} the onset of the 2SC phase, indicated by the appearance of a nonzero $\Delta_3$, as well as the 2SC$\to$CFL phase boundary, indicated by the appearance of nonzero $\Delta_1,\Delta_2$ and a drop in $\Delta_3$, move to lower chemical potentials. This effect is larger for the 2SC$\to$CFL transition than for the onset of the 2SC phase. This is expected, as the 2SC$\to$CFL transition happens at a larger $\mu/\Lambda'$-ratio which is more sensitive to cutoff artifacts. The largest effect, however, can be seen in the value of the diquark condensates themselves. 
In the CFL phase, at the highest chemical potential shown, they reach values that are almost 90\% higher in the RG-consistent scheme than for the conventional cutoff regularization ($\lambda = 1$). For $\Delta_3$ in the 2SC phase, this effect is again smaller (increase of up to 20\%). Qualitatively, we find that already for $\lambda\geq2$, all the diquark condensates keep increasing for the shown chemical potentials.

Related to the increase of the condensates at $T=0$, we find in Fig.~\ref{fig:convergence_at_cutoff} much higher melting temperatures for the RG-consistent case than for $\lambda=1$. Another interesting feature is the appearance of a small peak in $\Delta_3$ around the melting temperature of $\Delta_1$ and $\Delta_2$.
A similar behavior was also seen in Ref.~\cite{Buballa:2001gj} with conventional cutoff regularization, so it is not an entirely new feature of our RG-consistent scheme. 
Its nature is related to the fact that the 2SC gaps are generally larger than the CFL gaps for a fixed chemical potential and temperature \cite{Schmitt:2002sc}, which also explains the drop of $\Delta_3$ in Fig.~\ref{fig:convergence}. Considering $\Delta_3$ as a function of temperature, the peak which would develop due to the drop of  $\Delta_1$ and $\Delta_2$ can be washed out by the drop of $\Delta_3$ itself and therefore the existence of such a peak depends on details of the parameters. In the present example, however, the nonexistence of the peak for $\lambda=1$ is easily explained by the observation that in this case $\Delta_3$ vanishes \textit{earlier} than $\Delta_2$, giving rise to a 2SC$_{us}$ phase in this region. Since there is no physical reason for such a phase, this is clearly a cutoff artifact. We will come back to it in Sec.~\ref{sec:phasediagram}.

\subsection{Behavior at very large chemical potentials}
\label{sec:largemu}

The RG-consistent regularization allows us to study the model at arbitrarily large chemical potentials without running into cutoff artifacts. In particular, $\mu$ is not restricted to be lower than the scale $\Lambda'$ at which the model parameters were fixed in vacuum. Of course the model itself has a limited range of validity and does not capture the properties of perturbative QCD at ultra-high densities. However, besides being interesting in its own right, exactly for that reason it is important to know how the model behaves at very large chemical potentials. 

\begin{figure}[h]
		 		\begin{minipage}[t]{0.45\textwidth}
			 		\includegraphics[width=\textwidth]{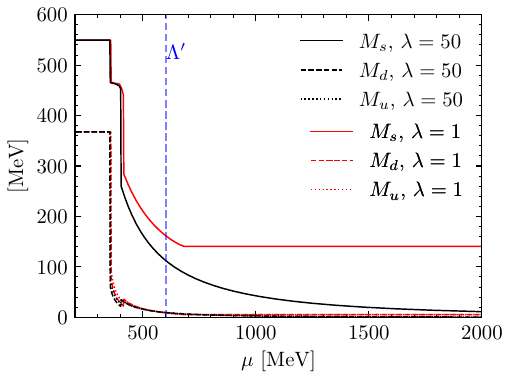}
			 	\end{minipage}
		 	\begin{minipage}[t]{0.45\textwidth}
		\includegraphics[width=\textwidth]{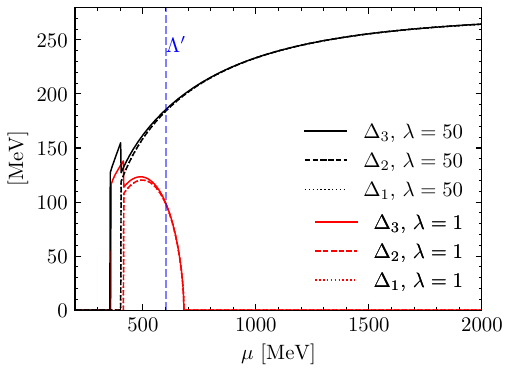}
			 	\end{minipage}
			 			\caption{Solutions of the mass gap equations (top) and diquark gap equations (bottom) as functions of the chemical potential $\mu$ at $T=0$.
       The red curves correspond to the model of Ref.~\cite{Ruester:2005jc} with conventional cutoff regularization, the black curves to the RG-consistent regularization (massless scheme) with $\lambda=50$. The blue dashed line indicates $\mu=\Lambda'$.}       

       \label{fig:gapsbeyond}
			 \end{figure}

Fig.~\ref{fig:gapsbeyond} shows the solutions of the gap equations at $T=0$ as functions of the chemical potential up to $\mu=2000$~MeV. The results with conventional regularization ($\lambda = 1$, red lines) are compared with the RG-consistent ones (black lines). The latter are obtained with $\lambda = 50$ in order to ensure that $\Lambda = \lambda\Lambda' \gg \mu$ in the entire region.  
For $\lambda=1$, we see that the diquark gaps (bottom panel) do not only decrease beyond a certain value of $\mu$ but eventually even vanish at $\mu \approx 680$~MeV. At the same point the dynamical quark masses (top panel) go to the current masses. In other words, the loop contributions are cut off completely and we are left with a non-interacting theory beyond that point. With the RG-consistent regularization, on the other hand, the diquark gaps keep growing while the masses drop below the values of current quark masses and approach zero as $\mu\to\infty$.

A detailed analysis of the asymptotic behavior of the masses and diquark gaps in our model is beyond the scope of the present paper. Here we just note that both the rise of $\Delta_A$ and the decrease of the masses are in qualitative agreement with their behavior in high-density QCD. Although the physical origin and, related to this, the functional form of the running are probably very different, this puts us into the position to compare our results at high densities with model independent predictions based on weak-coupling QCD in a meaningful way. We will come back to this at the end of Sec.~\ref{sec:CFL_melting}.

\begin{figure*}[t]
		\begin{minipage}[t]{0.32\textwidth}		 		
  \includegraphics[width=\textwidth]{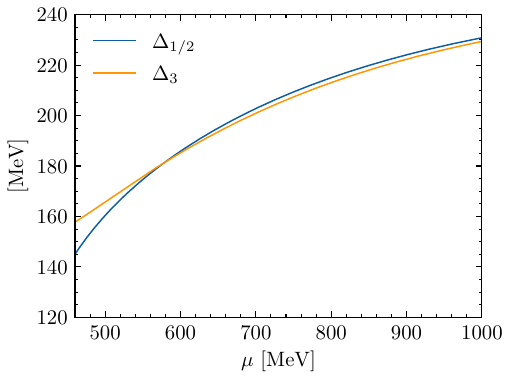}
			 	\end{minipage}
		 		\begin{minipage}[t]{0.32\textwidth}
			 		\includegraphics[width=\textwidth]{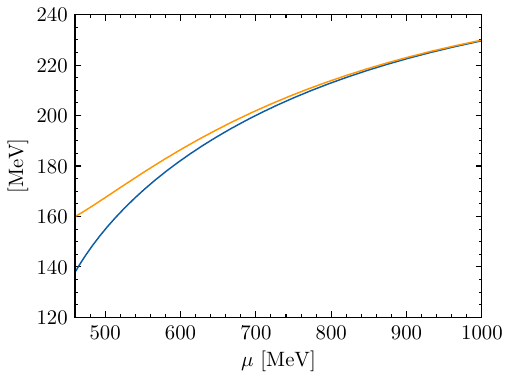}
			 	\end{minipage}
		 	\begin{minipage}[t]{0.32\textwidth}
		\includegraphics[width=\textwidth]{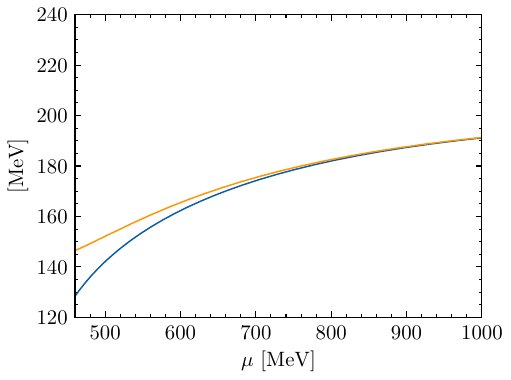}
			 	\end{minipage}
			 			\caption{Diquark condensates $\Delta_1 = \Delta_2$ and $\Delta_3$ as functions of a common chemical potential at $T=0$  in the massive (left), massless (middle) and minimal (right) scheme with $\lambda =10$.}
 
		\label{fig:scheme_comparison_neutral}	 	
     \end{figure*}	
     \begin{figure*}[t]
		\begin{minipage}[t]{0.32\textwidth}		 		
  \includegraphics[width=\textwidth]{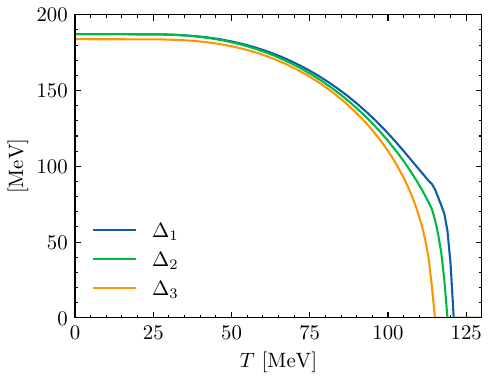}
			 	\end{minipage}
		 		\begin{minipage}[t]{0.32\textwidth}
			 		\includegraphics[width=\textwidth]{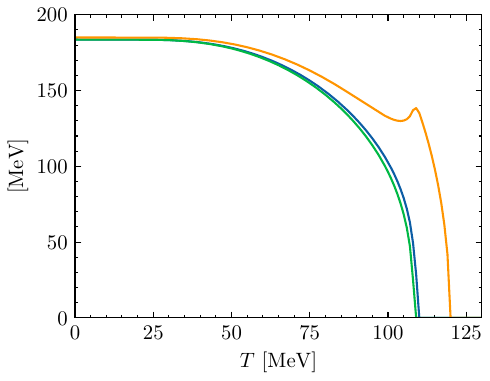}
			 	\end{minipage}
		 	\begin{minipage}[t]{0.32\textwidth}
		\includegraphics[width=\textwidth]{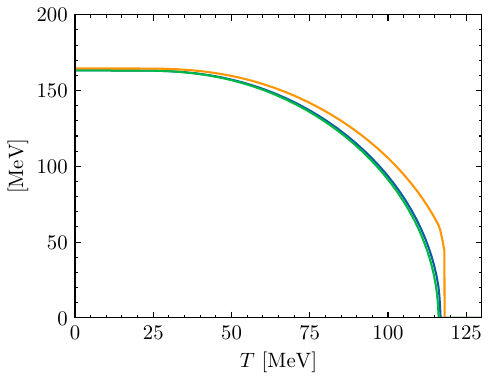}
			 	\end{minipage}
			 			\caption{Diquark condensates $\Delta_1,\Delta_2,\Delta_3$ for color and electrically neutral matter as functions of the temperature at $\mu=\Lambda'=602.3$\,MeV in the massive (left), massless (middle) and minimal (right) scheme with $\lambda =10$.} 

		\label{fig:scheme_comparison}	 	
     \end{figure*}	
     
\subsection{Comparison of different regularization schemes}

In this section we compare the three different schemes defined in Sec.~\ref{sec:renopro}, i.e., the massive, the massless and the minimal scheme. 
In Fig.~\ref{fig:scheme_comparison_neutral}	the diquark condensates $\Delta_1 = \Delta_2$ and $\Delta_3$ for the three schemes are plotted as functions of a common, i.e., color and flavor independent, chemical potential at $T=0$. Whereas the results of the massive (left) and massless (middle) schemes are close, the diquark condensates in the minimal scheme are significantly smaller (right). The difference increases with increasing chemical potential. This can be traced back to the fact that the difference arises from $\Delta_A$-dependent finite contributions to the counterterms (see Sec.~\ref{sec:schemes}), which grow as the diquark condensates also grow with increasing $\mu$.

However, most importantly, a small but qualitative difference is seen at large chemical potential. For the massive scheme (left panel) we find that at $\mu\approx 570 \text{ MeV}$ the lines cross and, above this point, $\Delta_3$ is smaller than $\Delta_1 = \Delta_2$. This behavior is clearly unphysical: 
Since $M_s$ is larger than $M_u$ and $M_d$, the strange quarks have a lower Fermi momentum than the nonstrange quarks at equal chemical potentials. Therefore the pairing of a strange with a nonstrange quark in $\Delta_1$ and $\Delta_2$ is slightly suppressed compared with $\Delta_3$, where only nonstrange quarks are involved. As a consequence we should always have $\Delta_1 = \Delta_2 < \Delta_3$. With increasing chemical potential the relative importance of the strange-quark mass decreases. We therefore expect that the curves approach each other but they should never cross.  In fact the expected behavior is exactly what we find in the massless (middle) and the minimal (right) scheme. From this we conclude that the mass-dependent finite contributions to the counterterms in the massive scheme (see Sec.~\ref{sec:schemes}) are excessive, leading to this unphysical behavior.

Fig.~\ref{fig:scheme_comparison} shows a comparison of the different schemes for the calculation of the diquark condensates as a function of $T$ at $\mu=\Lambda'=602.3\text{ MeV}$ under neutrality conditions. As we have seen before, the condensates at low temperature are smaller in the minimal scheme than in the two other schemes. However, this is not reflected in the critical temperatures which are more or less comparable, although there are important differences in detail. One of them has the effect that the peak in $\Delta_3$ which we encountered already in Fig.~\ref{fig:convergence_at_cutoff} only appears in the massless scheme (middle), where the difference between the critical temperature of $\Delta_3$ and the two others is largest.

The most important qualitative difference, however, is again between the massive (left) and the two other schemes. While both massless and minimal scheme have the melting pattern CFL$\,\to\,$dSC$\,\to\,$2SC, the massive scheme shows the melting pattern CFL$\,\to\,$sSC$\,\to\,$$\text{2SC}_{ds}$. 
As we will discuss in more details in Sec.~\ref{sec:CFL_melting},
the first melting pattern is in agreement with the GL predictions of Refs.~\cite{Iida:2003cc,Iida:2004cj} for high densities. However, there is no physical reason why the $u$-$d$ pairs of $\Delta_3$ should be disrupted first and the $d$-$s$ pairs of $\Delta_1$ survive the longest, as we find for the massive scheme.

Technically, this behavior seems to be related to the observation that already at zero temperature, $\Delta_3$ is suppressed compared to $\Delta_1$ and $\Delta_2$, similar to what we found in Fig.~\ref{fig:scheme_comparison_neutral}	and what we explained by the mass-dependent finite contributions to the counterterms in the massive scheme.  

\label{sec:phasediagram}
	 	\begin{figure*}[t]
                                \centering
			 			\includegraphics[width=0.95\textwidth]{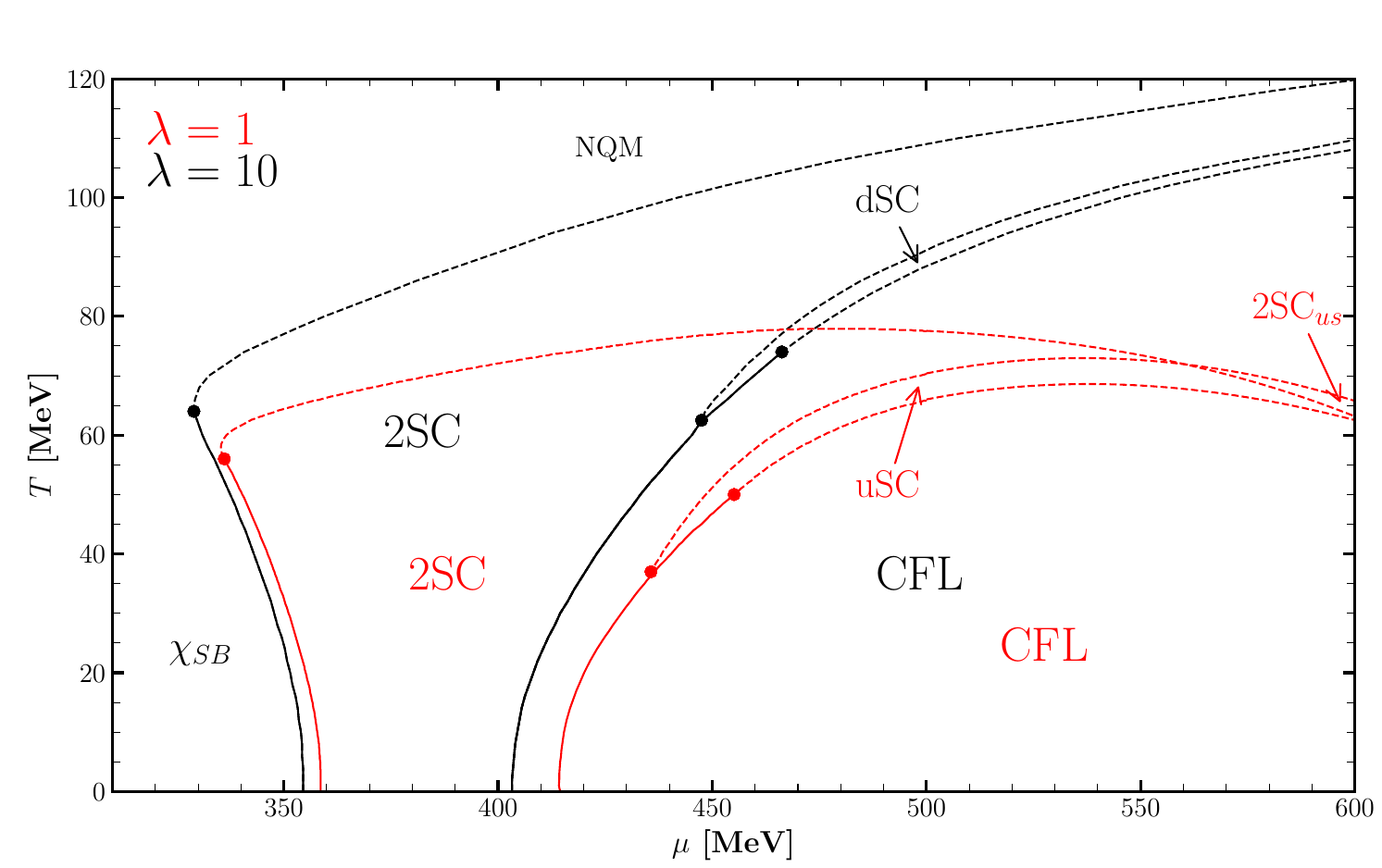}
			 			\caption{Phase diagram of neutral quark matter for $G_D/G_S=1$: RG-consistent calculation with $\lambda=10$ (black) and conventional cutoff regularization ($\lambda=1$) of Ref.~\cite{Ruester:2005jc} extended to higher chemical potential (red). Solid (dashed) lines indicate first (second) order phase transitions. 
       }
       \label{fig:pd}
			 \end{figure*}

In summary, we found unphysical features in the massive scheme in the high-density region, namely in Fig.~\ref{fig:scheme_comparison_neutral} the wrong ordering of the gap sizes at $T=0$ and in Fig.~\ref{fig:scheme_comparison} a wrong melting pattern. Therefore the massless and minimal schemes are preferable.

\subsection{RG-consistent phase diagram of neutral quark matter}
\label{sec:phasediagram}

By minimizing the effective potential of the RG-consistent model in the massless scheme with $G_D=G_S$ for different quark chemical potentials and temperatures under the neutrality constraints, we obtain the phase diagram shown in Fig.~\ref{fig:pd}. The UV scale is again set to $\Lambda=10\,\Lambda'$, which is large enough to eliminate cutoff artifacts, see Figs.~\ref{fig:convergence} and \ref{fig:convergence_at_cutoff}. In Fig.~\ref{fig:pd}, we also show the phase diagram for the same parameters with conventional cutoff regularization (corresponding to $\Lambda=\Lambda'$) from Ref.~\cite{Ruester:2005jc}. For better comparison, we extended the calculations of Ref.~\cite{Ruester:2005jc} up to $\mu=600\,$MeV.\footnote{For simplicity, we do not distinguish between gapped and gapless color-superconducting  phases \cite{Shovkovy_gapless,Alford:2003fq} (``Sarma phases'') in Fig.~\ref{fig:pd}.}

Comparing the two phase diagrams, we find several features qualitatively unchanged:  
At low temperatures and chemical potentials, there is a non-color-superconducting phase with large chiral-symmetry breaking ($M_{u,d} \gg m_{u,d}$).
With increasing temperature the chiral condensates melt, and we reach a region of normal (unpaired) quark matter where chiral symmetry is approximately restored. Although there is no true phase transition between those two regions, we follow Ref.~\cite{Ruester:2005jc} and denote them by $\chi_\text{SB}$ and NQ, respectively.
Starting in the $\chi_\text{SB}$ phase and increasing the chemical potential, there is a first-order phase transition to a 2SC phase, followed by a first-order transition to the CFL phase at higher chemical potentials. The first order phase boundary between the $\chi_\text{SB}$ phase and the 2SC phase ends in a critical point.\footnote{This is a remnant of the chiral critical endpoint which is found for weaker diquark couplings, see, e.g., the phase diagram with ``intermediate diquark coupling'' $G_D= \frac{3}{4} G_S$ in Ref.~\cite{Ruester:2005jc}.} 

At temperatures above this point, the phase transition is second order and turns around towards higher chemical potentials, separating the 2SC phase from the NQ phase.

On the other hand, there are quantitative and also qualitative differences between the two phase diagrams: 
As expected, with RG-consistent regularization, the critical temperatures for the melting of all color-superconducting phases keep increasing with the chemical potential, while this is not the case in the conventional regularization scheme. Related to this is also the unphysical appearance of the 2SC$_{us}$ phase in the conventional regularization scheme, which we already inferred from Fig.~\ref{fig:convergence_at_cutoff}.	Both cutoff artifacts are removed in the RG-consistent scheme. 
Furthermore, compared to the conventional regularization, the critical temperatures are enhanced significantly and the location of the low-temperature first-order phase boundaries of the $\chi_{\text{SB}}\,\to$ 2SC - transition and the 2SC $\to$ CFL - transition are shifted to lower chemical potentials. This can already be anticipated from the enhancement of the diquark condensates at $T=0$ in Fig.~\ref{fig:convergence}. 
Of course, the exact values of the diquark condensates depend sensitively on the choice of the coupling constant $G_D$. However, even if we adjust $G_D$ in such a way that, e.g., the value of $\Delta_3$ at the onset of the 2SC phase at $T=0$ is the same in both schemes, the RG-consistent scheme will lead to larger gaps and, hence, larger critical temperatures, the farther we move away from this point in temperature or chemical-potential direction.

An interesting qualitative difference is seen in the temperature region above the CFL phase.
In both schemes the first-order 2SC-CFL phase boundary splits into two phase boundaries at a certain chemical potential, above which the diquark condensates $\Delta_1$ and $\Delta_2$ melt successively with increasing temperature. This gives rise to a new intermediate phase in a narrow region.
In the RG-consistent scheme, $\Delta_2$ melts first, such that the CFL phase first gets replaced by a dSC phase, which then melts to a 2SC phase (``dSC-type melting pattern''). 
At the beginning, the CFL $ \to $ dSC transition is of first order and changes to second order at larger chemical potentials. 

With conventional regularization, on the other hand, instead of having a narrow dSC region, there is a narrow uSC region above the CFL phase (``uSC-type melting pattern''). One might be tempted to conclude that this is a cutoff artifact as well. However, as we will discuss next, the issue is by far less trivial.

\subsection{The CFL melting pattern at (moderately) high densities}
\label{sec:CFL_melting}

The question whether the phase diagram exhibits a dSC or a uSC phase in the temperature region above the CFL phase has been a subject of active research in the past decades \cite{Fukushima:2004zq,Iida:2003cc,Iida:2004cj,Abuki:2005ms}. 
The authors of Refs.~\cite{Iida:2003cc,Iida:2004cj} performed a GL expansion of the thermodynamic potential around the critical temperature $T_{c0}$ of the second-order CFL $\to$ NQ phase transition existing in the limit of three massless quark flavors. Taking into account effects of a nonzero strange mass and charge neutrality to  quadratic order in $M_s$, they found that at very high densities where $M_s \ll \mu$, the melting pattern of the CFL phase must be of dSC-type.
In Ref.~\cite{Abuki:2005ms}, the expansion was extended to quartic order in $M_s$.\footnote{The leading- and next-to-leading order $M_s$ corrections are proportional to $M_s^2/(\mu^2) \ln(\mu/T_{c0})$ and $M_s^4/(\mu^2 T_{c0}^2)$, respectively.} The authors found that, with this additional term, the dSC phase could turn into a uSC phase at lower chemical potential. More precisely, they showed that at this order the melting pattern is of dSC (uSC) type if the squared strange quark mass is less (greater) than
\begin{equation}
\label{eq:Mscrit}
    M_{s,\text{crit}}^2 = \frac{32\pi^2}{21\zeta(3)} T_{c0}^2 \ln(\mu/T_{c0}) .
\end{equation}
Since at asymptotic chemical potential the gaps and, hence, $T_{c0}$ become infinitely large in QCD, this means that at sufficiently large $\mu$ there must be a dSC phase, as already predicted in Refs.~\cite{Iida:2003cc,Iida:2004cj}. At lower chemical potential, on the other hand, a uSC phase might be realized, meaning that the presence of this phase in the conventional NJL studies is not necessarily unphysical. Qualitatively, the result is also consistent with the NJL-model-type calculation of Ref.~\cite{Fukushima:2004zq}, where $M_s^2/\mu$ was treated as a free parameter.

\begin{figure}[h!]
\centering
\begin{minipage}[t]{0.45\textwidth}
			 		\includegraphics[width=\textwidth]{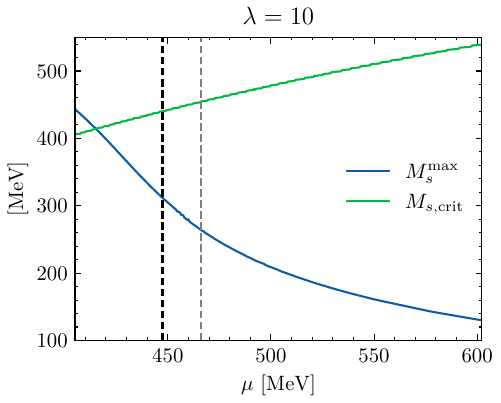}
			 	\end{minipage}
		 	\begin{minipage}[t]{0.45\textwidth}
		\includegraphics[width=\textwidth]{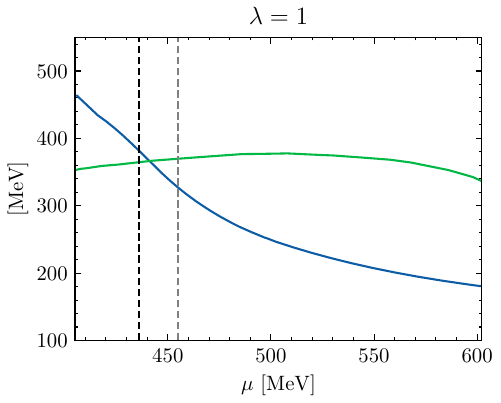}
			 	\end{minipage}
			 			\caption{Maximum and critical strange masses as functions of the chemical potential for the RG-consistent massless scheme with $\lambda=10$ (top) and with conventional cutoff regularization ($\lambda=1$, bottom). For each chemical potential, $M_s^\text{max}$ is defined as the maximum value of $M_s$ as a function of temperature. $M_{s,\text{crit}}$ is given in \Eq{eq:Mscrit}. The vertical dashed lines indicate the chemical potentials of the critical points of the CFL phase in the phase diagrams of Fig.~\ref{fig:pd}.
       \label{fig:Mscrit}
       } 	
     \end{figure}

In order to investigate this quantitatively, we compare in Fig.~\ref{fig:Mscrit} the strange-quark mass as a function of $\mu$ with $M_{s,\text{crit}}$. To calculate $M_{s,\text{crit}}$ with \Eq{eq:Mscrit} we calculated the critical temperature $T_{c0}$ of the CFL$\rightarrow$NQ transition in the NJL model for three massless quarks. Since $M_s$ is temperature and phase dependent in our model, we plot its maximum value with respect to temperature at given chemical potential and take this as an upper limit. The results for the massless scheme are shown in the top panel. As one can see, for most chemical potentials we have $M_s < M_{s,\text{crit}}$, predicting the existence of a dSC phase.
Only in the low-$\mu$ part of the figure we have $M_s > M_{s,\text{crit}}$, which would predict a uSC phase. However, this lies already in the region where we have a first-order CFL $\to$ 2SC phase transition in Fig.~\ref{fig:pd}, so the GL analysis is not valid in this regime. Our finding of a dSC-type melting pattern in Fig.~\ref{fig:pd} is therefore consistent with the GL predictions.

In the bottom panel of Fig.~\ref{fig:Mscrit} we make the same comparison for the conventional regularization scheme. Qualitatively the results are similar, meaning that the GL analysis also predicts a dSC-type melting pattern in this case. This is in contrast to the numerical results shown in Fig.~\ref{fig:pd}, where a uSC-type melting pattern is seen. This contradiction can have two reasons. One possibility is that the uSC phase is an artifact of the conventional regularization. In fact, the finding of a 2SC$_{us}$ phase above $\mu\simeq 560$~MeV, for which no physical explanation exists, demonstrates that there are cutoff artifacts which can change the melting pattern in an unphysical way. 

On the other hand, we cannot exclude that in the studied chemical-potential regime, $M_s$ is still too large for the GL analysis of Ref.~\cite{Abuki:2005ms} to be valid. For the conventional regularization scheme we then face the problem that the cutoff simply prevents us from going to high enough chemical potentials. This is different for the RG-consistent scheme, which allows us to extend the model studies mathematically to arbitrarily high densities. Although the validity of the NJL model itself eventually breaks down, its qualitative behavior at high densities is reasonable, as we have discussed in Sec.~\ref{sec:largemu}. In particular, the fact that $M_s$ goes to zero for $\mu \rightarrow \infty$ while the pairing gaps stay nonzero (see Fig.~\ref{fig:gapsbeyond}) ensures that at sufficiently high chemical potentials we must find a dSC phase in our model -- which is exactly what we do.

Finally we could ask whether we can also find a uSC-type melting pattern with RG-consistent regularization. As we have discussed, the fact that we do not see it in the phase diagram is consistent with \Eq{eq:Mscrit}, corresponding to the next-to-leading order GL analysis of Ref.~\cite{Abuki:2005ms}. Assuming the validity of this analysis at even lower chemical potentials, the top panel of Fig.~\ref{fig:Mscrit} suggests that there should be a uSC phase below $\mu \lesssim 420$~MeV. As pointed out, this is already in the regime of the first-order CFL $\to$ 2SC phase transition where the GL expansion is no longer valid. However, in that chemical-potential region but at temperatures just below the first-order phase boundary, we find the hierarchy of the gaps as $\Delta_3>\Delta_2>\Delta_1$. If there was a second-order phase transition, we would thus expect that $\Delta_1$ melts first, leading to a uSC phase. Therefore it seems that only the existence of the first-order transition, at which $\Delta_1$ and $\Delta_2$ jump to zero simultaneously, prevents this scenario from being realized. It would thus be interesting to investigate whether a uSC phase shows up if we vary the diquark coupling strength. In this case, there must then be a double-critical point  \cite{Fukushima:2004zq,Abuki:2005ms} in the phase diagram at which the uSC phase is connected to the dSC phase, which always exists at high densities, independently of the parameters.

\section{Conclusion}\label{sec:conclusion}
Understanding the phase structure of dense quark matter is essential to study astrophysical systems in which deconfined quark matter might exist, e.g., neutron stars and neutron-star mergers. Neutron-star mergers, as an example, are systems which probe temperatures up to $100\,$MeV and densities up to a few times the nuclear saturation density, see e.g. Ref. \cite{Perego:2019adq} and Refs. \cite{Baiotti:2016qnr,Burns:2019byj} for reviews on neutron-star mergers. The results of the present work suggest that the NJL model is plagued by cutoff artifacts in this region of the phase diagram. To cure these artifacts, we formulated and implemented an RG-consistent regularization for a three-flavor NJL model with color superconductivity in mean-field approximation. The concept of RG-consistency was introduced in Ref.~\cite{Braun:2018svj}. In the conventional cutoff-regularization, all momentum integrals are regularized with the same three-momentum cutoff $\Lambda'$. This leads to the unphysical situation that the diquark condensates and critical temperatures decrease as a function of the chemical potential for chemical potential close to $\Lambda'$. In the RG-consistent regularization, convergent integrals are regularized with a cutoff $\Lambda\gg\Lambda'$.
From a functional renormalization group perspective, $\Lambda$ is identified with the initial UV scale, from which the flow equation is integrated to the infrared. By choosing $\Lambda$ much larger than all external and internal scales of the system, the full quantum effective action in the infrared does not depend on the value of $\Lambda$ anymore, thus fulfilling
the criterion of RG-consistency. 

In an NJL model without diquark pairing, the separation of the effective action into a divergent vacuum contribution, which is regularized with the cutoff $\Lambda'$, and a convergent medium contribution, which is regularized with the cutoff $\Lambda$, is straightforward. However, with the appearance of quasiparticle modes with a color-superconducting gap in the spectrum, divergences of the form $\sim\mu^2\Delta^2 \ln (\Lambda)$ arise when the action is evaluated at nonzero chemical potential. In Ref.~\cite{Braun:2018svj}, this issue was addressed in the context of a two-flavor Quark-Meson-Diquark model in the chiral limit. There, the authors introduced a $\Delta-$dependent counterterm to mediate the problem of the divergence.

In the present work, we extended the analysis of Ref.~\cite{Braun:2018svj} to the more complicated case of three-flavor diquark pairing with nonzero quark masses and non-equal chemical potentials for different quark species. In this case, the gap-induced medium divergence takes the form $\sim\sum \mu_{\alpha a,\beta b}^2\Delta_{\alpha a,\beta b}^2 \ln(\Lambda)$, where $\Delta_{\alpha a,\beta b}$ denotes the gap for pairing quarks of flavor $\alpha$ and color $a$ with flavor $\beta$ and color $b$, and $\mu_{\alpha a,\beta b}=(\mu_{\alpha a}+\mu_{\beta b})/2$ is the average chemical potential of the paired quarks. Therefore, the counterterm of Ref. \cite{Braun:2018svj} has to be modified. We compared three different possible counterterms as three different renormalization schemes. In the \textit{massless} scheme, the counterterm is evaluated at zero quark masses whereas in the \textit{massive scheme}, the counterterm includes terms with quark mass contributions. Both schemes reduce to an analogue of the counterterm in Ref.~\cite{Braun:2018svj} in the case of two-flavor pairing of massless quarks. The counterterm in the \textit{minimal} scheme is given by an expansion of the counterterms in the massless scheme to quadratic order in $\Delta_{\alpha a,\beta b}$, which is sufficient to cancel all divergences.

All three schemes proof to be RG-consistent and remove the cutoff artifacts mentioned above.

However, in order to be consistent with general physical arguments and Ginzburg-Landau analysis, we must also request that in CFL quark matter with a common chemical potential the pairing between up and down quarks leads to a larger gap than the pairing involving strange quarks ($\Delta_3>\Delta_1=\Delta_2)$ and that in neutral CFL matter the $u$-$d$ pairs have the highest melting temperature.

We found that these additional requirements are violated in the massive scheme at high chemical potentials. 
This can be traced back to the observation that the asymptotic form of the divergences do not depend on quark masses, and therefore the inclusion of mass-dependent contributions in the counterterms is excessive. We conclude that the massive scheme must not be applied. 

We calculated the phase diagram of the massless scheme for $G_D/G_S=1$ and compared the result with the extension of the conventional cutoff-regularization in Ref.~\cite{Ruester:2005jc} up to $\mu=600\,$MeV. We showed how the RG-consistent calculation eliminates cutoff artifacts in the phase diagram: the diquark gaps and the critical temperatures keep increasing as a function of the chemical potential. Furthermore, the values of the gaps (and correspondingly the critical temperatures) for a fixed chemical potential are significantly enhanced in comparison with the conventional cutoff-regularization. These findings are consistent with the results of Ref.~\cite{Braun:2018svj} for the RG-consistent two-flavor Quark-Meson-Diquark model in mean field. 
Related to the larger gaps,
both the $\chi_{SB} \to $ 2SC and the 2SC $\to$ CFL transition along the chemical potential axis are moved to lower chemical potential.
Since, by construction the vacuum properties of the model are left unchanged, all these effects become larger with increasing chemical potential and temperature.

Interestingly, the phase diagram changes qualitatively from a uSC-type melting of the CFL phase for conventional cutoff regularization to a dSC-type melting pattern in the RG consistent scheme. From Ginzburg-Landau analysis it is known that CFL quark matter melts in a dSC-type pattern for small $M_s$ \cite{Iida:2003cc,Iida:2004cj}, but it can change to a uSC-type pattern at lower chemical potential when the strange mass exceeds a critical value \cite{Abuki:2005ms}. We showed that the dSC phase in the RG-consistent scheme is qualitatively consistent with the GL predictions while the uSC phase in the conventional framework is not. We could not fully resolve whether this inconsistency is caused by a cutoff artifact or by the breakdown of the GL analysis in the accessible chemical potential region. Important is that with conventional regularization we cannot extend the analysis to higher densities where we are sure that the GL analysis is valid. With RG consistent regularization, on the other hand, we can and we find the correct result.

An alternative regularization scheme based on Refs. \cite{Battistel:1999phd,Farias:2005cr} was used in Ref.~\cite{Duarte:2018kfd} for a color and charge neutral two flavor NJL model at vanishing temperature. There, the issue of the divergence is instead handled by isolating medium-independent divergences in the gap equations and fitting them to vacuum observables. The thermodynamic potential itself was obtained from integrating the gap equations. The resulting diquark condensate kept increasing as a function of the chemical potential. However, this approach is limited to chemical potentials  below an implicit regularization scale where the fitting to physical observables is done. To extend the model to higher densities and nonzero temperatures, a running of the coupling constant was introduced \cite{Farias:2006cs}. It would be interesting to compare their regularization scheme with the RG-consistent schemes studied in the present paper. 

The challenge of applying the RG-consistency framework in color-superconducting matter lies in the fact that the gapped modes lead to medium divergences. The procedure which we used to cancel these divergences is general. By investigating which gap parameter is associated with which (average) chemical potential, we subtracted appropriate counterterms. In the case of the massless and the minimal scheme, the obtained counterterm is given by a simple analytic expression, even for the case of a complicated pairing pattern like a neutral CFL phase with nonzero quark masses. This procedure can be applied to systems with other types of color-superconducting pairing patterns (e.g. a kaon-condensed CFL-phase \cite{Schafer:2000ew,Warringa:2006dk,Basler:2009vk}) or chirally imbalanced systems studied in Ref.~\cite{Pasqualotto:2023hho}. Furthermore, the present work can be extended to investigate the detailed dependence of the phase diagram on the diquark coupling strength, including gapless phases.\\

\section*{Acknowledgements}
The authors thank Laurin Pannullo for bringing our attention to the RG-consistency ideas for the first time during the HFHF Theory Retreat in 2022. We extend our gratitude to Lennart Kurth and Martin Steil for their discussions and guidance on the FRG perspective. We thank Jens Braun and Andreas Geißel for their feedback on our manuscript. We also thank Mark Winstel for useful discussions. H.G. thanks Lorenz von Smekal for discussions around the medium renormalization.  This work has been supported by the Deutsche Forschungsgemeinschaft (DFG) through the Collaborative Research Center TransRegio CRC-TR 211 ``Strong-interaction matter under extreme conditions''.
M.H. is supported by the GSI F\&E.

\onecolumngrid
\appendix

\section{Color-flavor structure of quark pairing}\label{sec:appendix_pairing_pattern}
The full inverse propagator in Nambu-Gorkov space, as shown in Ref.~\cite{Ruester:2005jc}, can be symbolically represented as

\begin{eqnarray}
   S^{-1}= \left(
\begin{array}{cc}
 [G_0^+]^{-1}  & \Phi^- \\
 \Phi^+ & [G_0^-]^{-1}  \\
\end{array}
\right),
\end{eqnarray}
where 
\begin{eqnarray}
        [G_0^\pm]^{-1}&=&\gamma^\mu P_\mu\pm\hat{\mu}\gamma_0-\hat M,\\
        \left(\Phi^-\right)_{ab}^{\alpha\beta}&=&-\sum_{c}\epsilon^{\alpha\beta c}\epsilon_{abc}\Delta_c\gamma_5,\\
        \Phi^+&=&\gamma^0(\Phi^-)^\dag\gamma^0,
\end{eqnarray}
  and $P=(p_0,\vect p)\equiv (i\omega_n,\vect p)$.  For the calculation of the effective potential, we are interested in calculating $\det{S^{-1}}=\det{\gamma_0 S^{-1}}$. Defining the spin projectors
    \begin{align}
        \mathcal{P}_\pm=\frac{1}{2}\left(1\pm\vect\sigma\cdot\hat{\vect p}\right),
    \end{align}
    where $\hat{\vect{p}}=\frac{\vect p}{|\vect p|}$ and using the above relation, $\gamma_0 S^{-1}$ can be written as
    \begin{align}
        \gamma_0 S^{-1}=\sum_{s=\pm} \hat{S}_s^{-1}\mathcal{P}_s,
    \end{align}
    with $\hat{S}_s^{-1}=\mathds{1}p_0-\mathcal{M}_s$ and
\begin{eqnarray}
   \mathcal{M}_s= \left(
\begin{array}{cccc}
 M-\hat\mu  & s p  & 0 & \gamma^5\Phi^+ \\
 sp  & -\hat\mu -M & \gamma^5\Phi^- & 0 \\
 0 & \gamma^5\Phi^- & \hat\mu +M & sp \\
 \gamma^5\Phi^+ & 0 & sp  & \hat\mu -M \\
\end{array}
\right).
\end{eqnarray}

Because of spin degeneracy, $\mathcal{M}_+$ and $\mathcal{M}_-$ have identical eigenvalues (see \cite{Ruester:2005jc} for the details). In the following we therefore consider only $\mathcal{M}_+$, which is a 36$\times$36 matrix and consists of the following seven blocks:

\begin{eqnarray*}
    \begin{array}{cc}
    \text{M1=} & 
    \begin{minipage}{0.45\linewidth}
    \scalemath{0.9}{
    \left(
    \begin{array}{cccc}
        M_d - \mu _{dr} & p & 0 & -\Delta_{3} \\
        p & -M_d - \mu _{dr} & \Delta_{3} & 0 \\
        0 & \Delta_{3} & \mu _{ug} + M_u & p \\
        -\Delta_{3} & 0 & p & \mu _{ug} - M_u \\
    \end{array}
    \right)
    }
    \end{minipage}
    \end{array}
    \begin{array}{cc}
    \text{M2=} & 
    \begin{minipage}{0.45\linewidth}
    \scalemath{0.9}{
    \left(
    \begin{array}{cccc}
        \mu _{dr} - M_d & p & 0 & -\Delta_{3} \\
        p & M_d + \mu _{dr} & \Delta_{3} & 0 \\
        0 & \Delta_{3} & -\mu _{ug} - M_u & p \\
        -\Delta_{3} & 0 & p & M_u - \mu _{ug} \\
    \end{array}
    \right)
    }
    \end{minipage}
    \end{array}
\end{eqnarray*}

\begin{eqnarray*}
    \begin{array}{cc}
    \text{M3=} & 
    \begin{minipage}{0.45\linewidth}
    \scalemath{0.9}{
    \left(
    \begin{array}{cccc}
        M_{\text{s}} - \mu _{sr} & p & 0 & -\Delta_{2} \\
        p & -M_{\text{s}} - \mu _{sr} & \Delta_{2} & 0 \\
        0 & \Delta_{2} & \mu _{ub} + M_u & p \\
        -\Delta_{2} & 0 & p & \mu _{ub} - M_u \\
    \end{array}
    \right)
    }
    \end{minipage}
    \end{array}
    \begin{array}{cc}
    \text{M4=} & 
    \begin{minipage}{0.45\linewidth}
    \scalemath{0.9}{
    \left(
    \begin{array}{cccc}
        \mu _{sr} - M_{\text{s}} & p & 0 & -\Delta_{2} \\
        p & M_{\text{s}} + \mu _{sr} & \Delta_{2} & 0 \\
        0 & \Delta_{2} & -\mu _{ub} - M_u & p \\
        -\Delta_{2} & 0 & p & M_u - \mu _{ub} \\
    \end{array}
    \right)
    }
    \end{minipage}
    \end{array}
\end{eqnarray*}

\begin{eqnarray*}
    \begin{array}{cc}
    \text{M5=} & 
    \begin{minipage}{0.45\linewidth}
    \scalemath{0.9}{
    \left(
    \begin{array}{cccc}
        M_{\text{s}} - \mu _{sg} & p & 0 & -\Delta_{1} \\
        p & -\mu _{sg} - M_{\text{s}} & \Delta_{1} & 0 \\
        0 & \Delta_{1} & \mu _{db} + M_d & p \\
        -\Delta_{1} & 0 & p & \mu _{db} - M_d \\
    \end{array}
    \right)
    }
    \end{minipage}
    \end{array}
    \begin{array}{cc}
    \text{M6=} & 
    \begin{minipage}{0.45\linewidth}
    \scalemath{0.9}{
    \left(
    \begin{array}{cccc}
        \mu _{sg} - M_{\text{s}} & p & 0 & -\Delta_{1} \\
        p & \mu _{sg} + M_{\text{s}} & \Delta_{1} & 0 \\
        0 & \Delta_{1} & -\mu _{db} - M_d & p \\
        -\Delta_{1} & 0 & p & M_d - \mu _{db} \\
    \end{array}
    \right)
    }
    \end{minipage}
    \end{array}
\end{eqnarray*}

\begin{eqnarray*}
    \!\!\begin{array}{cc}
 \text{M7=} & \left(
 \scalemath{0.7}{
\begin{array}{cccccccccccc}
 -M_u-\mu _{ur} & p & 0 & 0 & 0 & 0 & 0 & -\Delta_{3} & 0 & 0 & 0 & -\Delta_{2} \\
 p & M_u-\mu _{ur} & 0 & 0 & 0 & 0 & \Delta_{3} & 0 & 0 & 0 & \Delta_{2} & 0 \\
 0 & 0 & \mu _{ur}-M_u & p & 0 & \Delta_{3} & 0 & 0 & 0 & \Delta_{2} & 0 & 0 \\
 0 & 0 & p & M_u+\mu _{ur} & -\Delta_{3} & 0 & 0 & 0 & -\Delta_{2} & 0 & 0 & 0 \\
 0 & 0 & 0 & -\Delta_{3} & -\mu _{dg}-M_d & p & 0 & 0 & 0 & 0 & 0 & -\Delta_{1} \\
 0 & 0 & \Delta_{3} & 0 & p & M_d-\mu _{dg} & 0 & 0 & 0 & 0 & \Delta_{1} & 0 \\
 0 & \Delta_{3} & 0 & 0 & 0 & 0 & \mu _{dg}-M_d & p & 0 & \Delta_{1} & 0 & 0 \\
 -\Delta_{3} & 0 & 0 & 0 & 0 & 0 & p & M_d+\mu _{dg} & -\Delta_{1} & 0 & 0 & 0 \\
 0 & 0 & 0 & -\Delta_{2} & 0 & 0 & 0 & -\Delta_{1} & -\mu _{sb}-M_{\text{s}} & p & 0 & 0 \\
 0 & 0 & \Delta_{2} & 0 & 0 & 0 & \Delta_{1} & 0 & p & M_{\text{s}}-\mu _{sb} & 0 & 0 \\
 0 & \Delta_{2} & 0 & 0 & 0 & \Delta_{1} & 0 & 0 & 0 & 0 & \mu _{sb}-M_{\text{s}} & p \\
 -\Delta_{2} & 0 & 0 & 0 & -\Delta_{1} & 0 & 0 & 0 & 0 & 0 & p & M_{\text{s}}+\mu _{sb} \\
\end{array}}
\right) \\
\end{array}
\end{eqnarray*}

These blocks reflect the pairing structure in the condensates $\Delta_A$, as defined in \Eq{eq:DeltaA}. Specifically, the $(ug)$-$(dr)$-pairs of $\Delta_3$ contribute to $\text{M1}$ and $\text{M2}$, the $(ub)$-$(sr)$-pairs of $\Delta_2$ to  $\text{M3}$ and $\text{M4}$, and the $(db)$-$(sg)$-pairs of $\Delta_1$ to $\text{M5}$ and $\text{M6}$. 
The remaining pairs, i.e., $(ur)$-$(dg)$ in $\Delta_3$, $(ur)$-$(sb)$ in $\Delta_2$ and $(dg)$-$(sb)$ in $\Delta_1$ all contribute to $\text{M7}$, since each of those quark species is involved in two condensates.

\section{Medium divergences}\label{app:medium_divergences}

The general form of the effective action is (see Eqs.~\eqref{eq:Omega_eff}, \eqref{eq:Adef}, \eqref{eq:Omeff_Gamma})
\begin{equation}\label{eq:quarkpot}
  \frac{\Gamma}{V_4}(\vect\mu,T,\vect\chi)=-\int \frac{d^3p}{(2\pi)^3}\, T \sum_n \frac{1}{2}\text{tr}\ln{\left(\frac{1}{T}S^{-1}(i\omega_n,\vect{p};\vect{\mu},\vect\chi)\right)}+\mathcal{V}(\vect\chi). 
\end{equation}
Since the nonzero temperature part is always finite, we restrict ourselves to T=0 while keeping the chemical potential in the inverse propagator structure. The Matsubara sum then gets replaced by an integral over the Euclidean energy $p_4 = i p_0$,
\begin{equation}
     \int \frac{d^3p}{(2\pi)^3}\, T \sum_n    
    \, \rightarrow \, 
    \int \frac{d^3p}{(2\pi)^3}  \int_{-\infty}^{\infty}\frac{dp_4}{2\pi}  
    \, =  \,
    \int \frac{p^2 dp}{2\pi^2} \int_{-\infty}^{\infty}\frac{dp_4}{2\pi},  
\end{equation}
where in the last step we used that the integrand is spherically symmetric in the spatial momentum directions.
Since our focus is solely on the medium divergences, we subtract the vacuum contribution. The remaining part is then given by
\begin{equation}
  \mathcal{F}(\vect\mu,\vect\chi)=-\frac{1}{2}  \int_{0}^{\infty}\frac{p^2\,dp}{2\pi^2} \int_{-\infty}^{\infty}\frac{dp_4}{2\pi}\ln\det{\left( \frac{S^{-1}(ip_4,\vect{p};\vect{\mu},\vect{\chi})}{S^{-1}(ip_4,\vect{p};\vect{0},\vect{\chi})}\right)}.
\end{equation}
In order to identify the divergences arising from the integral above, we examine its asymptotic behavior in powers of a momentum cutoff $\Lambda$. It is advantageous to scale the temporal part as 
\begin{equation}
    p_4=p\cdot x  ,
\end{equation}
so that the $p_4$ integration gets replaced by an integration over the dimensionless variable $x$. We then have $\mathcal{F} = \lim_{\Lambda\to\infty} \mathcal{F}_\Lambda$ with 
\begin{eqnarray}
      \mathcal{F}_\Lambda(\vect\mu,\vect\chi)&=&-\frac{1}{2}  \int_{0}^{\Lambda}\frac{dp\,p^3}{4\pi^3}\int_{-\infty}^{\infty}\,dx\ln\det{\left( \frac{S^{-1}(ix\,p,\vect{p};\vect{\mu},\vect{\chi})}{S^{-1}(ix\,p,\vect{p};\vect{0},\vect{\chi})}\right)}\nonumber\\
      &\equiv&-\frac{1}{2}  \int_{0}^{\Lambda}\frac{dp}{4\pi^3}\int_{-\infty}^{\infty}\,dx\;\zeta(x,p;\vect\mu,\vect\chi) .
\end{eqnarray}
The divergence structure for $\Lambda \rightarrow \infty$ can then be obtained by an asymptotic expansion of the function $\zeta$ for large $p$ and integrating it over $x$. This can be done separately for each of the blocks of the inverse propagator listed in Appendix~\ref{sec:appendix_pairing_pattern}.

The asymptotic behavior analysis for the blocks $\text{M1}$ and $\text{M2}$ yields the following contribution to the integrand:
\begin{eqnarray}
    \zeta_{1\&2}&=&\frac{
   \left(x^2-1\right) \left(\mu _{dr}^2+\mu    _{ug}^2\right)}{\pi  \left(x^2+1\right)^2}\;p\nonumber\\&&+\bigg(-\frac{2 \Delta_3^2 \left(-\left(x^2+1\right) \mu _{dr} \mu    _{ug}+\left(x^2-1\right) \mu _{dr}^2+\left(x^2-1\right) \mu
   _{ug}^2\right)}{\pi   \left(x^2+1\right)^3}\nonumber\\&&-\frac{\left(3 x^4+2 x^2-1\right) \left(M_d^2 \mu _{dr}^2+ M_u^2 \mu    _{ug}^2\right)}{\pi   \left(x^2+1\right)^4}-\frac{\left(x^4-6 x^2+1\right)\left(\mu _{dr}^4+\mu
   _{ug}^4\right)}{2 \pi 
    \left(x^2+1\right)^4}\bigg)\frac{1}{p}\nonumber\\&&+\mathcal{O}\left(\frac{1}{p^3}\right).
\end{eqnarray}
Naively, the first line, which is of the order $\mathcal{O}(p)$, leads to a quadratic divergence upon momentum integration. It turns out, however, that the coefficient vanishes upon integration over $x$. Since the $\mathcal{O}(\frac{1}{p^3})$-terms yield nondivergent contributions,  
the only remaining relevant terms for this analysis are the $\mathcal{O}(\frac{1}{p})$-terms (second and third line), which yield
\begin{eqnarray}
    -\frac{1}{2}  \int\limits_{0}^{\Lambda}\frac{dp}{4\pi^3}\int\limits_{-\infty}^{\infty}\,dx\;\zeta_{1\&2}\simeq -\frac{1}{2\pi^2}\int\limits^{\Lambda}dp\,\frac{\Delta_{3}^2 \left(\mu _{dr}+\mu _{ug}\right){}^2}{2 p}=-\frac{1}{4\pi^2}\Delta_3^2\left(\mu_{dr}+\mu_{ug}\right)^2\ln(\Lambda),
\end{eqnarray}
where we dropped the finite terms and we disregarded the lower bound $p=0$, where the asymptotic expansion is not valid and which is irrelevant for the UV divergences.

The contributions of the blocks $\text{M3}$ to $\text{M6}$ are analogous, and therefore the combination of the first six blocks gives the contribution
\begin{eqnarray}
    -\frac{1}{2}  \int\limits_{0}^{\Lambda}\frac{dp}{4\pi^3}\int\limits_{-\infty }^{\infty}\,dx\; \zeta_{1-6}\simeq -\frac{1}{4\pi^2} \ln (\Lambda ) \left(\Delta_{1}^2 \left(\mu _{db}+\mu _{sg}\right){}^2+\Delta
   _{2}^2 \left(\mu _{ub}+\mu _{sr}\right){}^2+\Delta_{3}^2 \left(\mu _{dr}+\mu
   _{ug}\right){}^2\right).
   \nonumber\\
\label{eq:zeta1-6}   
\end{eqnarray}
Employing the same procedure on the seventh block yields
\begin{eqnarray}
 -\frac{1}{2}  \int\limits_{0}^{\Lambda}\frac{dp}{4\pi^3}\int\limits_{-\infty}^{\infty}\,dx\;\zeta_7\simeq -\frac{1}{4\pi^2} \ln (\Lambda ) \left(\Delta_{1}^2 \left(\mu _{sb}+\mu _{dg}\right){}^2+\Delta
   _{2}^2 \left(\mu _{sb}+\mu _{ur}\right){}^2+\Delta_{3}^2 \left(\mu _{dg}+\mu
   _{ur}\right){}^2\right).
   \nonumber\\ 
\label{eq:zeta7}     
\end{eqnarray}
Hence, taking the sum of Eqs.~\eqref{eq:zeta1-6} and \eqref{eq:zeta7}, we obtain
\begin{align}
\mathcal{F}_\Lambda(\vect\mu,\vect\chi)\simeq - \frac{1}{4\pi^2} \bigg(& ((\mu_{ur}+\mu_{dg})^2+(\mu_{ug}+\mu_{dr})^2)\Delta_{3}^2+\nonumber\\
 &((\mu_{ur}+\mu_{sb})^2+(\mu_{ub}+\mu_{sr})^2)\Delta_{2}^2+
      ((\mu_{dg}+\mu_{sb})^2+(\mu_{sg}+\mu_{db})^2)\Delta_{1}^2
      \bigg)\ln\Lambda,      
\end{align}
which is the divergence structure quoted in \Eq{cflgammainfinite}.

\section{Counterterms}
\label{app:CT}
Here we prove that the counterterm satisfies the requirements \eqref{req:1.1} and \eqref{req:1.2}. We begin with the \textit{massive scheme} defined by \Eq{3f:CT}.
\subsection{Massive scheme}\label{app:massschemeproof}
By writing
\begin{equation}
\delta \mathcal{A}\equiv\sum\frac{1}{2} \mu_{\alpha a,\beta b}^2\left(\frac{\partial^2}{\partial\mu_{\alpha a,\beta b}^2}\mathcal{A}(\vect\mu,0,\vect\chi)\right)\bigg|_{\vect\mu=0;\Delta_{{\alpha a,\beta b}}\neq0}\bigg)
\end{equation} and using 
\begin{align}
    \frac{\partial^2}{\partial\mu_{\alpha a,\beta b}^2}=\left(\frac{\partial}{\partial\mu_{\alpha a}}+\frac{\partial}{\partial\mu_{\beta b}}\right)^2
\end{align}
the momentum integrand of the counterterm reads
\begin{align}{\label{h:8}}
   \delta\mathcal{A}=&\frac{\mu_{ud;rg}^2}{2}\left(\left(\frac{\partial^2}{\partial\mu_{ur}^2}+\frac{\partial^2}{\partial\mu_{dg}^2}+2\frac{\partial^2}{\partial\mu_{ur}\partial\mu_{dg}}+\frac{\partial^2}{\partial\mu_{ug}^2}+\frac{\partial^2}{\partial\mu_{dr}^2}+2\frac{\partial^2}{\partial\mu_{ug}\partial\mu_{dr}}\right)\mathcal{A}(\mu,0,\vect\chi)\right)\Big|_{\vect\mu=\Delta_1=\Delta_2=0}\nonumber\\
   +&\frac{\mu_{us;rb}^2}{2}\left(\left(\frac{\partial^2}{\partial\mu_{ur}^2}+\frac{\partial^2}{\partial\mu_{sb}^2}+2\frac{\partial^2}{\partial\mu_{ur}\partial\mu_{sb}}+\frac{\partial^2}{\partial\mu_{ub}^2}+\frac{\partial^2}{\partial\mu_{sr}^2}+2\frac{\partial^2}{\partial\mu_{ub}\partial\mu_{sr}}\right)\mathcal{A}(\mu,0,\vect\chi)\right)\Big|_{\vect\mu=\Delta_1=\Delta_3=0}\nonumber\\
   +&\frac{\mu_{ds;gb}^2}{2}\left(\left(\frac{\partial^2}{\partial\mu_{dg}^2}+\frac{\partial^2}{\partial\mu_{sb}^2}+2\frac{\partial^2}{\partial\mu_{dg}\partial\mu_{sb}}+\frac{\partial^2}{\partial\mu_{db}^2}+\frac{\partial^2}{\partial\mu_{sg}^2}+2\frac{\partial^2}{\partial\mu_{db}\partial\mu_{sg}}\right)\mathcal{A}(\mu,0,\vect\chi)\right)\Big|_{\vect\mu=\Delta_2=\Delta_3=0}.
\end{align}
For the first requirement \eqref{req:1.1}, in case that $\vect{\mu}=0$, it can be easily seen that the whole counterterm is zero. Assuming that all diquark condensates $\{\Delta_i\}$ are zero, i.e., $\Delta_1=\Delta_2=\Delta_3=0$, the first line in the expression is calculated as 
\begin{eqnarray}
&&\frac{\mu_{ud;rg}^2}{2}\left(\left(\frac{\partial^2}{\partial\mu_{ur}^2}+\frac{\partial^2}{\partial\mu_{dg}^2}+2\frac{\partial^2}{\partial\mu_{ur}\partial\mu_{dg}}+\frac{\partial^2}{\partial\mu_{ug}^2}+\frac{\partial^2}{\partial\mu_{dr}^2}+2\frac{\partial^2}{\partial\mu_{ug}\partial\mu_{dr}}\right)\mathcal{A}(\mu,0,\vect\chi)\right)\Big|_{\vect\mu=\Delta_1=\Delta_2=0}\nonumber=\nonumber\\
&&\int_{-\infty}^\infty dx\,\frac{16 p^3}{\pi } \left(-\frac{2 \left(p^2+M_d^2\right)}{\left(p^2 \left(x^2+1\right)+M_d^2\right)^2}+\frac{1}{p^2 \left(x^2+1\right)+M_d^2}+\frac{1}{p^2
   \left(x^2+1\right)+M_u^2}-\frac{2 \left(p^2+M_u^2\right)}{\left(p^2 \left(x^2+1\right)+M_u^2\right)^2}\right)=0.\nonumber\\&&
\end{eqnarray}
where integration is over $x=p_4/p$.
For $\vect{\Delta}=0$, analogous expressions from the second and third lines in \eqref{h:8} differ only by the appearance of different masses. Consequently, $\delta\mathcal{A}\Big|_{\vect\Delta=0}$=0 and requirement \Eq{req:1.1} is fulfilled. Assuming, e.g. that $\Delta_1,\Delta_2>0$ but $\Delta_3=0$, the counterterm associated to up-down pariring (first line in \Eq{h:8}) vanishes. Analogously this is true in the absence of up-strange pairing ($\Delta_2$) and down-strange pairing ($\Delta_1$). Thus, requirement \Eq{req:1.2} is satisfied as well.

\subsection{Massless scheme}\label{app:masslessscheme}
In the massless scheme with the counterterm defined in equation \eqref{masslessCT}, one can apply the conditions $\vect M=0 \text{ and } \Delta_i=\Delta_j=0$ before taking the second derivative with respect to the associated chemical potentials. 

With the dispersion relations
\begin{eqnarray}{\label{eq:energymasslesscase}} 
&&\epsilon_{\pm,\Delta_{\alpha a,\beta b}}^+ =  \epsilon_{\pm,\Delta_{\alpha a,\beta b}} + \frac{\delta\mu_{\alpha a,\beta b}}{2},
\quad
\epsilon_{\pm,\Delta_{\alpha a,\beta b}}^- =  \epsilon_{\pm,\Delta_{\alpha a,\beta b}} - \frac{\delta\mu_{\alpha a,\beta b}}{2}
\\ \text{with}\quad&& 
\epsilon_{\pm,\Delta_{\alpha a,\beta b}}=\sqrt{(p\pm\mu_{\alpha a,\beta b})^2+\Delta_{\alpha a,\beta b}^2},
\end{eqnarray}
the counterterm is given by
\begin{eqnarray}\label{TempdeltaL}
\delta\mathcal{L} &=& \frac{2}{\pi^2} \Bigg[ \mu_{ds;gb}^2 \Delta_1^2 \Bigg( \frac{\Lambda'}{\sqrt{\Lambda'^2 + \Delta_1^2}} - \frac{\Lambda}{\sqrt{\Lambda^2 + \Delta_1^2}} \nonumber + \ln{\left( \frac{\Lambda + \sqrt{\Lambda^2 + \Delta_1^2}}{\Lambda' + \sqrt{\Lambda'^2 + \Delta_1^2}} \right)} \Bigg) \nonumber \\
&&+ \mu_{us;rb}^2 \Delta_2^2 \Bigg( \frac{\Lambda'}{\sqrt{\Lambda'^2 + \Delta_2^2}} - \frac{\Lambda}{\sqrt{\Lambda^2 + \Delta_2^2}} \nonumber + \ln{\left( \frac{\Lambda + \sqrt{\Lambda^2 + \Delta_2^2}}{\Lambda' + \sqrt{\Lambda'^2 + \Delta_2^2}} \right)} \Bigg) \nonumber \\
&&+ \mu_{ud;rg}^2 \Delta_3^2 \Bigg( \frac{\Lambda'}{\sqrt{\Lambda'^2 + \Delta_3^2}} - \frac{\Lambda}{\sqrt{\Lambda^2 + \Delta_3^2}}  + \ln{\left( \frac{\Lambda + \sqrt{\Lambda^2 + \Delta_3^2}}{\Lambda' + \sqrt{\Lambda'^2 + \Delta_3^2}} \right)} \Bigg) \Bigg].
\end{eqnarray}
For asymptotically large $\Lambda$, this is the same expression as \Eq{cflgammainfinite} with a relative negative sign such that it cancels the medium divergence and ensures RG-consistency. It is easy to check that requirements \eqref{req:1.1} and \eqref{req:1.2} are fulfilled.

\section{Explicit form of the gap equations in the massive scheme}\label{app:gap}
Here we show how the effective potential in the massive scheme as well as the gap equations and neutrality constraints
can be calculated explicitly. The gap equations and neutrality constraints are given by partial derivatives of the effective potential
\begin{align}
   \partial_X\Omega_{\text{eff}}(\mu,T;\vect\chi)= &\partial_X\mathcal{V}
    -\frac{1}{2\pi^2}\bigg(\int_0^{\Lambda} dp\,p^2 \partial_X\mathcal{A}(\vect{\mu},T,\vect\chi) -\int_{\Lambda'}^{\Lambda} dp\,p^2 \partial_X\mathcal{A}_{\text{vac}}(\vect\chi)\nonumber\\
    &\qquad-\int_{\Lambda'}^{\Lambda} dp\,p^2 \partial_X\sum\frac{1}{2} \mu_{\alpha a,\beta b}^2\left(\frac{\partial^2}{\partial\mu_{\alpha a,\beta b}^2}\mathcal{A}(\vect\mu,0,\vect\chi)\right)\bigg|_{\vect\mu=0;\Delta_{{\alpha a,\beta b}}\neq0}\bigg)\label{eq:gapeq_explicit}.
\end{align}

Therefore we consider general partial derivatives with respect to quantities $X_1,X_2,X_3$ which stand for condensates or chemical potentials. The second partial derivatives of $\mathcal{A}$ are calculated as
\begin{align}
    \frac{\partial^2\mathcal{A}}{\partial X_2\partial X_1}=&
    T\frac{\partial}{\partial X_2}\sum_n \text{tr}\left(S_+\left(i\omega_n,p\right)\frac{\partial}{\partial X_1}S^{-1}_+\left(i\omega_n,p\right)\right)\nonumber\\
    =&T\sum_n \text{tr}\left(\left(\frac{\partial}{\partial X_2}S_+\left(i\omega_n,p\right)\right)\Gamma_{X_1}(p)+S_+(i\omega_n,p)\frac{\partial}{\partial X_2}\Gamma_{X_1}(p)\right).
\end{align}

Here we introduced the derivative of the inverse propagator $\Gamma_X(p) \equiv \partial_X S_+^{-1}$.
The inverse propagator contains the condensates and chemical potentials to at most linear order. Therefore, second derivatives $\partial_{X_2} \Gamma_X$ vanish.
Using
\begin{align}
    0=\frac{\partial}{\partial X_2}(S_+^{-1}S_+)
    =\left(\frac{\partial S_+^{-1}}{\partial X_2}\right)S_+ + S_+^{-1}\frac{\partial S_+}{\partial X_2}
\end{align}
we can express the derivative of $S_+$ in terms of first derivatives of $S_{+}^{-1}$ and $S_{+}$ itself:
\begin{align}\label{eq:deriv_propagator}
    \frac{\partial S_+}{\partial X_2}=-S_+\left(\frac{\partial S_+^{-1}}{\partial X_2}\right)S_+ = -S_+ \Gamma_{X_2}S_+.
\end{align}
This yields
\begin{align}
    \frac{\partial^2A}{\partial X_2\partial X_1}=&
    -T\sum_n \text{tr}\left(S_+(i\omega_n,p)\Gamma_{X_2}(p)S_+(i\omega_n,p)\Gamma_{X_1}(p)\right).
\end{align}
The propagator can be diagonalized in the form $S(p_0,\Vec{p})=UDU^T$ with the diagonal matrix $D=(i\omega_n-\mathcal{D}^{-1})$ where $\mathcal{D}=\text{diag}(\epsilon_1,\epsilon_2,...)$ is the diagonal matrix conatining the dispersion relations $\epsilon_i$ on the diagonal and $U$ is a unitary matrix:
\begin{align}
    \frac{\partial^2A}{\partial X_2\partial X_1}=&
    -T\sum_n \text{tr}\left(UDU^T\Gamma_{X_2}(p)UDU^T\Gamma_{X_1}(p)\right)
    \nonumber\\
    &=
    -T\sum_n \text{tr}\left(\tilde{\Gamma}_{X_1}(p)D\tilde{\Gamma}_{X_2}(p)D\right)\nonumber\\
    &=-T\sum_n\sum_{i,j} 
\Tilde{\Gamma}_{X_1,ij}\Tilde{\Gamma}_{X_2,ji}\left[(i\omega_n-\epsilon_j)(i\omega_n-\epsilon_i)\right]^{-1}.
\end{align}
Here, we employ the notation $\tilde{\Gamma}_X=U^T\Gamma_X U^{-1}$ and make use of the cyclic property of the trace. Evaluating the Matsubara sum gives
\begin{align}
    \frac{\partial^2A}{\partial X_2\partial X_1} = -\sum_{ij}\Tilde{\Gamma}_{X_1,ij}\Tilde{\Gamma}_{X_2,ji}\frac{n(\epsilon_i)-n(\epsilon_j)}{\epsilon_i-\epsilon_j}.
\end{align}
with the Fermi-Dirac factors $n(x)=(1+e^x)^{-1}$.

In vacuum, this expression becomes
\begin{equation}\label{eq:Aderiv2}
    \frac{\partial^2A}{\partial X_2\partial X_1}\Bigg\vert_{\vect{\mu}=T=0}=-\sum_{ij}\Tilde{\Gamma}_{X_1,ij}\Tilde{\Gamma}_{X_2,ji}\frac{\theta(-\epsilon_i)-\theta(-\epsilon_j)}{\epsilon_i-\epsilon_j}.
\end{equation}

The gap equations and neutrality constraints are given by derivatives of the effective potential, see \Eq{eq:gapeq_explicit}. The contribution of the counterterm integrand is
\begin{align}
    &\partial_X\sum\frac{1}{2} \mu_{\alpha a,\beta b}^2\left(\frac{\partial^2}{\partial\mu_{\alpha a,\beta b}^2}\mathcal{A}(\vect\mu,0,\vect\chi)\right)\bigg|_{\vect\mu=0;\Delta_{{\alpha a,\beta b}}\neq0}\bigg)\nonumber\\
    =&\sum\frac{1}{2} \mu_{\alpha a,\beta b}^2\;\partial_X\left(\frac{\partial^2}{\partial\mu_{\alpha a,\beta b}^2}\mathcal{A}(\vect\mu,0,\vect\chi)\right)\bigg|_{\vect\mu=0;\Delta_{{\alpha a,\beta b}}\neq0}\bigg)\nonumber\\
&+\sum\frac{1}{2} \mu_{\alpha a,\beta b}\partial_X(\mu_{\alpha a,\beta b}) \left(\frac{\partial^2}{\partial\mu_{\alpha a,\beta b}^2}\mathcal{A}(\vect\mu,0,\vect\chi)\right)\bigg|_{\vect\mu=0;\Delta_{{\alpha a,\beta b}}\neq0}\bigg).
\label{eq:third_der_counterterm}
\end{align}
For the neutrality constraints, $X$ is a chemical potential and the first term vanishes. For the gap equations ($X=\phi_u,\phi_d,\phi_s,\Delta_1,\Delta_2,\Delta_3$) the second term in \Eq{eq:third_der_counterterm} vanishes but the first term is nonzero. Thus we need to calculate third order derivatives of $\mathcal{A}$.
Using again \Eq{eq:deriv_propagator} and $\partial_{X_3}\Gamma_X=0$ we obtain 

\begin{align}
    \frac{\partial^3\mathcal{A}}{\partial X_3\partial X_2 \partial X_1}=&
     -T\sum_n\partial_{X_3}\text{tr}[S_+\Gamma_{X_2}S_+\Gamma_{X_1}]\nonumber\\
     =&T\sum_n(\text{tr}[S_+\Gamma_{X_3}S_+\Gamma_{X_2}S_+\Gamma_{X_1}]
    +\text{tr}[S_+\Gamma_{X_2}S_+\Gamma_{X_3}S_+\Gamma_{X_1}])\label{eq:traces}\\
    =&T\sum_n([UD\tilde{\Gamma}_{X_3}D\tilde{\Gamma}_{X_2}DU^{-1}\Gamma_{X_1}]
    +\text{tr}[UD\tilde{\Gamma}_{X_2}D\tilde{\Gamma}_{X_3}DU^{-1}\Gamma_{X_1}])\nonumber\\
    =&T\sum_n\sum_{i,j,k}(\tilde{\Gamma}_{X_1,ij}\tilde{\Gamma}_{X_3,jk}\tilde{\Gamma}_{X_2,ki}+\tilde{\Gamma}_{X_1,ij}\tilde{\Gamma}_{X_3,jk}\tilde{\Gamma}_{X_2,ki})[(i\omega_n-\epsilon_i)(i\omega_n-\epsilon_j)(i\omega_n-\epsilon_k)]^{-1}\nonumber.
\end{align}
The order of the derivatives must not matter. This is guaranteed by the sum of the cyclic term $\text{tr} (\tilde{\Gamma}_{X_1}\tilde{\Gamma}_{X_2}\tilde{\Gamma}_{X_3})$ and the anti-cyclic term $\text{tr} (\tilde{\Gamma}_{X_1}\tilde{\Gamma}_{X_3}\tilde{\Gamma}_{X_2})$ in \Eq{eq:traces}.

Evaluation of the Matsubara sum gives for the vacuum expression
\begin{align}\label{eq:Aderiv3}
\frac{\partial^3\mathcal{A}}{\partial X_3\partial X_2 \partial X_1}\Bigg\vert_{\vect{\mu}=T=0}=\sum_{ijk} 
(\Tilde{\Gamma}_{X_1,ij}\Tilde{\Gamma}_{X_2,jk}\Tilde{\Gamma}_{X_3,ki}+\Tilde{\Gamma}_{X_1,ij}\Tilde{\Gamma}_{X_3,jk}\Tilde{\Gamma}_{X_2,ki})h(\epsilon_i,\epsilon_j,\epsilon_k).
\end{align}
The function $h$ is defined as
\begin{align}
    h(x_1,x_2,x_3)\coloneqq &\lim_{T\to 0}T\sum_n[(i\omega_n-x_1)(i\omega_n-x_2)(i\omega_n-x_3)]^{-1}\nonumber\\
    =&
    \begin{cases}
    0 \quad \text{for $x_1=x_2=x_3$},\\
    -\frac{(\theta(-x_i)-\theta(-x_j))}{(x_i-x_j)^2}\quad \text{for $x_i\neq x_j$ and $x_i=x_k$ or $x_j=x_k$},\\
    \sum_{i=1}^3\theta(-x_i)\Pi_{j\neq i}(x_i-x_j)^{-1}\quad\text{else}.
    \end{cases}
\end{align}
With the second order derivatives \Eq{eq:Aderiv2} and third order derivatives \Eq{eq:Aderiv3} the gap equations and neutrality constraints \Eq{eq:gapeq_explicit} can be calculated.
\\
\twocolumngrid

\bibliography{bib.bib}

\begin{thebibliography}{54}%
\makeatletter
\providecommand \@ifxundefined [1]{%
 \@ifx{#1\undefined}
}%
\providecommand \@ifnum [1]{%
 \ifnum #1\expandafter \@firstoftwo
 \else \expandafter \@secondoftwo
 \fi
}%
\providecommand \@ifx [1]{%
 \ifx #1\expandafter \@firstoftwo
 \else \expandafter \@secondoftwo
 \fi
}%
\providecommand \natexlab [1]{#1}%
\providecommand \enquote  [1]{``#1''}%
\providecommand \bibnamefont  [1]{#1}%
\providecommand \bibfnamefont [1]{#1}%
\providecommand \citenamefont [1]{#1}%
\providecommand \href@noop [0]{\@secondoftwo}%
\providecommand \href [0]{\begingroup \@sanitize@url \@href}%
\providecommand \@href[1]{\@@startlink{#1}\@@href}%
\providecommand \@@href[1]{\endgroup#1\@@endlink}%
\providecommand \@sanitize@url [0]{\catcode `\\12\catcode `\$12\catcode
  `\&12\catcode `\#12\catcode `\^12\catcode `\_12\catcode `\%12\relax}%
\providecommand \@@startlink[1]{}%
\providecommand \@@endlink[0]{}%
\providecommand \url  [0]{\begingroup\@sanitize@url \@url }%
\providecommand \@url [1]{\endgroup\@href {#1}{\urlprefix }}%
\providecommand \urlprefix  [0]{URL }%
\providecommand \Eprint [0]{\href }%
\providecommand \doibase [0]{http://dx.doi.org/}%
\providecommand \selectlanguage [0]{\@gobble}%
\providecommand \bibinfo  [0]{\@secondoftwo}%
\providecommand \bibfield  [0]{\@secondoftwo}%
\providecommand \translation [1]{[#1]}%
\providecommand \BibitemOpen [0]{}%
\providecommand \bibitemStop [0]{}%
\providecommand \bibitemNoStop [0]{.\EOS\space}%
\providecommand \EOS [0]{\spacefactor3000\relax}%
\providecommand \BibitemShut  [1]{\csname bibitem#1\endcsname}%
\let\auto@bib@innerbib\@empty
\bibitem [{\citenamefont {Son}(1999)}]{Son:1998uk}%
  \BibitemOpen
  \bibfield  {author} {\bibinfo {author} {\bibfnamefont {D.~T.}\ \bibnamefont
  {Son}},\ }\href {\doibase 10.1103/PhysRevD.59.094019} {\bibfield  {journal}
  {\bibinfo  {journal} {Phys. Rev. D}\ }\textbf {\bibinfo {volume} {59}},\
  \bibinfo {pages} {094019} (\bibinfo {year} {1999})},\ \Eprint
  {http://arxiv.org/abs/hep-ph/9812287} {arXiv:hep-ph/9812287} \BibitemShut
  {NoStop}%
\bibitem [{\citenamefont {Sch\"afer}\ and\ \citenamefont
  {Wilczek}(1999)}]{Schafer:1999jg}%
  \BibitemOpen
  \bibfield  {author} {\bibinfo {author} {\bibfnamefont {T.}~\bibnamefont
  {Sch\"afer}}\ and\ \bibinfo {author} {\bibfnamefont {F.}~\bibnamefont
  {Wilczek}},\ }\href {\doibase 10.1103/PhysRevD.60.114033} {\bibfield
  {journal} {\bibinfo  {journal} {Phys. Rev. D}\ }\textbf {\bibinfo {volume}
  {60}},\ \bibinfo {pages} {114033} (\bibinfo {year} {1999})},\ \Eprint
  {http://arxiv.org/abs/hep-ph/9906512} {arXiv:hep-ph/9906512} \BibitemShut
  {NoStop}%
\bibitem [{\citenamefont {Pisarski}\ and\ \citenamefont
  {Rischke}(2000)}]{Pisarski:1999tv}%
  \BibitemOpen
  \bibfield  {author} {\bibinfo {author} {\bibfnamefont {R.~D.}\ \bibnamefont
  {Pisarski}}\ and\ \bibinfo {author} {\bibfnamefont {D.~H.}\ \bibnamefont
  {Rischke}},\ }\href {\doibase 10.1103/PhysRevD.61.074017} {\bibfield
  {journal} {\bibinfo  {journal} {Phys. Rev. D}\ }\textbf {\bibinfo {volume}
  {61}},\ \bibinfo {pages} {074017} (\bibinfo {year} {2000})},\ \Eprint
  {http://arxiv.org/abs/nucl-th/9910056} {arXiv:nucl-th/9910056} \BibitemShut
  {NoStop}%
\bibitem [{\citenamefont {Alford}\ \emph {et~al.}(2008)\citenamefont {Alford},
  \citenamefont {Schmitt}, \citenamefont {Rajagopal},\ and\ \citenamefont
  {Sch\"afer}}]{Alford:2007xm}%
  \BibitemOpen
  \bibfield  {author} {\bibinfo {author} {\bibfnamefont {M.~G.}\ \bibnamefont
  {Alford}}, \bibinfo {author} {\bibfnamefont {A.}~\bibnamefont {Schmitt}},
  \bibinfo {author} {\bibfnamefont {K.}~\bibnamefont {Rajagopal}}, \ and\
  \bibinfo {author} {\bibfnamefont {T.}~\bibnamefont {Sch\"afer}},\ }\href
  {\doibase 10.1103/RevModPhys.80.1455} {\bibfield  {journal} {\bibinfo
  {journal} {Rev. Mod. Phys.}\ }\textbf {\bibinfo {volume} {80}},\ \bibinfo
  {pages} {1455} (\bibinfo {year} {2008})},\ \Eprint
  {http://arxiv.org/abs/0709.4635} {arXiv:0709.4635 [hep-ph]} \BibitemShut
  {NoStop}%
\bibitem [{\citenamefont {Sch\"afer}(2000{\natexlab{a}})}]{Schafer:1999fe}%
  \BibitemOpen
  \bibfield  {author} {\bibinfo {author} {\bibfnamefont {T.}~\bibnamefont
  {Sch\"afer}},\ }\href {\doibase 10.1016/S0550-3213(00)00063-8} {\bibfield
  {journal} {\bibinfo  {journal} {Nucl. Phys. B}\ }\textbf {\bibinfo {volume}
  {575}},\ \bibinfo {pages} {269} (\bibinfo {year} {2000}{\natexlab{a}})},\
  \Eprint {http://arxiv.org/abs/hep-ph/9909574} {arXiv:hep-ph/9909574}
  \BibitemShut {NoStop}%
\bibitem [{\citenamefont {Shovkovy}\ and\ \citenamefont
  {Wijewardhana}(1999)}]{Shovkovy:1999mr}%
  \BibitemOpen
  \bibfield  {author} {\bibinfo {author} {\bibfnamefont {I.~A.}\ \bibnamefont
  {Shovkovy}}\ and\ \bibinfo {author} {\bibfnamefont {L.~C.~R.}\ \bibnamefont
  {Wijewardhana}},\ }\href {\doibase 10.1016/S0370-2693(99)01297-6} {\bibfield
  {journal} {\bibinfo  {journal} {Phys. Lett. B}\ }\textbf {\bibinfo {volume}
  {470}},\ \bibinfo {pages} {189} (\bibinfo {year} {1999})},\ \Eprint
  {http://arxiv.org/abs/hep-ph/9910225} {arXiv:hep-ph/9910225} \BibitemShut
  {NoStop}%
\bibitem [{\citenamefont {Alford}\ \emph
  {et~al.}(1999{\natexlab{a}})\citenamefont {Alford}, \citenamefont
  {Rajagopal},\ and\ \citenamefont {Wilczek}}]{Alford:1998mk}%
  \BibitemOpen
  \bibfield  {author} {\bibinfo {author} {\bibfnamefont {M.~G.}\ \bibnamefont
  {Alford}}, \bibinfo {author} {\bibfnamefont {K.}~\bibnamefont {Rajagopal}}, \
  and\ \bibinfo {author} {\bibfnamefont {F.}~\bibnamefont {Wilczek}},\ }\href
  {\doibase 10.1016/S0550-3213(98)00668-3} {\bibfield  {journal} {\bibinfo
  {journal} {Nucl. Phys. B}\ }\textbf {\bibinfo {volume} {537}},\ \bibinfo
  {pages} {443} (\bibinfo {year} {1999}{\natexlab{a}})},\ \Eprint
  {http://arxiv.org/abs/hep-ph/9804403} {arXiv:hep-ph/9804403} \BibitemShut
  {NoStop}%
\bibitem [{\citenamefont {Alford}\ \emph
  {et~al.}(1999{\natexlab{b}})\citenamefont {Alford}, \citenamefont {Berges},\
  and\ \citenamefont {Rajagopal}}]{AlfordUnlocking}%
  \BibitemOpen
  \bibfield  {author} {\bibinfo {author} {\bibfnamefont {M.}~\bibnamefont
  {Alford}}, \bibinfo {author} {\bibfnamefont {J.}~\bibnamefont {Berges}}, \
  and\ \bibinfo {author} {\bibfnamefont {K.}~\bibnamefont {Rajagopal}},\ }\href
  {\doibase https://doi.org/10.1016/S0550-3213(99)00410-1} {\bibfield
  {journal} {\bibinfo  {journal} {Nuclear Physics B}\ }\textbf {\bibinfo
  {volume} {558}},\ \bibinfo {pages} {219} (\bibinfo {year}
  {1999}{\natexlab{b}})}\BibitemShut {NoStop}%
\bibitem [{\citenamefont {Alford}\ and\ \citenamefont
  {Rajagopal}(2002)}]{Alford:2002kj}%
  \BibitemOpen
  \bibfield  {author} {\bibinfo {author} {\bibfnamefont {M.}~\bibnamefont
  {Alford}}\ and\ \bibinfo {author} {\bibfnamefont {K.}~\bibnamefont
  {Rajagopal}},\ }\href {\doibase 10.1088/1126-6708/2002/06/031} {\bibfield
  {journal} {\bibinfo  {journal} {JHEP}\ }\textbf {\bibinfo {volume} {06}},\
  \bibinfo {pages} {031} (\bibinfo {year} {2002})},\ \Eprint
  {http://arxiv.org/abs/hep-ph/0204001} {arXiv:hep-ph/0204001} \BibitemShut
  {NoStop}%
\bibitem [{\citenamefont {Rajagopal}\ and\ \citenamefont
  {Schmitt}(2006)}]{Rajagopal_stressed_pairing}%
  \BibitemOpen
  \bibfield  {author} {\bibinfo {author} {\bibfnamefont {K.}~\bibnamefont
  {Rajagopal}}\ and\ \bibinfo {author} {\bibfnamefont {A.}~\bibnamefont
  {Schmitt}},\ }\href {\doibase 10.1103/PhysRevD.73.045003} {\bibfield
  {journal} {\bibinfo  {journal} {Phys. Rev. D}\ }\textbf {\bibinfo {volume}
  {73}},\ \bibinfo {pages} {045003} (\bibinfo {year} {2006})}\BibitemShut
  {NoStop}%
\bibitem [{\citenamefont {Nickel}\ \emph
  {et~al.}(2006{\natexlab{a}})\citenamefont {Nickel}, \citenamefont {Wambach},\
  and\ \citenamefont {Alkofer}}]{Nickel:2006vf}%
  \BibitemOpen
  \bibfield  {author} {\bibinfo {author} {\bibfnamefont {D.}~\bibnamefont
  {Nickel}}, \bibinfo {author} {\bibfnamefont {J.}~\bibnamefont {Wambach}}, \
  and\ \bibinfo {author} {\bibfnamefont {R.}~\bibnamefont {Alkofer}},\ }\href
  {\doibase 10.1103/PhysRevD.73.114028} {\bibfield  {journal} {\bibinfo
  {journal} {Phys. Rev. D}\ }\textbf {\bibinfo {volume} {73}},\ \bibinfo
  {pages} {114028} (\bibinfo {year} {2006}{\natexlab{a}})},\ \Eprint
  {http://arxiv.org/abs/hep-ph/0603163} {arXiv:hep-ph/0603163} \BibitemShut
  {NoStop}%
\bibitem [{\citenamefont {Nickel}\ \emph
  {et~al.}(2006{\natexlab{b}})\citenamefont {Nickel}, \citenamefont {Alkofer},\
  and\ \citenamefont {Wambach}}]{Nickel:2006kc}%
  \BibitemOpen
  \bibfield  {author} {\bibinfo {author} {\bibfnamefont {D.}~\bibnamefont
  {Nickel}}, \bibinfo {author} {\bibfnamefont {R.}~\bibnamefont {Alkofer}}, \
  and\ \bibinfo {author} {\bibfnamefont {J.}~\bibnamefont {Wambach}},\ }\href
  {\doibase 10.1103/PhysRevD.74.114015} {\bibfield  {journal} {\bibinfo
  {journal} {Phys. Rev. D}\ }\textbf {\bibinfo {volume} {74}},\ \bibinfo
  {pages} {114015} (\bibinfo {year} {2006}{\natexlab{b}})},\ \Eprint
  {http://arxiv.org/abs/hep-ph/0609198} {arXiv:hep-ph/0609198} \BibitemShut
  {NoStop}%
\bibitem [{\citenamefont {Nickel}\ \emph {et~al.}(2008)\citenamefont {Nickel},
  \citenamefont {Alkofer},\ and\ \citenamefont {Wambach}}]{Nickel:2008ef}%
  \BibitemOpen
  \bibfield  {author} {\bibinfo {author} {\bibfnamefont {D.}~\bibnamefont
  {Nickel}}, \bibinfo {author} {\bibfnamefont {R.}~\bibnamefont {Alkofer}}, \
  and\ \bibinfo {author} {\bibfnamefont {J.}~\bibnamefont {Wambach}},\ }\href
  {\doibase 10.1103/PhysRevD.77.114010} {\bibfield  {journal} {\bibinfo
  {journal} {Phys. Rev. D}\ }\textbf {\bibinfo {volume} {77}},\ \bibinfo
  {pages} {114010} (\bibinfo {year} {2008})},\ \Eprint
  {http://arxiv.org/abs/0802.3187} {arXiv:0802.3187 [hep-ph]} \BibitemShut
  {NoStop}%
\bibitem [{\citenamefont {M\"uller}\ \emph {et~al.}(2013)\citenamefont
  {M\"uller}, \citenamefont {Buballa},\ and\ \citenamefont
  {Wambach}}]{Muller:2013pya}%
  \BibitemOpen
  \bibfield  {author} {\bibinfo {author} {\bibfnamefont {D.}~\bibnamefont
  {M\"uller}}, \bibinfo {author} {\bibfnamefont {M.}~\bibnamefont {Buballa}}, \
  and\ \bibinfo {author} {\bibfnamefont {J.}~\bibnamefont {Wambach}},\ }\href
  {\doibase 10.1140/epja/i2013-13096-5} {\bibfield  {journal} {\bibinfo
  {journal} {Eur. Phys. J. A}\ }\textbf {\bibinfo {volume} {49}},\ \bibinfo
  {pages} {96} (\bibinfo {year} {2013})},\ \Eprint
  {http://arxiv.org/abs/1303.2693} {arXiv:1303.2693 [hep-ph]} \BibitemShut
  {NoStop}%
\bibitem [{\citenamefont {M\"uller}\ \emph {et~al.}(2016)\citenamefont
  {M\"uller}, \citenamefont {Buballa},\ and\ \citenamefont
  {Wambach}}]{Muller:2016fdr}%
  \BibitemOpen
  \bibfield  {author} {\bibinfo {author} {\bibfnamefont {D.}~\bibnamefont
  {M\"uller}}, \bibinfo {author} {\bibfnamefont {M.}~\bibnamefont {Buballa}}, \
  and\ \bibinfo {author} {\bibfnamefont {J.}~\bibnamefont {Wambach}},\
  }\href@noop {} {\  (\bibinfo {year} {2016})},\ \Eprint
  {http://arxiv.org/abs/1603.02865} {arXiv:1603.02865 [hep-ph]} \BibitemShut
  {NoStop}%
\bibitem [{\citenamefont {Gunkel}\ and\ \citenamefont
  {Fischer}(2021)}]{Gunkel:2021oya}%
  \BibitemOpen
  \bibfield  {author} {\bibinfo {author} {\bibfnamefont {P.~J.}\ \bibnamefont
  {Gunkel}}\ and\ \bibinfo {author} {\bibfnamefont {C.~S.}\ \bibnamefont
  {Fischer}},\ }\href {\doibase 10.1103/PhysRevD.104.054022} {\bibfield
  {journal} {\bibinfo  {journal} {Phys. Rev. D}\ }\textbf {\bibinfo {volume}
  {104}},\ \bibinfo {pages} {054022} (\bibinfo {year} {2021})},\ \Eprint
  {http://arxiv.org/abs/2106.08356} {arXiv:2106.08356 [hep-ph]} \BibitemShut
  {NoStop}%
\bibitem [{\citenamefont {Fischer}(2019)}]{Fischer:2018sdj}%
  \BibitemOpen
  \bibfield  {author} {\bibinfo {author} {\bibfnamefont {C.~S.}\ \bibnamefont
  {Fischer}},\ }\href {\doibase 10.1016/j.ppnp.2019.01.002} {\bibfield
  {journal} {\bibinfo  {journal} {Prog. Part. Nucl. Phys.}\ }\textbf {\bibinfo
  {volume} {105}},\ \bibinfo {pages} {1} (\bibinfo {year} {2019})},\ \Eprint
  {http://arxiv.org/abs/1810.12938} {arXiv:1810.12938 [hep-ph]} \BibitemShut
  {NoStop}%
\bibitem [{\citenamefont {Buballa}(2005)}]{BUBALLAHABIL}%
  \BibitemOpen
  \bibfield  {author} {\bibinfo {author} {\bibfnamefont {M.}~\bibnamefont
  {Buballa}},\ }\href {\doibase https://doi.org/10.1016/j.physrep.2004.11.004}
  {\bibfield  {journal} {\bibinfo  {journal} {Physics Reports}\ }\textbf
  {\bibinfo {volume} {407}},\ \bibinfo {pages} {205} (\bibinfo {year}
  {2005})}\BibitemShut {NoStop}%
\bibitem [{\citenamefont {Ruester}\ \emph {et~al.}(2005)\citenamefont
  {Ruester}, \citenamefont {Werth}, \citenamefont {Buballa}, \citenamefont
  {Shovkovy},\ and\ \citenamefont {Rischke}}]{Ruester:2005jc}%
  \BibitemOpen
  \bibfield  {author} {\bibinfo {author} {\bibfnamefont {S.~B.}\ \bibnamefont
  {Ruester}}, \bibinfo {author} {\bibfnamefont {V.}~\bibnamefont {Werth}},
  \bibinfo {author} {\bibfnamefont {M.}~\bibnamefont {Buballa}}, \bibinfo
  {author} {\bibfnamefont {I.~A.}\ \bibnamefont {Shovkovy}}, \ and\ \bibinfo
  {author} {\bibfnamefont {D.~H.}\ \bibnamefont {Rischke}},\ }\href {\doibase
  10.1103/PhysRevD.72.034004} {\bibfield  {journal} {\bibinfo  {journal} {Phys.
  Rev. D}\ }\textbf {\bibinfo {volume} {72}},\ \bibinfo {pages} {034004}
  (\bibinfo {year} {2005})},\ \Eprint {http://arxiv.org/abs/hep-ph/0503184}
  {arXiv:hep-ph/0503184} \BibitemShut {NoStop}%
\bibitem [{\citenamefont {Blaschke}\ \emph {et~al.}(2005)\citenamefont
  {Blaschke}, \citenamefont {Fredriksson}, \citenamefont {Grigorian},
  \citenamefont {Oztas},\ and\ \citenamefont {Sandin}}]{Blaschke:2005uj}%
  \BibitemOpen
  \bibfield  {author} {\bibinfo {author} {\bibfnamefont {D.}~\bibnamefont
  {Blaschke}}, \bibinfo {author} {\bibfnamefont {S.}~\bibnamefont
  {Fredriksson}}, \bibinfo {author} {\bibfnamefont {H.}~\bibnamefont
  {Grigorian}}, \bibinfo {author} {\bibfnamefont {A.~M.}\ \bibnamefont
  {Oztas}}, \ and\ \bibinfo {author} {\bibfnamefont {F.}~\bibnamefont
  {Sandin}},\ }\href {\doibase 10.1103/PhysRevD.72.065020} {\bibfield
  {journal} {\bibinfo  {journal} {Phys. Rev. D}\ }\textbf {\bibinfo {volume}
  {72}},\ \bibinfo {pages} {065020} (\bibinfo {year} {2005})},\ \Eprint
  {http://arxiv.org/abs/hep-ph/0503194} {arXiv:hep-ph/0503194} \BibitemShut
  {NoStop}%
\bibitem [{\citenamefont {Abuki}\ and\ \citenamefont
  {Kunihiro}(2006)}]{Abuki:2005ms}%
  \BibitemOpen
  \bibfield  {author} {\bibinfo {author} {\bibfnamefont {H.}~\bibnamefont
  {Abuki}}\ and\ \bibinfo {author} {\bibfnamefont {T.}~\bibnamefont
  {Kunihiro}},\ }\href {\doibase 10.1016/j.nuclphysa.2005.12.019} {\bibfield
  {journal} {\bibinfo  {journal} {Nucl. Phys. A}\ }\textbf {\bibinfo {volume}
  {768}},\ \bibinfo {pages} {118} (\bibinfo {year} {2006})},\ \Eprint
  {http://arxiv.org/abs/hep-ph/0509172} {arXiv:hep-ph/0509172} \BibitemShut
  {NoStop}%
\bibitem [{\citenamefont {Buballa}\ and\ \citenamefont
  {Oertel}(2002)}]{Buballa:2001gj}%
  \BibitemOpen
  \bibfield  {author} {\bibinfo {author} {\bibfnamefont {M.}~\bibnamefont
  {Buballa}}\ and\ \bibinfo {author} {\bibfnamefont {M.}~\bibnamefont
  {Oertel}},\ }\href {\doibase 10.1016/S0375-9474(01)01674-8} {\bibfield
  {journal} {\bibinfo  {journal} {Nucl. Phys. A}\ }\textbf {\bibinfo {volume}
  {703}},\ \bibinfo {pages} {770} (\bibinfo {year} {2002})},\ \Eprint
  {http://arxiv.org/abs/hep-ph/0109095} {arXiv:hep-ph/0109095} \BibitemShut
  {NoStop}%
\bibitem [{\citenamefont {Klevansky}(1992)}]{Klevansky:1992qe}%
  \BibitemOpen
  \bibfield  {author} {\bibinfo {author} {\bibfnamefont {S.~P.}\ \bibnamefont
  {Klevansky}},\ }\href {\doibase 10.1103/RevModPhys.64.649} {\bibfield
  {journal} {\bibinfo  {journal} {Rev. Mod. Phys.}\ }\textbf {\bibinfo {volume}
  {64}},\ \bibinfo {pages} {649} (\bibinfo {year} {1992})}\BibitemShut
  {NoStop}%
\bibitem [{\citenamefont {Farias}\ \emph {et~al.}(2006)\citenamefont {Farias},
  \citenamefont {Dallabona}, \citenamefont {Krein},\ and\ \citenamefont
  {Battistel}}]{Farias:2005cr}%
  \BibitemOpen
  \bibfield  {author} {\bibinfo {author} {\bibfnamefont {R.~L.~S.}\
  \bibnamefont {Farias}}, \bibinfo {author} {\bibfnamefont {G.}~\bibnamefont
  {Dallabona}}, \bibinfo {author} {\bibfnamefont {G.}~\bibnamefont {Krein}}, \
  and\ \bibinfo {author} {\bibfnamefont {O.~A.}\ \bibnamefont {Battistel}},\
  }\href {\doibase 10.1103/PhysRevC.73.018201} {\bibfield  {journal} {\bibinfo
  {journal} {Phys. Rev. C}\ }\textbf {\bibinfo {volume} {73}},\ \bibinfo
  {pages} {018201} (\bibinfo {year} {2006})},\ \Eprint
  {http://arxiv.org/abs/hep-ph/0510145} {arXiv:hep-ph/0510145} \BibitemShut
  {NoStop}%
\bibitem [{\citenamefont {Iida}\ \emph {et~al.}(2004)\citenamefont {Iida},
  \citenamefont {Matsuura}, \citenamefont {Tachibana},\ and\ \citenamefont
  {Hatsuda}}]{Iida:2003cc}%
  \BibitemOpen
  \bibfield  {author} {\bibinfo {author} {\bibfnamefont {K.}~\bibnamefont
  {Iida}}, \bibinfo {author} {\bibfnamefont {T.}~\bibnamefont {Matsuura}},
  \bibinfo {author} {\bibfnamefont {M.}~\bibnamefont {Tachibana}}, \ and\
  \bibinfo {author} {\bibfnamefont {T.}~\bibnamefont {Hatsuda}},\ }\href
  {\doibase 10.1103/PhysRevLett.93.132001} {\bibfield  {journal} {\bibinfo
  {journal} {Phys. Rev. Lett.}\ }\textbf {\bibinfo {volume} {93}},\ \bibinfo
  {pages} {132001} (\bibinfo {year} {2004})},\ \Eprint
  {http://arxiv.org/abs/hep-ph/0312363} {arXiv:hep-ph/0312363} \BibitemShut
  {NoStop}%
\bibitem [{\citenamefont {Iida}\ \emph {et~al.}(2005)\citenamefont {Iida},
  \citenamefont {Matsuura}, \citenamefont {Tachibana},\ and\ \citenamefont
  {Hatsuda}}]{Iida:2004cj}%
  \BibitemOpen
  \bibfield  {author} {\bibinfo {author} {\bibfnamefont {K.}~\bibnamefont
  {Iida}}, \bibinfo {author} {\bibfnamefont {T.}~\bibnamefont {Matsuura}},
  \bibinfo {author} {\bibfnamefont {M.}~\bibnamefont {Tachibana}}, \ and\
  \bibinfo {author} {\bibfnamefont {T.}~\bibnamefont {Hatsuda}},\ }\href
  {\doibase 10.1103/PhysRevD.71.054003} {\bibfield  {journal} {\bibinfo
  {journal} {Phys. Rev. D}\ }\textbf {\bibinfo {volume} {71}},\ \bibinfo
  {pages} {054003} (\bibinfo {year} {2005})},\ \Eprint
  {http://arxiv.org/abs/hep-ph/0411356} {arXiv:hep-ph/0411356} \BibitemShut
  {NoStop}%
\bibitem [{\citenamefont {Fukushima}\ \emph {et~al.}(2005)\citenamefont
  {Fukushima}, \citenamefont {Kouvaris},\ and\ \citenamefont
  {Rajagopal}}]{Fukushima:2004zq}%
  \BibitemOpen
  \bibfield  {author} {\bibinfo {author} {\bibfnamefont {K.}~\bibnamefont
  {Fukushima}}, \bibinfo {author} {\bibfnamefont {C.}~\bibnamefont {Kouvaris}},
  \ and\ \bibinfo {author} {\bibfnamefont {K.}~\bibnamefont {Rajagopal}},\
  }\href {\doibase 10.1103/PhysRevD.71.034002} {\bibfield  {journal} {\bibinfo
  {journal} {Phys. Rev. D}\ }\textbf {\bibinfo {volume} {71}},\ \bibinfo
  {pages} {034002} (\bibinfo {year} {2005})},\ \Eprint
  {http://arxiv.org/abs/hep-ph/0408322} {arXiv:hep-ph/0408322} \BibitemShut
  {NoStop}%
\bibitem [{\citenamefont {Pasqualotto}\ \emph {et~al.}(2023)\citenamefont
  {Pasqualotto}, \citenamefont {Farias}, \citenamefont {Tavares}, \citenamefont
  {Avancini},\ and\ \citenamefont {Krein}}]{Pasqualotto:2023hho}%
  \BibitemOpen
  \bibfield  {author} {\bibinfo {author} {\bibfnamefont {A.~E.~B.}\
  \bibnamefont {Pasqualotto}}, \bibinfo {author} {\bibfnamefont {R.~L.~S.}\
  \bibnamefont {Farias}}, \bibinfo {author} {\bibfnamefont {W.~R.}\
  \bibnamefont {Tavares}}, \bibinfo {author} {\bibfnamefont {S.~S.}\
  \bibnamefont {Avancini}}, \ and\ \bibinfo {author} {\bibfnamefont {G.~a.}\
  \bibnamefont {Krein}},\ }\href {\doibase 10.1103/PhysRevD.107.096017}
  {\bibfield  {journal} {\bibinfo  {journal} {Phys. Rev. D}\ }\textbf {\bibinfo
  {volume} {107}},\ \bibinfo {pages} {096017} (\bibinfo {year} {2023})},\
  \Eprint {http://arxiv.org/abs/2301.10721} {arXiv:2301.10721 [hep-ph]}
  \BibitemShut {NoStop}%
\bibitem [{\citenamefont {Duarte}\ \emph {et~al.}(2019)\citenamefont {Duarte},
  \citenamefont {Farias},\ and\ \citenamefont {Ramos}}]{Duarte:2018kfd}%
  \BibitemOpen
  \bibfield  {author} {\bibinfo {author} {\bibfnamefont {D.~C.}\ \bibnamefont
  {Duarte}}, \bibinfo {author} {\bibfnamefont {R.~L.~S.}\ \bibnamefont
  {Farias}}, \ and\ \bibinfo {author} {\bibfnamefont {R.~O.}\ \bibnamefont
  {Ramos}},\ }\href {\doibase 10.1103/PhysRevD.99.016005} {\bibfield  {journal}
  {\bibinfo  {journal} {Phys. Rev. D}\ }\textbf {\bibinfo {volume} {99}},\
  \bibinfo {pages} {016005} (\bibinfo {year} {2019})},\ \Eprint
  {http://arxiv.org/abs/1811.10598} {arXiv:1811.10598 [hep-ph]} \BibitemShut
  {NoStop}%
\bibitem [{\citenamefont {Braun}\ \emph {et~al.}(2019)\citenamefont {Braun},
  \citenamefont {Leonhardt},\ and\ \citenamefont {Pawlowski}}]{Braun:2018svj}%
  \BibitemOpen
  \bibfield  {author} {\bibinfo {author} {\bibfnamefont {J.}~\bibnamefont
  {Braun}}, \bibinfo {author} {\bibfnamefont {M.}~\bibnamefont {Leonhardt}}, \
  and\ \bibinfo {author} {\bibfnamefont {J.~M.}\ \bibnamefont {Pawlowski}},\
  }\href {\doibase 10.21468/SciPostPhys.6.5.056} {\bibfield  {journal}
  {\bibinfo  {journal} {SciPost Phys.}\ }\textbf {\bibinfo {volume} {6}},\
  \bibinfo {pages} {056} (\bibinfo {year} {2019})},\ \Eprint
  {http://arxiv.org/abs/1806.04432} {arXiv:1806.04432 [hep-ph]} \BibitemShut
  {NoStop}%
\bibitem [{\citenamefont {Braun}\ and\ \citenamefont
  {Schallmo}(2022)}]{Braun:2022olp}%
  \BibitemOpen
  \bibfield  {author} {\bibinfo {author} {\bibfnamefont {J.}~\bibnamefont
  {Braun}}\ and\ \bibinfo {author} {\bibfnamefont {B.}~\bibnamefont
  {Schallmo}},\ }\href {\doibase 10.1103/PhysRevD.106.076010} {\bibfield
  {journal} {\bibinfo  {journal} {Phys. Rev. D}\ }\textbf {\bibinfo {volume}
  {106}},\ \bibinfo {pages} {076010} (\bibinfo {year} {2022})},\ \Eprint
  {http://arxiv.org/abs/2204.00358} {arXiv:2204.00358 [nucl-th]} \BibitemShut
  {NoStop}%
\bibitem [{\citenamefont {Rehberg}\ \emph {et~al.}(1996)\citenamefont
  {Rehberg}, \citenamefont {Klevansky},\ and\ \citenamefont
  {Hufner}}]{Rehberg:1995kh}%
  \BibitemOpen
  \bibfield  {author} {\bibinfo {author} {\bibfnamefont {P.}~\bibnamefont
  {Rehberg}}, \bibinfo {author} {\bibfnamefont {S.~P.}\ \bibnamefont
  {Klevansky}}, \ and\ \bibinfo {author} {\bibfnamefont {J.}~\bibnamefont
  {Hufner}},\ }\href {\doibase 10.1103/PhysRevC.53.410} {\bibfield  {journal}
  {\bibinfo  {journal} {Phys. Rev. C}\ }\textbf {\bibinfo {volume} {53}},\
  \bibinfo {pages} {410} (\bibinfo {year} {1996})},\ \Eprint
  {http://arxiv.org/abs/hep-ph/9506436} {arXiv:hep-ph/9506436} \BibitemShut
  {NoStop}%
\bibitem [{\citenamefont {Kobayashi}\ and\ \citenamefont
  {Maskawa}(1970)}]{Kobayashi:1970ji}%
  \BibitemOpen
  \bibfield  {author} {\bibinfo {author} {\bibfnamefont {M.}~\bibnamefont
  {Kobayashi}}\ and\ \bibinfo {author} {\bibfnamefont {T.}~\bibnamefont
  {Maskawa}},\ }\href {\doibase 10.1143/PTP.44.1422} {\bibfield  {journal}
  {\bibinfo  {journal} {Prog. Theor. Phys.}\ }\textbf {\bibinfo {volume}
  {44}},\ \bibinfo {pages} {1422} (\bibinfo {year} {1970})}\BibitemShut
  {NoStop}%
\bibitem [{\citenamefont {'t~Hooft}(1976)}]{tHooft:1976rip}%
  \BibitemOpen
  \bibfield  {author} {\bibinfo {author} {\bibfnamefont {G.}~\bibnamefont
  {'t~Hooft}},\ }\href {\doibase 10.1103/PhysRevLett.37.8} {\bibfield
  {journal} {\bibinfo  {journal} {Phys. Rev. Lett.}\ }\textbf {\bibinfo
  {volume} {37}},\ \bibinfo {pages} {8} (\bibinfo {year} {1976})}\BibitemShut
  {NoStop}%
\bibitem [{\citenamefont {Buballa}\ and\ \citenamefont
  {Shovkovy}(2005)}]{Buballa:2005bv}%
  \BibitemOpen
  \bibfield  {author} {\bibinfo {author} {\bibfnamefont {M.}~\bibnamefont
  {Buballa}}\ and\ \bibinfo {author} {\bibfnamefont {I.~A.}\ \bibnamefont
  {Shovkovy}},\ }\href {\doibase 10.1103/PhysRevD.72.097501} {\bibfield
  {journal} {\bibinfo  {journal} {Phys. Rev. D}\ }\textbf {\bibinfo {volume}
  {72}},\ \bibinfo {pages} {097501} (\bibinfo {year} {2005})},\ \Eprint
  {http://arxiv.org/abs/hep-ph/0508197} {arXiv:hep-ph/0508197} \BibitemShut
  {NoStop}%
\bibitem [{\citenamefont {Pannullo}\ \emph {et~al.}(2024)\citenamefont
  {Pannullo}, \citenamefont {Wagner},\ and\ \citenamefont
  {Winstel}}]{Pannullo:2024sov}%
  \BibitemOpen
  \bibfield  {author} {\bibinfo {author} {\bibfnamefont {L.}~\bibnamefont
  {Pannullo}}, \bibinfo {author} {\bibfnamefont {M.}~\bibnamefont {Wagner}}, \
  and\ \bibinfo {author} {\bibfnamefont {M.}~\bibnamefont {Winstel}},\
  }\href@noop {} {\  (\bibinfo {year} {2024})},\ \Eprint
  {http://arxiv.org/abs/2406.11312} {arXiv:2406.11312 [hep-ph]} \BibitemShut
  {NoStop}%
\bibitem [{\citenamefont {Wetterich}(1993)}]{Wetterich:1992yh}%
  \BibitemOpen
  \bibfield  {author} {\bibinfo {author} {\bibfnamefont {C.}~\bibnamefont
  {Wetterich}},\ }\href {\doibase 10.1016/0370-2693(93)90726-X} {\bibfield
  {journal} {\bibinfo  {journal} {Phys. Lett. B}\ }\textbf {\bibinfo {volume}
  {301}},\ \bibinfo {pages} {90} (\bibinfo {year} {1993})},\ \Eprint
  {http://arxiv.org/abs/1710.05815} {arXiv:1710.05815 [hep-th]} \BibitemShut
  {NoStop}%
\bibitem [{\citenamefont {Gies}(2012)}]{Gies:2006wv}%
  \BibitemOpen
  \bibfield  {author} {\bibinfo {author} {\bibfnamefont {H.}~\bibnamefont
  {Gies}},\ }\href {\doibase 10.1007/978-3-642-27320-9_6} {\bibfield  {journal}
  {\bibinfo  {journal} {Lect. Notes Phys.}\ }\textbf {\bibinfo {volume}
  {852}},\ \bibinfo {pages} {287} (\bibinfo {year} {2012})},\ \Eprint
  {http://arxiv.org/abs/hep-ph/0611146} {arXiv:hep-ph/0611146} \BibitemShut
  {NoStop}%
\bibitem [{\citenamefont {Pawlowski}\ \emph {et~al.}(2017)\citenamefont
  {Pawlowski}, \citenamefont {Scherer}, \citenamefont {Schmidt},\ and\
  \citenamefont {Wetzel}}]{Pawlowski:2015mlf}%
  \BibitemOpen
  \bibfield  {author} {\bibinfo {author} {\bibfnamefont {J.~M.}\ \bibnamefont
  {Pawlowski}}, \bibinfo {author} {\bibfnamefont {M.~M.}\ \bibnamefont
  {Scherer}}, \bibinfo {author} {\bibfnamefont {R.}~\bibnamefont {Schmidt}}, \
  and\ \bibinfo {author} {\bibfnamefont {S.~J.}\ \bibnamefont {Wetzel}},\
  }\href {\doibase 10.1016/j.aop.2017.06.017} {\bibfield  {journal} {\bibinfo
  {journal} {Annals Phys.}\ }\textbf {\bibinfo {volume} {384}},\ \bibinfo
  {pages} {165} (\bibinfo {year} {2017})},\ \Eprint
  {http://arxiv.org/abs/1512.03598} {arXiv:1512.03598 [hep-th]} \BibitemShut
  {NoStop}%
\bibitem [{\citenamefont {Steil}(2024)}]{Steil:2024phd}%
  \BibitemOpen
  \bibfield  {author} {\bibinfo {author} {\bibfnamefont {M.~J.}\ \bibnamefont
  {Steil}},\ }\emph {\bibinfo {title} {{From zero-dimensional theories to
  inhomogeneous phases with the functional renormalization group}}},\ \href
  {\doibase 10.26083/tuprints-00027380} {\bibinfo {type} {Phd thesis}},\
  \bibinfo  {school} {Technische Universität Darmstadt} (\bibinfo {year}
  {2024})\BibitemShut {NoStop}%
\bibitem [{\citenamefont {Bedaque}(2002)}]{Bedaque:1999nu}%
  \BibitemOpen
  \bibfield  {author} {\bibinfo {author} {\bibfnamefont {P.~F.}\ \bibnamefont
  {Bedaque}},\ }\href {\doibase 10.1016/S0375-9474(01)01234-9} {\bibfield
  {journal} {\bibinfo  {journal} {Nucl. Phys. A}\ }\textbf {\bibinfo {volume}
  {697}},\ \bibinfo {pages} {569} (\bibinfo {year} {2002})},\ \Eprint
  {http://arxiv.org/abs/hep-ph/9910247} {arXiv:hep-ph/9910247} \BibitemShut
  {NoStop}%
\bibitem [{\citenamefont {Gholami}\ \emph {et~al.}(ming)\citenamefont
  {Gholami}, \citenamefont {Kurth},\ and\ \citenamefont {Buballa}}]{QMDFUTURE}%
  \BibitemOpen
  \bibfield  {author} {\bibinfo {author} {\bibfnamefont {H.}~\bibnamefont
  {Gholami}}, \bibinfo {author} {\bibfnamefont {L.}~\bibnamefont {Kurth}}, \
  and\ \bibinfo {author} {\bibfnamefont {M.}~\bibnamefont {Buballa}},\
  }\href@noop {} {\  (\bibinfo {year} {forthcoming})}\BibitemShut {NoStop}%
\bibitem [{\citenamefont {Andersen}\ and\ \citenamefont
  {N\o{}dtvedt}(2024)}]{Andersen:2024qus}%
  \BibitemOpen
  \bibfield  {author} {\bibinfo {author} {\bibfnamefont {J.~O.}\ \bibnamefont
  {Andersen}}\ and\ \bibinfo {author} {\bibfnamefont {M.~P.}\ \bibnamefont
  {N\o{}dtvedt}},\ }\href@noop {} {\  (\bibinfo {year} {2024})},\ \Eprint
  {http://arxiv.org/abs/2408.12361} {arXiv:2408.12361 [hep-ph]} \BibitemShut
  {NoStop}%
\bibitem [{\citenamefont {Schmitt}\ \emph {et~al.}(2002)\citenamefont
  {Schmitt}, \citenamefont {Wang},\ and\ \citenamefont
  {Rischke}}]{Schmitt:2002sc}%
  \BibitemOpen
  \bibfield  {author} {\bibinfo {author} {\bibfnamefont {A.}~\bibnamefont
  {Schmitt}}, \bibinfo {author} {\bibfnamefont {Q.}~\bibnamefont {Wang}}, \
  and\ \bibinfo {author} {\bibfnamefont {D.~H.}\ \bibnamefont {Rischke}},\
  }\href {\doibase 10.1103/PhysRevD.66.114010} {\bibfield  {journal} {\bibinfo
  {journal} {Phys. Rev. D}\ }\textbf {\bibinfo {volume} {66}},\ \bibinfo
  {pages} {114010} (\bibinfo {year} {2002})},\ \Eprint
  {http://arxiv.org/abs/nucl-th/0209050} {arXiv:nucl-th/0209050} \BibitemShut
  {NoStop}%
\bibitem [{\citenamefont {Shovkovy}\ and\ \citenamefont
  {Huang}(2003)}]{Shovkovy_gapless}%
  \BibitemOpen
  \bibfield  {author} {\bibinfo {author} {\bibfnamefont {I.}~\bibnamefont
  {Shovkovy}}\ and\ \bibinfo {author} {\bibfnamefont {M.}~\bibnamefont
  {Huang}},\ }\href {\doibase https://doi.org/10.1016/S0370-2693(03)00748-2}
  {\bibfield  {journal} {\bibinfo  {journal} {Physics Letters B}\ }\textbf
  {\bibinfo {volume} {564}},\ \bibinfo {pages} {205} (\bibinfo {year}
  {2003})}\BibitemShut {NoStop}%
\bibitem [{\citenamefont {Alford}\ \emph {et~al.}(2004)\citenamefont {Alford},
  \citenamefont {Kouvaris},\ and\ \citenamefont {Rajagopal}}]{Alford:2003fq}%
  \BibitemOpen
  \bibfield  {author} {\bibinfo {author} {\bibfnamefont {M.}~\bibnamefont
  {Alford}}, \bibinfo {author} {\bibfnamefont {C.}~\bibnamefont {Kouvaris}}, \
  and\ \bibinfo {author} {\bibfnamefont {K.}~\bibnamefont {Rajagopal}},\ }\href
  {\doibase 10.1103/PhysRevLett.92.222001} {\bibfield  {journal} {\bibinfo
  {journal} {Phys. Rev. Lett.}\ }\textbf {\bibinfo {volume} {92}},\ \bibinfo
  {pages} {222001} (\bibinfo {year} {2004})},\ \Eprint
  {http://arxiv.org/abs/hep-ph/0311286} {arXiv:hep-ph/0311286} \BibitemShut
  {NoStop}%
\bibitem [{\citenamefont {Perego}\ \emph {et~al.}(2019)\citenamefont {Perego},
  \citenamefont {Bernuzzi},\ and\ \citenamefont {Radice}}]{Perego:2019adq}%
  \BibitemOpen
  \bibfield  {author} {\bibinfo {author} {\bibfnamefont {A.}~\bibnamefont
  {Perego}}, \bibinfo {author} {\bibfnamefont {S.}~\bibnamefont {Bernuzzi}}, \
  and\ \bibinfo {author} {\bibfnamefont {D.}~\bibnamefont {Radice}},\ }\href
  {\doibase 10.1140/epja/i2019-12810-7} {\bibfield  {journal} {\bibinfo
  {journal} {Eur. Phys. J. A}\ }\textbf {\bibinfo {volume} {55}},\ \bibinfo
  {pages} {124} (\bibinfo {year} {2019})},\ \Eprint
  {http://arxiv.org/abs/1903.07898} {arXiv:1903.07898 [gr-qc]} \BibitemShut
  {NoStop}%
\bibitem [{\citenamefont {Baiotti}\ and\ \citenamefont
  {Rezzolla}(2017)}]{Baiotti:2016qnr}%
  \BibitemOpen
  \bibfield  {author} {\bibinfo {author} {\bibfnamefont {L.}~\bibnamefont
  {Baiotti}}\ and\ \bibinfo {author} {\bibfnamefont {L.}~\bibnamefont
  {Rezzolla}},\ }\href {\doibase 10.1088/1361-6633/aa67bb} {\bibfield
  {journal} {\bibinfo  {journal} {Rept. Prog. Phys.}\ }\textbf {\bibinfo
  {volume} {80}},\ \bibinfo {pages} {096901} (\bibinfo {year} {2017})},\
  \Eprint {http://arxiv.org/abs/1607.03540} {arXiv:1607.03540 [gr-qc]}
  \BibitemShut {NoStop}%
\bibitem [{\citenamefont {Burns}(2020)}]{Burns:2019byj}%
  \BibitemOpen
  \bibfield  {author} {\bibinfo {author} {\bibfnamefont {E.}~\bibnamefont
  {Burns}},\ }\href {\doibase 10.1007/s41114-020-00028-7} {\bibfield  {journal}
  {\bibinfo  {journal} {Living Rev. Rel.}\ }\textbf {\bibinfo {volume} {23}},\
  \bibinfo {pages} {4} (\bibinfo {year} {2020})},\ \Eprint
  {http://arxiv.org/abs/1909.06085} {arXiv:1909.06085 [astro-ph.HE]}
  \BibitemShut {NoStop}%
\bibitem [{\citenamefont {Battistel}(1999)}]{Battistel:1999phd}%
  \BibitemOpen
  \bibfield  {author} {\bibinfo {author} {\bibfnamefont {O.~A.}\ \bibnamefont
  {Battistel}},\ }\emph {\bibinfo {title} {{A New Strategy to Manipulate and
  Calculate Divergencies in TQC}}},\ \href@noop {} {\bibinfo {type} {Phd
  thesis}},\ \bibinfo  {school} {Universidade Federal de Minas Gerais}
  (\bibinfo {year} {1999})\BibitemShut {NoStop}%
\bibitem [{\citenamefont {Farias}\ \emph {et~al.}(2008)\citenamefont {Farias},
  \citenamefont {Dallabona}, \citenamefont {Krein},\ and\ \citenamefont
  {Battistel}}]{Farias:2006cs}%
  \BibitemOpen
  \bibfield  {author} {\bibinfo {author} {\bibfnamefont {R.~L.~S.}\
  \bibnamefont {Farias}}, \bibinfo {author} {\bibfnamefont {G.}~\bibnamefont
  {Dallabona}}, \bibinfo {author} {\bibfnamefont {G.}~\bibnamefont {Krein}}, \
  and\ \bibinfo {author} {\bibfnamefont {O.~A.}\ \bibnamefont {Battistel}},\
  }\href {\doibase 10.1103/PhysRevC.77.065201} {\bibfield  {journal} {\bibinfo
  {journal} {Phys. Rev. C}\ }\textbf {\bibinfo {volume} {77}},\ \bibinfo
  {pages} {065201} (\bibinfo {year} {2008})},\ \Eprint
  {http://arxiv.org/abs/hep-ph/0604203} {arXiv:hep-ph/0604203} \BibitemShut
  {NoStop}%
\bibitem [{\citenamefont {Sch\"afer}(2000{\natexlab{b}})}]{Schafer:2000ew}%
  \BibitemOpen
  \bibfield  {author} {\bibinfo {author} {\bibfnamefont {T.}~\bibnamefont
  {Sch\"afer}},\ }\href {\doibase 10.1103/PhysRevLett.85.5531} {\bibfield
  {journal} {\bibinfo  {journal} {Phys. Rev. Lett.}\ }\textbf {\bibinfo
  {volume} {85}},\ \bibinfo {pages} {5531} (\bibinfo {year}
  {2000}{\natexlab{b}})},\ \Eprint {http://arxiv.org/abs/nucl-th/0007021}
  {arXiv:nucl-th/0007021} \BibitemShut {NoStop}%
\bibitem [{\citenamefont {Warringa}(2006)}]{Warringa:2006dk}%
  \BibitemOpen
  \bibfield  {author} {\bibinfo {author} {\bibfnamefont {H.~J.}\ \bibnamefont
  {Warringa}},\ }\href@noop {} {\  (\bibinfo {year} {2006})},\ \Eprint
  {http://arxiv.org/abs/hep-ph/0606063} {arXiv:hep-ph/0606063} \BibitemShut
  {NoStop}%
\bibitem [{\citenamefont {Basler}\ and\ \citenamefont
  {Buballa}(2010)}]{Basler:2009vk}%
  \BibitemOpen
  \bibfield  {author} {\bibinfo {author} {\bibfnamefont {H.}~\bibnamefont
  {Basler}}\ and\ \bibinfo {author} {\bibfnamefont {M.}~\bibnamefont
  {Buballa}},\ }\href {\doibase 10.1103/PhysRevD.81.054033} {\bibfield
  {journal} {\bibinfo  {journal} {Phys. Rev. D}\ }\textbf {\bibinfo {volume}
  {81}},\ \bibinfo {pages} {054033} (\bibinfo {year} {2010})},\ \Eprint
  {http://arxiv.org/abs/0912.3411} {arXiv:0912.3411 [hep-ph]} \BibitemShut
  {NoStop}%
\end{thebibliography}%
\bibliographystyle{apsrev4-1}

\end{document}